\documentclass[12pt,a4paper]{{article}}
\usepackage{amssymb}
\usepackage[T2A]{fontenc}
\usepackage[cp866]{inputenc}
\usepackage[russian,english]{babel}
\normalbaselines

\usepackage[unicode,linktocpage,colorlinks,linkcolor=blue]{hyperref}
\begin{document}

\voffset=-2cm

\def\emline#1#2#3#4#5#6{%
      \put(#1,#2){\special{em:moveto}}
     \put(#4,#5){\special{em:lineto}}}
\def\newpic#1{}

\begin{center}
{\Large Self-interaction in classical gauge theories and gravitation}\\[3mm]
B. P. Kosyakov\\[3mm]
{\small  Russian Federal Nuclear Center--VNIIEF, Sarov, 607188 Nizhny Novgorod Region, Russia \\
and Moscow Institute of Physics {\&} Technology, Dolgoprudny, 141700 Moscow Region, Russia}\\[2mm]
{\small Electronic address: ${\rm kosyakov.boris@gmail.com}$}
\end{center}
\tableofcontents

\begin{abstract}
\noindent
{To develop a systematic treatment of the self-interaction problem in classical gauge
theories and general relativity, we study tenable manifestations of self-interaction: topological 
phases, and rearrangements of degrees of freedom appearing in the action.
We outline the occurrence of topological phases in pure field systems. 
We show that the rearranged Maxwell--Lorentz electrodynamics is a mathematically consistent and 
physically satisfactory theory which describes new entities, dressed charged particles and radiation. 
We extend this analysis to cover different modifications of the Maxwell--Lorentz electrodynamics and 
the SU$(N)$ Yang--Mills--Wong theory. 
We take a brief look at a subtle mechanism of self-interaction in classical strings. 
Turning to general relativity, we note that the total energy and momentum of a system with
nontrivial topological content, such as a black hole, are ambiguous, coordinatization-dependent 
quantities, which resembles the situation with paradoxical decompositions in the Banach--Tarski 
theorem.}
\end{abstract}


\section{INTRODUCTION}
\label
{Introduction}
The self-interaction problem was a central subject for study in fundamental physics of the 20th 
century.
This problem was especially pressing over a period from the late 1930s to the early 1970s in relation 
to the discovery of ultraviolet divergences in quantum field theory and subsequent effort to remove 
or avoid them.
The next page of this story was the use of the gained experience of handling theories which are 
suffered from the ultraviolet disease to elaborate criteria for discriminating between appropriate 
and inappropriate theories.
The quest culminated in establishing the Standard Model of particle physics.
With the advent of string theory things calmed down.
There comes a time when lessons from those stupendous developments could be drawn. 
Since there is an extensive literature covering quantum aspects of the problem, attention may be 
get to less analyzed classical aspects. 

Traditionally one inquires into the properties of classical self-interaction which share a number 
of traits with those of quantum self-interaction because the way for eliminating the ultraviolet 
problems from a classical system may hopefully suggest a cure for such troubles in the quantum 
incarnation of this system.
There are, however, notable classical properties of this phenomenon that bear no relation to the 
corresponding quantum properties. 
For example, the behavior of many classical self-interacting systems is irreversible, while the
associated quantum regime of evolution is reversible.
Both common and distinguishing properties of classical and quantum self-interacting systems are a 
major preoccupation of the present work.

Eighty years ago, Dirac \cite{Dirac1938} offered a thorough study of a classical radiating electron, which 
later furnished the most influential paradigm of self-interaction.
Technically, it is much easier to cope with this classical problem than with its quantum counterpart.
The classical theory of relativistic charged particles, the Maxwell--Lorentz electrodynamics, can be 
exactly solved. 
In contrast, four-dimensional relativistic quantum field theories, specifically quantum 
electrodynamics, defy all attempts to solve them exactly. 
The only reliable analytical tool to attack such theories is perturbation series in coupling 
constants, which, however, fails to grasp non-perturbative effects.
Meanwhile it is just these effects which are essential for the proper understanding of the 
self-interaction mechanisms.

\subsection{Elusive renditions of self-interaction}
\label
{rendition}
A distinctive feature of the classical picture is the coexistence of particles and fields 
mediating interactions between these particles.
For comparison, the fundamental notion of quantum field theory is a quantized field.
Excitations of quantum fields act as particles. 
Our interest here is with classical systems containing both particles and fields, as well as with 
pure field systems, that is, systems devoid of particles.
We begin with the former.

Abraham \cite{Abraham1903}, \cite{Abraham1904}, \cite{Abraham1905} and Lorentz \cite{Lorentz1904}, 
\cite{Lorentz1909} pioneered in applying the idea of self-interaction to a nonrelativistic model of 
the electron as a rigid body of finite extent. 
They tried to find the force of the electron on itself, that is, the resultant force due to 
different parts of the charge distribution, acting on one another.
In doing so they conceived of the electron, affected by this ``self-force'', as a warp-free body, 
namely a sphere of diameter $d$.

However, the very concept of {continuous} charged matter with a reasonably steady charge distribution 
is inconsistent in the classical context.
Each part of a lump of charged matter exerts a repulsive force on other parts, which cause the 
lump to become a rarefied medium.  
A homogeneous mixture of two oppositely charged fluids is also unstable: part of the mixture 
would collapse forming a neutral cluster, while the remainder possessing an uncompensated residual 
charge would spread.
Poincar\'e \cite{Poincare1906} conjectured that stable existence of continuous charged matter is ensured by 
{nonelectromagnetic cohesive forces}.             
A striking implication of this conjecture is that electrodynamics is {fundamentally unclosed}.
Among other things, electromagnetic self-interaction eludes analyzing separately from the 
Poincar\'e force contribution.

One may be inclined to think that the joint action of the cohesive and electromagnetic forces 
manifests itself in a rigidity condition. 
Alternatively, the joint action can be realized by calling into play an equation of state adapted to a 
deformable model of the electron. 

However, the existence of rigid bodies is contrary to the theory of relativity (see, {\it e.~g.}, 
\cite{LandauLifshitz1971}, ${\S\,15}$, where the results of a long discussion of this issue at the 
dawn of the age 
of this theory are briefly summarized), and therefore the rendition of self-interaction proposed by Abraham 
and Lorentz must be abandoned. 
More recent attempt to reinstate the rigid body model \cite{Caldirola1956} was based on the belief 
that there is a fundamental length $\ell$, about the same value as $d$, so that acausal signals are 
allowable in regions of size comparable with $\ell$.
There are two characteristic lengths related to the electron, the classical radius of the 
electron $r_0=e^2/mc^2=2.8\cdot 10^{-13}$ cm, and the Compton wave length of the electron
$\lambda_e=\hbar/mc=3.9\cdot 10^{-11}$ cm.
Both are not as small as is necessary to mark that a new physics is triggered.
The smallest length constructed from constants of nature (the velocity of light $c$, Planck's 
constant $\hbar$, and Newton's constant $G_{\rm N}$) is the Planck length $l_{\rm P}=(\hbar G_{\rm N}/c^3)^{1/2}=
1.6\cdot 10^{-33}$ cm.
In regions of size $\sim l_{\rm P}$, quantum fluctuations of the metric are expected to become 
significant, and the usual relations between cause and effect may not apply.
However, $l_{\rm P}$ is most likely to be unrelated to the electron size.
In addition, we are in the dark about the explicit form of violation of causality in the interior
of the electron; any argument of this kind appears highly speculative.

As for the deformable electron models, one has to resort to thermodynamical variables (pressure, 
temperature, {\it etc.}) which form the equation of state, but are foreign to the laws of 
microscopic dynamics. 
Hence, this approach is almost unfailingly accompanied by an {\it ad hoc} phenomenology, and 
furthermore, a tolerable equation of state is uncertain.

It remains to see whether $p$-branes, flexible extended objects with $p$ spatial dimensions, can be 
adapted for use as a pertinent classical model.
The habitat of $p$-branes is said to be sub-Planckian regions, alien to classical physics.
However, this type of extended objects is of great theoretical interest, and must be mentioned if
only for completeness of our discussion.
The points of a $p$-brane in ambient spacetime are given by $X^\mu(u^0,u^1,\ldots,u^p)$, and the 
action of a free $p$-brane is proportional to the  $(p+1)$-dimensional {world volume} swept out by 
this $p$-brane,
\begin{equation}
S=-T_p\int d^{p+1}\!u\,\sqrt{-h}\,.
\label
{action-p-brane}
\end{equation}                        
Here, $T_p$ is a constant necessary for rendering the action dimensionless, 
$h_{ab}=\partial_a X^\mu\,\partial_b X_\mu$, $a,b=0,1,\ldots,p$, is a metric on the world volume 
induced by the Lorentz metric of the ambient spacetime, and $h=\det\left(h_{ab}\right)$.
The dynamics of $p$-branes dispenses with the need for arbitrary, unjustifiable phenomenological 
assumptions; it is uniquely determined by the requirement that the action be the simplest action 
invariant under reparametrizations
\begin{equation}
u^a=f^a({\bar u})\,,
\quad
{h}_{ab}(u)=
\frac{\partial{\bar u}^{c}}{\partial u^a}\,\frac{\partial{\bar u}^{d}}{\partial u^b}\,{\bar h}_{cd}
({\bar u})\,,
\label
{diffeomorph-brane}
\end{equation}                        
where $f^a$ are arbitrary smooth functions.
The action (\ref{action-p-brane}) meets this requirement.

In four-dimensional spacetime, there are two species of $p$-branes: 1-branes (strings) and 2-branes 
(membranes).
Both are systems with infinite degrees of freedom.
Their dynamics share many features of field theory.
The study of these objects may be combined with that of pure field systems defined on curved  
$(p+1)$-dimensional manifolds.
 
We will see in Sec.~\ref{string} that classical strings exhibit a specific form of self-interaction: 
a free charged closed string is capable of spontaneous splitting into two such strings.
This phenomenon might be naturally interpreted as a manifestation of self-interaction of these 
extended classical objects.

Returning to the history, Frenkel \cite{Frenkel1925} was the first to argue that electromagnetism may be 
accounted for {by itself}, without resort to Poincar\'e cohesive forces.
He deemed the electron as a {point} in the precise geometric sense.
A point particle can be envisioned as a sphere of radius $r$ in the limit $r\to 0$.
Electrostatic repulsive forces are put to {distinct points} of the sphere, and, therefore, each part 
of the sphere tends to move away from other parts. 
However, all the repulsive forces are brought into a {single point} and {cancel} as $r\to 0$.
Therefore, a point charged particle is free from the explosion tendency, and the stable existence 
of such objects has no need of the cohesive force conjecture.

Inspired by Frenkel's idea, Dirac \cite{Dirac1938} gave its adequate mathematical formulation through the 
delta-function.
The idea of a point source of a field was a useful guide in the development of quantum field theory, 
and came up with the present paradigm of {local interactions} of quantized fields.
However, the problem of infinite self-energy was the price to pay for the conceptual simplicity and 
mathematical elegancy. 

Another impact of this idea is the necessity to sacrifice the rendition of self-interaction.
Taking the view of the electron as a structureless point particle, we have to abandon all attempts 
to conceive of the putative ``self-force'' which would combine infinitesimal repulsive forces 
contributed by different elements of the electron, and exclude this term from the pedagogical usage 
and physics folklore.
Since we do not have at our disposal a pictorial rendition of self-interaction, we are forced 
to content ourselves with the study of noticeable manifestations of self-interaction in the system 
``a charged point particle plus electromagnetic field''.

Turning to pure field systems, we find that the notion of self-interaction 
is far from clear.
The line of demarcation between interacting and free systems is often fuzzy. 
The behavior of any {free} field is believed to be governed by a linear equation with constant 
coefficients. 
A simple example is given by a real scalar field $\phi$ obeying the Klein--Gordon equation
\begin{equation}
\left(\Box+\mu^2\right)\phi=0\,,
\label
{KG-normal}
\end{equation}                          
which is derived from the Lagrangian quadratic in $\phi$,
\begin{equation}
{\cal L}=\frac12\left(\partial_\mu\phi\,\partial^\mu\phi\right)-\frac{\mu^2}{2}\,\phi^2\,.
\label
{free-lagr-scalar}
\end{equation}                          
However, the linearity is sometimes a matter of {convention}, which can be eliminated as the need 
arises.
To illustrate, we refer to the equation of motion for points of a string that becomes either linear or 
nonlinear according to which gauge condition is adopted.

A generic solution to Eq.~(\ref{KG-normal}) tells us that $\phi$ executes simple harmonic oscillations 
at every point of space.
On the other hand, if the Lagrangian involves powers of $\phi$ higher than quadratic, such as 
\begin{equation}
{\cal L}=\frac12\left({\partial}_\alpha{\phi}\,{\partial}^\alpha{\phi}\right)-
\frac{\mu^2}{2}\,\phi^2-\frac{\lambda^2}{4}\,\phi^4\,,
\label
{Klein-Gordon-plus-quartic}
\end{equation}                        
then the system is generally taken to be self-interacting because the Euler--Lagrange equations are 
{nonlinear}. 
The system governed by the Lagrangian (\ref{Klein-Gordon-plus-quartic}) executes anharmonic 
oscillations.
This behavior is qualitatively the same as that of the free system.
The only difference is that the period of harmonic oscillations is independent of their amplitudes,
while the period of anharmonic oscillations is amplitude-dependent.

The next complication concerning this notion follows from the fact that the system with quadratic 
Lagrangians can be converted to an ostensibly self-interacting system by a nonlinear field 
transformation,
\begin{equation}
{\phi}={\Phi}+{\Phi}^2F\left(\Phi\right).
\label
{Chisholm}
\end{equation}                          
Such transformations are very important in quantum field theory because they provide us with a 
subclass of nonrenormalizable field theories physically equivalent to renormalizable ones.
The main point, unveiled in \cite{Chisholm1961}, \cite{EfimovRutenberg1973}, and \cite{BergereLam1976}, 
is that two quantum field theories related by transformation (\ref{Chisholm}) have the same $S$ 
matrix. 
This statement is known in the literature as the ``equivalence theorem''.
A simple proof of this theorem, proposed in \cite{Blasi1999}, shows that the supposedly 
nonrenormalizable part of the resulting theory is actually a kind of gauge fixing, attributable to 
the cohomologically trivial sector of the theory.

Note also the absence of a clear-cut distinction between the notions of ``self-interaction'' and
``interaction between two fields'', as exemplified by the quartic term of a complex field 
$\phi=A+iB$ whose role can be understood as a single self-interaction term
$\frac14\,\lambda^2\left({\phi}^\ast{\phi}\right)^2$, or, alternatively, as the sum of two 
self-interaction terms $\frac14\, \lambda^2\left(A^4+B^4\right)$ and the term $\frac12\,\lambda^2 A^2B^2$ 
which contains mixed contribution of two real fields $A$ and $B$ and corresponds to their coupling.
Until the early 1970s the subnuclear zoo was divided into two classes: ``matter'', represented by
fermions, and ``fields'', represented by bosons, carriers of the fundamental forces of nature.
This classification might be substantiated by the statement that, in the classical limit, bosons are 
susceptible to the Bose--Einstein condensation, and hence the behavior of their collection bears 
a general resemblance to that of classical fields, while fermions,
which follow the Pauli blocking principle, share many traits with classical particles.
The advent of supersymmetry produced a dramatic change in that order.
A supersymmetric system accommodates field degrees of freedom of different kind, as well as forces 
between them, in a single self-interacting entity. 
 
\subsection{Manifestations of self-interaction}
\label
{heart}
A central idea of this paper is that self-interaction of a classical system shows itself in two
significant manifestations.
Those will be treated under the names ``topological phases'' and ``rearrangements of initial 
degrees of freedom appearing in the action'', or, shortly, ``rearrangement''.
To explicate these notions, consider a classical system whose states are described by generalized
variables $\phi_i$, and the dynamics is encoded by the action $S[\phi]$.
The behavior of the system is governed by the Euler--Lagrange equations resulted from the principle
of least action,
\begin{equation}
{\cal E}_i\left({\phi}\right)=\frac{\delta S}{\delta \phi_i}=0\,.
\label
{least_action}
\end{equation}                          

Suppose we are aware of joint solutions to the entire set of these equations, and among them 
there are physically relevant solutions, namely such that the energy of the system is finite.
If two or more solutions describe configurations of distinct topological structures,
then we are entitled to claim that the space of states has different phases, and relate the
existence of the topological phases to self-interaction of the system.
Section \ref{higgs_goldsone} outlines some simple pure field systems exhibiting topological phases.

While on the subject of systems which involves point particles and fields mediating interactions 
between these particles, we should recognize that an attempt to find a joint solution to the set of 
the Euler--Lagrange equations (\ref{least_action}) will in most cases end in a fiasco.
Section~\ref{rearrangement_ML} demonstrates that the Maxwell--Lorentz electrodynamics, 
formulated in terms of mechanical variables $z^\mu(s)$ describing world lines of bare charged point
particles and the electromagnetic vector potentials $A^\mu(x)$, experiences a blowup, which can 
be construed, in the spirit of quantum field theory, as a kind of ultraviolet divergence.
It will transpire that an interplay between degrees of freedom of bare particles and 
electromagnetic field rearranges the system giving rise to new entities, {dressed particles} and 
{radiation}, and that the rearranged dynamics is mathematically well-defined and physically reasonable.

Both manifestations of self-interaction can be combined in systems which contain point particles 
and non-Abelian gauge fields.
For example, in the Yang--Mills--Wong theory which describes $K$ particles carrying non-Abelian 
charges and interacting with the SU$(N)$ Yang--Mills field, $N\ge K+1$, we will observe two phases.
We will learn from Sec.~\ref{YMW} that these phases are invariant under different gauge groups,
SU$(N)$ and SL$(N,{\mathbb R})$, which are respectively the compact and a noncompact real forms of 
the complex group SL$(N,{\mathbb C})$, and that the system is rearranged differently in different 
phases.
In contrast, self-interaction of some systems may reveal itself by one of two manifestations. 
Classical gravitating systems are a good case in point.
They will be shown in Sec.~\ref{GR} to be capable of developing infinitely large number of phases 
but are unaffected by the rearrangement. 
 
Where do these manifestations of self-interaction come from? 
The general reason for their occurrence is that the system is unstable.
To be more specific, once $\phi_i$ is a joint solution to the set of the Euler--Lagrange equations 
(\ref{least_action}), the condition
\begin{equation}
\frac{\delta^2 S}{\delta \phi_i\,\delta \phi_j }<0\,
\label
{unstable_action}
\end{equation}                          
holds for some $i$ and $j$. 
Unstable modes tend to assemble into new stable modes.
Of course, for systems suffered from the ultraviolet disease,  the fact of their instability cannot 
be established directly through the use of (\ref{unstable_action}), and circumstantial evidence is 
required. 

It is interesting to compare this mechanism for displaying self-interaction of classical systems 
with what happens in the quantum picture.
The behavior of a quantum system can be described by the Feynman path integral.
Whatever the path with appropriate end points, it contributes to the path integral.
The principle of least action, Eq.~(\ref{least_action}), implying that the contribution of an 
extremal path dominates the path integral, is irrelevant to the quantum regime of evolution. 
Therefore, it is beyond reason to take the condition of instability (\ref{unstable_action}) as a 
prerequisite for rendering quantum self-interaction manifest.
In general, it would be wrong to place quantum systems into one of two categories, stable and 
unstable, because the notions of stability and instability make no sense outside the scope of the 
principle of least action.
Note also that the quantum and classical dressings are unrelated, even though they bear similar 
names.
We recall the reader that the vacuum polarization is of decisive importance for the quantum dressing, 
and that a cloud of virtual pairs of particles and antiparticles is dragged by a dressed quantum 
particle.
These phenomena are absent from the classical picture where the processes of creations and
annihilations of pairs of particles and antiparticles are strongly forbidden.
Perhaps the most outstanding distinction between the classical and quantum dynamics is that the 
latter is reversible, as exemplified by an electron which emits and absorbs photons with comparable 
probability amplitudes of these processes, whereas the former becomes irreversible after the 
rearrangement.

And yet the quantum picture shares a common trait with its classical relative, that of having vacuum 
expectation values of quantum variables $\langle 0|\phi|0\rangle$ governed by the action principle 
 \cite{Schwinger1951}.
Further still the condition of instability for quantum systems described in terms of $\langle 0|\phi|0\rangle$, 
Eq.~(\ref{unstable_action}), falls into the classical pattern \cite{Goldstone1961}. 

The term ``rearrangement'' was coined by Umezawa \cite{Umezawa1965} who looked at spontaneous symmetry breaking 
in the quantum context.
The mechanism for rearranging classical gauge field systems was then studied in a series of papers
\cite{Kosyakov1992}, \cite{Kosyakov1994}, \cite{Kosyakov1998}, \cite{Kosyakov1999}, \cite{Kosyakov2005}, 
\cite{Kosyakov2007}, \cite{Kosyakov2008a}, \cite{Kosyakov2008b}, and \cite{Chubykalo2017}, 
which are relied heavily on findings by Teitelboim \cite{Teitelboim1970}.

\subsection{Plan of the review}
\label
{layout}
This paper is written in a pedagogical manner. 
We restrict our attention to the simplest examples of classical self-interacting systems.
For those who wish to learn more about particular issues we provide links to original articles and 
other useful sources.
No attempt has been made to prepare a complete bibliography because this is a formidable task.

Our prime interest here is with conceptual aspects of the subject matter, rather than mathematical 
rigor, generality, and phenomenological utility.

The structure of the review is clear from the table of contents.
We briefly run through pure field systems exhibiting the availability of their topological phases in 
Secs.~\ref{higgs_goldsone} and \ref{pure}. 
A thorough analysis of such systems, in the light of spontaneous breakdown of symmetry, can be found 
in the existing literature.

We do not discuss self-interacting charged particles in curved manifolds \cite{DeWittBrehme1960},
\cite{Hobbs1968}, \cite{BarutVillaroel1975b}. 
Such a discussion would require rather sophisticated techniques, even though it has met with only 
limited success in gaining new insight into the self-interaction problem.

We do not mention many current studies, in particular, those related to gravitational self-interaction,  
which fall outside the purpose of this review for nonexperts.

We use units in which the speed of light and Planck's constant are taken to be unit throughout.
In Secs.~\ref{higgs_goldsone}--\ref{string} in which our concern is with the picture in Minkowski 
spacetime ${\mathbb R}_{1,3}$ we adopt the metric $\eta_{\mu\nu}={\rm diag}\left(1-1-1-1\right)$. 
When turning to pseudo-Riemannian manifolds in Sect.~\ref{GR}, we use the metric $g_{\mu\nu}(x)$ 
with the same signature.
In order to keep the conformity with the presentations of original research papers, we use 
interchangeably Gaussian and Heaviside units.

\section{TOPOLOGICAL PHASES}
\label
{higgs_goldsone}
To trace the advent of topological phases, we can conveniently discuss a simple prototype of the 
Goldstone model.
Consider a single real scalar field $\phi(t,x)$ in two dimensions whose dynamics is encoded by 
\begin{equation}
{\cal L}=\frac12\,(\partial_t\phi)^2-\frac12\,(\partial_x\phi)^2+\frac{\mu^2}{2}\,\phi^2
-\frac{\lambda^2}{4}\,\phi^4-{U}_0\,.
\label
{phi-4-real}
\end{equation}                          
The constant 
\begin{equation}
{U}_0=\frac14\,{\mu^2}\,\phi_0^2\,,
\label
{U_0}
\end{equation}                          
in which 
\begin{equation}
\phi_0=\frac{\mu}{\lambda}\,,
\label
{phi_0}
\end{equation}                          
is suitable for writing the Lagrangian in a succinct form:
\begin{equation}
{\cal L}=\frac12\,(\partial_t\phi)^2-\frac12\,(\partial_x\phi)^2-U(\phi)\,,
\label
{phi-4-real-symb}
\end{equation}                          
where
\begin{equation}
U(\phi)=\frac{\lambda^2}{4}\left(\phi^2-\phi_0^2\right)^2\,.
\label
{V-phi-4-reduc-om}
\end{equation}                          

The Lagrangian (\ref{phi-4-real-symb}) is invariant under reflection $\phi\to-\phi$.
However, the state $\phi=0$ realizing this symmetry is found to be unstable as soon as the principle of least 
action comes into effect.
Every minuscule perturbation remove the system from this state.
To put it differently, this state personifies a tachyon which, in the weak coupling limit $\lambda\to 0$, 
is governed by
\begin{equation}
\left(\frac{\partial^2}{\partial t^2}-\frac{\partial^2}{\partial x^2}-\mu^2\right)\phi=0\,.
\label
{eq-free-tachyon-}
\end{equation}                          

Since $\phi=0$ corresponds to unstable equilibrium, the {ground state} of the system, associated 
with the absolute minimum of the energy functional 
\begin{equation}
E=\int dx\,\left[\frac12\,(\partial_t\phi)^2+\frac12\,(\partial_x\phi)^2+{U}(\phi)\right],
\label
{energy-phi-4-real}
\end{equation}                       
is the state afforded by either of two solutions
\begin{equation}
\phi(t,x)=\pm\phi_0\,.
\label
{minima-phi}
\end{equation}                          
To see this, we note that the derivative terms in (\ref{energy-phi-4-real}) are minimized when 
$\phi$ is a constant.
This constant is specified by the minimum of ${U}(\phi)$.
For both of the solutions (\ref{minima-phi}), the energy (\ref{energy-phi-4-real}) is vanishing.
Assume that $\phi=+\phi_0$ furnishes the ground state.
Let $\chi$ be a small perturbation about $\phi_0$, 
\begin{equation}
\phi=\phi_0+\chi\,.
\label
{disturb-phi}
\end{equation}                          
Substituting (\ref{disturb-phi}) in (\ref{phi-4-real-symb}), we observe that the resulting Lagrangian 
\begin{equation}
{\cal L}=\frac12\,(\partial_t\chi)^2-\frac12\,(\partial_x\chi)^2-{\mu^2}\chi^2-
{\lambda}\mu\,\chi^3-\frac{\lambda^2}{4}\,\chi^4
\label
{phi-4-real-squ}
\end{equation}                          
exhibits an oscillatory mode with mass
\begin{equation}
m=\sqrt{2}\mu\,,
\label
{m=2mu}
\end{equation}                          
instead of the tachyon mode.
A similar mode appears in the phase associated with the solution $\phi=-\phi_0$. 

Therefore, the system executes almost periodic motions about either of two stable equilibrium 
points (\ref{minima-phi}).
The price for the stability is that the Lagrangian (\ref{phi-4-real-squ}) is not invariant 
under  reflection $\chi\to-\chi$.
This phenomenon, known as {spontaneous symmetry breaking}, is inherently classical because the 
criterion for discriminating between stable and unstable states stipulates that the principle of 
least action holds. 

There are two further topological phases corresponding to the so-called ``kink'' and ``antikink'' 
static solutions
\begin{equation}
\phi(x)=\pm\phi_0\,\tanh\!\left[\frac{\mu\left(x-x_0\right)}{\sqrt{2}}\right],
\label
{kink-antikink}
\end{equation}                          
which asymptotically approach either $\phi_0$ or $-\phi_0$ as $x\to\pm\infty$. 
The energy density of these configurations is localized near $x_0$:
\begin{equation}
\varepsilon(x)=\mu^2\,\frac{\phi_0^2}{2}\,{\rm sech}^{4}\!\left[\frac{\mu\left(x-x_0\right)}{\sqrt{2}}\right].
\label
{kink-energy-density}
\end{equation}                          
Accordingly, the total kink energy is finite,
\begin{equation}
E_{\rm kink}=\int^\infty_{-\infty}dx\,\varepsilon(x)=\mu\,\frac{2\sqrt{2}}{3}\,\phi_0^2\,.
\label
{kink-energy}
\end{equation}                          
The solutions (\ref{kink-antikink}) realize a local minimum of the energy functional 
(\ref{energy-phi-4-real}) in the sense that small perturbation about the kink (or antikink) are 
oscillatory modes associated with a bound state and scattering states, and a translation mode.
For a detailed discussion of the derivation and properties of these solutions see  
\cite{Coleman1975}, \cite{Rajaraman1975}, \cite{Rajaraman1982}, \cite{MantonSutcliffe2004}.

The field configuration space of all finite-energy solutions can be divided into four sectors, 
labelled by two indices 
\begin{equation}
\aleph_{\rm i}=\frac{\phi(x)}{\phi_0}\biggl|_{x=-\infty}\,,\quad 
\aleph_{\rm f}=\frac{\phi(x)}{\phi_0}\biggl|_{x=\infty}\,. 
\label
{indices}
\end{equation}                          
These sectors are topologically unconnected.
The trivial solutions ${\pm\phi_0}$ are characterized by $\aleph_{\rm i}=\aleph_{\rm f}=1$
and $\aleph_{\rm i}=\aleph_{\rm f}=-1$, respectively, the kink is marked by $\aleph_{\rm i}=-1$,
$\aleph_{\rm f}=1$, and the antikink by $\aleph_{\rm i}=1$, $\aleph_{\rm f}=-1$.
Fields of one sector cannot be distorted continuously into another. 
To switch between such sectors, it is necessary to leap over the potential barrier of height 
$\sim U_0L$, where $U_0$ is given by (\ref{U_0}), and the system is assumed to be in a large box 
whose length $L$ tends to $\infty$.  
Since time evolution is an example of continuous distortion, a field configuration from any one 
sector stays within that sector as time passes. 

Another way for identifying the topological phases is to use the topological charge
\begin{equation}
{\cal Q}= \aleph_{\rm f}-\aleph_{\rm i}\,.
\label
{topological_charge_df}
\end{equation}                          
The corresponding topological current
\begin{equation}
{\cal J}^\mu=\phi_0^{-1} \epsilon^{\mu\nu}\partial_\nu\phi\,,
\label
{topological_current}
\end{equation}                          
where $\epsilon^{\mu\nu}$ is the two-dimensional Levi-Civita symbol, obeys the local conservation law 
\begin{equation}
\partial_\mu {\cal J}^\mu=0\,
\label
{topological_current_conserv}
\end{equation}                          
for any $\phi$.
Therefore, the spatial integral of ${\cal J}^0$ is a conserved quantity identical to ${\cal Q}$,
\begin{equation}
\int^\infty_{-\infty} dx\,{\cal J}^0(x)
=\phi_0^{-1} \int^\infty_{-\infty} dx\,\frac{\partial\phi}{\partial x}
= \aleph_{\rm f}-\aleph_{\rm i}\,.
\label
{topological_charge}
\end{equation}                          
The kink and antikink phases are endowed with ${\cal Q}=2$ and ${\cal Q}=-2$, respectively, and both
trivial solutions (\ref{minima-phi}), realizing the degenerate ground state, have ${\cal Q}=0$.

Surprisingly, the four-dimensional analog of the system (\ref{phi-4-real}) does not display nontrivial 
topological phases associated with kink-like solutions, even though the system is unstable, and 
hence experiences spontaneous breakdown of symmetry. 
A general statement \cite{Derrick1964} is that localized static solutions of the system governed by
\begin{equation}
{\cal L}=\frac12\left(\partial_\mu\phi\, \partial^\mu\phi\right)-U(\phi)\,,
\label
{derrick_phi-4-real}
\end{equation}                          
are unstable for nonnegative smooth functions $U(\phi)$ with $U'(0)=0$.
This argument is extendable to higher dimensions. 
Furthermore, exponential instability of localized static
solutions of these systems was established in \cite{KarageorgisStrauss2007}.

In going from the above systems with a real scalar field $\phi$ to systems with a {complex} 
scalar field $\Phi$, the discrete symmetry $\phi\to -\phi$ is changed for a continuous symmetry
$\Phi\to e^{i\theta}\Phi$.
Accordingly the topological charge ${\cal Q}$ is substituted for the so-called 
winding number $n$ distinguishing different homotopy classes in mapping circles into circles.

Kink-like solutions are lacking in some of these systems.
To illustrate, we refer to the {Goldstone model}, 
\begin{equation}
{\cal L}=\frac12\left(\partial_\mu{\Phi}\right)^\ast\partial^\mu\Phi
+\frac{\mu^2}{2}\left({\Phi}^\ast\Phi\right)
-\frac{\lambda^2}{4}\left({\Phi}^\ast\Phi\right)^2-U_0\,,
\label
{phi-4-comp}
\end{equation}                          
where $\Phi$ is a {complex} scalar field.
Although the sign of the term $\frac12\,\mu^2\left({\Phi}^\ast\Phi\right)$ is such that $\Phi$ 
behaves as a tachyon, the model is devoid of kink-like solutions.

In contrast, solutions of this kind are peculiar to the {Higgs model} 
\cite{Higgs1964}
\begin{equation}
{\cal L}=-\frac{1}{16\pi}\left(F^{\mu\nu}F_{\mu\nu}\right)+\frac12\left({D_\mu\Phi}\right)^\ast\,
D^\mu\Phi+
\frac{\mu^2}{2}\left({\Phi}^\ast\Phi\right)-
\frac{\lambda^2}{4}\left({\Phi}^\ast\Phi\right)^2-U_0\,.
\label
{lagr-Higgs}
\end{equation}                          
Here $F_{\mu\nu}=\partial_\mu A_\nu-\partial_\nu A_\mu$, $D_\mu\phi=(\partial_\mu-ie A_\mu)\,\Phi$,
that is, (\ref{lagr-Higgs}) describes the interaction of a complex scalar field $\Phi$ with 
electromagnetic field.
Vortex-line solutions of this model, similar to the vortex line in a superconductor, were shown 
\cite{NielsenOlesen1973} to form topologically nontrivial phases labelled by winding number $n$. 
The reader interested in this topics would do best to consult the books \cite{Rajaraman1982}, 
\cite{Schwarz1991}, \cite{MantonSutcliffe2004}.

\section{SELF-INTERACTION IN ELECTRODYNAMICS}
\label
{self-interaction_ED}

\subsection{The Maxwell--Lorentz theory}
\label
{rearrangement_ML}
The system ``a charged particle plus electromagnetic field'' is described by the action
\begin{equation}
S=S_{\rm PP}+S_{\rm S}+S_{\rm L}\,,
\label
{S-Maxwell-Lorentz}
\end{equation}                        
where the Poincar\'e--Planck term 
\begin{equation}
S_{\rm PP}=-m_0\int d\tau\,\sqrt{{\dot z}^\mu\,{\dot z}_\mu}\,
\label
{S-Planck-general}
\end{equation}                        
governs the particle; the Schwarzschild term
\begin{equation}
S_{\rm S}=-e\int d\tau\,{\dot z}^\mu A_\mu(z)\,
\label
{S-Schwarzs}
\end{equation}                        
is responsible for the interaction of the particle and the field; and the Larmor term 
\begin{equation}
S_{\rm L}=-\frac{1}{16\pi}\int d^4x\, F_{\mu\nu} F^{\mu\nu}\,
\label
{S-em-field}
\end{equation}                        
encodes the field dynamics. 
The parameter $\tau$ is associated with evolution of the particle.
Derivatives with respect to $\tau$ are denoted by dots. 
$m_0$ stands for {mechanical} mass of the particle.

Does the extremalization of this action make the system unstable?
A direct way for tackling the question would be to obtain a joint solution to the Euler--Lagrange 
equations 
\begin{equation}
{\cal E}_\mu(x)=\partial^\nu F_{\mu\nu}(x)+4\pi e\int^\infty_{-\infty}\!ds\,{v}_\mu(s)\,
\delta^4\left[x-z(s)\right]=0\, 
\label
{Eulerian-em-II}
\end{equation}                        
and
\begin{equation}
\varepsilon^\lambda(z)=m_0 {a}^\lambda-e {v}_\mu F^{\lambda\mu}(z)=0\,
\label
{Eulerian-part}
\end{equation}                        
(where $s$ denotes the proper time, $v^\mu={dz^\mu}/{ds}$ is the four-velocity, 
and $a^\mu={dv^\mu}/{ds}$ is the four-acceleration), supplemented with the Bianchi identity
\begin{equation}
{\cal E}^{\lambda\mu\nu}=\partial^\lambda F^{\mu\nu}+
\partial^\nu F^{\lambda\mu}+\partial^\mu F^{\nu\lambda}=0\,,
\label
{Eulerian-em-I}
\end{equation}                        
and examine the variation of Eqs.~(\ref{Eulerian-em-II}) and (\ref{Eulerian-part}) about this 
solution.
To accomplish this plan, one should first find the joint solution, that is, taking the retarded 
condition and assuming that $z^\mu(s)$ is an arbitrary smooth timelike curve, solve Eqs.~(\ref{Eulerian-em-I}) 
and (\ref{Eulerian-em-II}); substitute the retarded solution $F^{\lambda\mu}$
into Eq.~(\ref{Eulerian-part}); and solve the resulting equation.
But applying the solution $F^{\lambda\mu}$ to (\ref{Eulerian-part}), we obtain a divergent expression.
The occurrence of this divergency can be thought of as if the field degrees of freedom are induced 
by the extremality of the action to attain a singularity on the world line, and the particle tends 
to blow up due to infinite concentration of the energy of its own field.
But if it is granted that the divergency is eliminated, through the renormalization procedure, the 
blowup appears to be suppressed. 

Therefore, care is required in analyzing the self-interaction problem in field theories with
delta-function sources.
An appropriate starting point for this analysis is a Noether identity \cite{Noether1918}, which, 
as applied to the action (\ref{S-Maxwell-Lorentz})--(\ref{S-em-field}), takes the form: 
\begin{equation}
\left(\partial_\mu T^{\lambda\mu}\right)(x)
=
{1\over 8\pi}\left({\cal E}^{\lambda\mu\nu}F_{\mu\nu}\right)(x)
+
{1\over 4\pi}\left({\cal E}_\mu F^{\lambda\mu}\right)(x)
+
\int^\infty_{-\infty}ds\,\varepsilon^\lambda(z)\,\delta^4\left[x-z(s)\right].
\label
{Noether-1-id-em}
\end{equation}                         
Here, $T^{\mu\nu}$ is the symmetric stress-energy tensor of this system, 
\begin{equation}
T^{\mu\nu}
=
\frac{2}{\sqrt{-g}}\frac{\delta S}{\delta g_{\mu\nu}}\biggl|_{g_{\mu\nu}=\eta_{\mu\nu}}
=
\Theta^{\mu\nu}+t^{\mu\nu}\,,
\label
{T=Theta+t}
\end{equation}                         
\begin{equation}
\Theta^{\mu\nu}={1\over 4\pi}\left(F^{\mu\alpha}F_\alpha^{~\nu}
+
{1\over 4}\,\eta^{\mu\nu}F_{\alpha\beta}F^{\alpha\beta}\right),
\label
{Theta-mu-nu}
\end{equation}                           
\begin{equation}
t^{\mu\nu}= m_0\int^\infty_{-\infty}\! ds\,{v}^\mu(s)\,{v}^\nu(s)\,\delta^4\left[x-z(s)\right],
\label
{t-mu-nu}
\end{equation}                         
and ${\cal E}^{\lambda\mu\nu}$, ${\cal E}_\mu$, $\varepsilon^\lambda$ are, 
respectively, the left-hand sides of equations (\ref{Eulerian-em-I}), (\ref{Eulerian-em-II}), 
(\ref{Eulerian-part}).
The derivation of Eqs.~(\ref{Noether-1-id-em})--(\ref{t-mu-nu}) has been detailed, {\it e.~g.,} in
\cite{Kosyakov2007}.

Note that the Noether identity (\ref{Noether-1-id-em}) does not stipulate that the action is
extremal.
The responsibility for the fulfilment of Eq.~(\ref{Noether-1-id-em}) only rests with translation 
invariance.

Were it not for the divergency, the equation 
\begin{equation}
\partial_\mu T^{\lambda\mu}=0 
\label
{Total-conserv-law}
\end{equation}                           
would imply ${\cal E}^{\lambda\mu\nu}=0$, ${\cal E}_\mu=0$, and $\varepsilon^\lambda=0$, that is, 
the local conservation law for the stress-energy tensor is formally equivalent to the equation of 
motion for a bare particle (\ref{Eulerian-part}) in which a solution to the field equations 
(\ref{Eulerian-em-I}) and (\ref{Eulerian-em-II}) is used. 
However, Eq.~(\ref{Total-conserv-law}) provides a more penetrating insight into the rearrangement of 
the Maxwell--Lorentz theory because $\Theta^{\mu\nu}$ can be segregated into terms with {integrable} and 
{nonintegrable} singularities.

\subsubsection{Radiation}
\label
{Radiation}
Let a point charge be moving along a smooth timelike world line $z^\mu(s)$.
The retarded field $F^{\mu\nu}$ generated by this charge can be written \cite{Dirac1938} as 
\begin{equation}
F^{\mu\nu}=\frac{e}{(R\cdot v)^3}\left({R}^\mu{U}^\nu-{R}^\nu{U}^\mu\right),
\label
{F-mu-nu-LW-III}
\end{equation}                          
\begin{equation}
U^\mu=\left(1-{a}\cdot R\right){v}^\mu+\left(R\cdot{v}\right){a}^\mu\,,
\label
{U-df}
\end{equation}                       
Here, 
\begin{equation}
R^\mu=x^\mu-{z}^\mu(s_{\rm ret}) 
\label
{R-mu-df}
\end{equation}                         
is a null vector drawn from a point ${z}^\mu(s_{\rm ret})$ on the world line, where the electromagnetic 
signal was emitted, to the point $x^\mu$, in which the signal was received;  the four-velocity 
$v^\mu$ and four-acceleration $a^\mu$ refer to the retarded instant 
$s_{\rm ret}$.
Further retarded covariant variables:  invariant distance $\rho$ between 
$x^\mu$ and ${z}^\mu(s_{\rm ret})$, which is actually the distance measured in the instantaneously 
comoving Lorentz frame at $s=s_{\rm ret}$, 
\begin{equation}
\rho=R\cdot v\,,
\label
{rho-df}
\end{equation}                         
a retarded scalar 
\begin{equation}
\lambda=a\cdot R-1\,,
\label
{lambda-def}
\end{equation}         
and a null vector $c^\mu$ aligned with $R^\mu$, 
\begin{equation}
R^\mu=\rho c^\mu\,, 
\label
{c-mu-df}
\end{equation}                         
are convenient to use for the present discussion.
$c^\mu$ can be represented as the sum of two orthogonal to each other normalized vectors,
\begin{equation}
c^\mu=v^\mu+u^\mu\,, 
\label
{u-mu-df}
\end{equation}                         
where $u^\mu$ is an imaginary-unit vector directed from ${z}^\mu(s_{\rm ret})$ to $x^\mu$, 
\begin{equation}
c^2=0\,,
\quad
v^2=-u^2=1\,,
\quad
u\cdot v=0\,, 
\quad
c\cdot v=-c\cdot u=1\,.
\label
{c,u,v_df}
\end{equation}                          
With these definitions, Eqs.~(\ref{F-mu-nu-LW-III}) and (\ref{U-df}) become 
\begin{equation}
F^{\mu\nu}=\frac{e}{\rho^2}\left(c^\mu U^\nu-c^\nu U^\mu\right), 
\label
{F-LW}
\end{equation}                          
\begin{equation}
U^\mu=-\lambda{v^\mu}+{\rho}{a^\mu}\,.
\label
{U}
\end{equation}                       

It is common to decompose this field into two parts, $F=F_{\rm\hskip0.3mm I}+F_{\rm\hskip0.3mm II}$, where
\begin{equation}
F_{\rm\hskip0.3mm I}=\frac{e}{\rho^2}\,{c}\wedge{v}\,,
\label
{F-I}
\end{equation}                          
\begin{equation}
F_{\rm\hskip0.3mm II}=\frac{e}{\rho}\,
{c}\wedge{\left[a-v\left(a\cdot c\right)\right]}\,,
\label
{F-II}
\end{equation}                       
and regard $F_{\rm\hskip0.3mm I}$ as a ``generalized Coulomb field'', and $F_{\rm\hskip0.3mm II}$ as 
the ``radiation field''.
However, this separation is of no utility: whatever the motion of the charge, there is a Lorentz 
frame, special for each point $x^\mu$, in which $F_{\rm\hskip0.3mm II}$ is completely eliminated, 
and only $F_{\rm\hskip0.3mm I}$ persists.
This is clear from the mere fact that ${\cal P}=\frac12 F_{\mu\nu}{}^\ast\!F^{\mu\nu}=0$,  
${\cal S}=\frac12 F_{\mu\nu}F^{\mu\nu}=-e^2/\rho^4$ for the field $F^{\mu\nu}$ defined by 
Eqs.~(\ref{F-LW}) and (\ref{U}). 
  
To indicate explicitly the frame of reference in which $F_{\rm\hskip0.3mm II}=0$, rewrite
(\ref{F-LW}) as
\begin{equation}
F=\frac{e}{\rho^2}\,\varpi\,,
\quad
\varpi={c}\wedge{U}\,.
\label
{F-LW-III}
\end{equation}                          
A rendition of the bivector $\varpi$ is the parallelogram of the vectors $c^\mu$ and $U^\mu$, with
the area of the parallelogram being equal to 1.  
The bivector $\varpi$ is invariant under the special linear group of real unimodular transformations 
SL$(2,{\mathbb R})$ which rotate and deform the initial parallelogram in the plane spanned by 
$c^\mu$ and $U^\mu$, converting it to parallelograms of unit area.  
Therefore, $\varpi$ is independent of directions and magnitudes of the constituent vectors, it 
depends only on the parallelogram's orientation.
The parallelogram can always be built from a timelike unit vector $e_0^\mu$ and a spacelike 
imaginary-unit vector $e_1^\mu$ perpendicular to $e_0^\mu$, $\varpi={e}_0\wedge{e}_1$.
There are three different cases: 

\noindent
(a) $U^2>0$,  
\begin{equation}
e_0^{\mu}=\frac{U^{\mu}}{\sqrt{U^2}}\,,
\quad 
e_1^{\mu}=\sqrt{U^2}\left(-c^{\mu}+\frac{U^\mu}{U^2}\right),
\label
{1.}
\end{equation}                         
(b) $U^2<0$, 
\begin{equation}
e_0^{\mu}=\sqrt{-U^2}\left(c^{\mu}-\frac{U^\mu}{U^2}\right),
\quad 
e_1^\mu=\frac{U^{\mu}}{\sqrt{-U^2}}\,,
\label
{2.}
\end{equation}                          
(c) $U^2=0$,
\begin{equation}
e_0^{\mu}=\frac{1}{\sqrt{2}}\left(c^{\mu}-U^{\mu}\right), 
\quad 
e_1^{\mu}=\frac{1}{\sqrt{2}}\left(c^{\mu}+U^{\mu}\right).
\label
{3.}
\end{equation}                          
In the Lorentz frame with the time axis parallel to $e^\mu_0$, all components of the
$F^{\mu\nu}$ are vanishing, except for $F^{\hskip0.2mm 01}$ which behaves as $\rho^{-2}$.
Equations (\ref{1.})--(\ref{3.}) explicitly specify a frame in which the retarded electromagnetic 
field generated by a single arbitrarily moving charge appears as a pure Coulomb field at each 
observation point.
With a curved world line, this frame is noninertial.

The SL$(2,{\mathbb R})$ transformations can be carried out independently at any spacetime point.
We are thus dealing with {local} transformations.
The invariance of $F^{\mu\nu}$ is not pertinent to electrodynamics as a whole, and hence gives rise 
to no Noether identities.
Rather, this is a property of the {retarded solution} to Maxwell's equations $F_{\rm ret}$.
The advanced solution $F_{\rm adv}$ can also be put in the form similar to (\ref{F-LW})--(\ref{U}),
that is, $F_{\rm adv}$ is decomposable, whereas combinations $\alpha F_{\rm ret}+\beta F_{\rm adv}$ 
are not.

It is thus seen that the notion of radiation field is problematic: the segregation between 
parts of the retarded field scaling as $\rho^{-2}$ and $\rho^{-1}$ is disavowed by the local 
SL$(2,{\mathbb R})$ invariance of the Li\'enard--Wiechert solution (\ref{F-LW})--(\ref{U}).
Under these circumstances one may look at the stress-energy tensor $\Theta^{\mu\nu}$ for clues.
A motivation for this is that $\Theta^{\mu\nu}$ is frame-dependent.
It is natural to accommodate $\Theta^{\mu\nu}$ to Lorentz frames with rectangular coordinates.
Substituting (\ref{F-LW})--(\ref{U}) into (\ref{Theta-mu-nu}) gives
\begin{equation}
\Theta^{\mu\nu}= {e^2\over 4\pi\rho^4}\left(c^\mu U^\nu +
c^\nu U^\mu -U^2 c^\mu c^\nu-{1\over 2}\,\eta^{\mu\nu}\right),
\label
{Theta-via-LW}
\end{equation}                         
which is split into nonintegrable and integrable parts, 
$\Theta^{\mu\nu}=\Theta^{\mu\nu}_{\rm\hskip0.3mm I}+\Theta^{\mu\nu}_{\rm\hskip0.3mm II}$, 
\begin{equation}
\Theta^{\mu\nu}_{\rm\hskip0.3mm I}={e^2\over 4\pi\rho^4}\left(c^\mu U^\nu +
c^\nu U^\mu -c^\mu c^\nu-{1\over 2}\,\eta^{\mu\nu}\right),
\label
{Theta-I}
\end{equation}                         
\begin{equation}
\Theta^{\mu\nu}_{\rm\hskip0.3mm II}=-{e^2\over 4\pi\rho^2}\left[a^2+(a\cdot c)^2\right]
c^\mu c^\nu\,.
\label
{Theta-II}
\end{equation}                           

One can show \cite{Teitelboim1970} that two local conservation laws hold outside the world line: 
\begin{equation}
\partial_\mu\Theta^{\mu\nu}_{\rm\hskip0.3mm I}=0\,,
\quad
\partial_\mu \Theta_{{\rm\hskip0.3mm II}}^{\mu\nu}=0\,,
\label
{cons-Theta-I-II}
\end{equation}                                       
which suggest that $\Theta^{\mu\nu}_{\rm\hskip0.3mm I}$ and $\Theta^{\mu\nu}_{\rm\hskip0.3mm II}$
are dynamically independent off the world line.
Let us compare the properties of $\Theta^{\mu\nu}_{\rm\hskip0.3mm I}$ and $\Theta^{\mu\nu}_{\rm\hskip0.3mm II}$.
We begin with the latter. 

$\Theta^{\mu\nu}_{{\rm\hskip0.3mm II}}$ leaves the source at the speed of light.
Indeed, the surface element of the future light cone $C_+$ drawn from ${z}^\mu(s_{\rm ret})$ is 
$d\sigma^\mu=c^\mu\rho^2 d\rho\,d\Omega$.
Since $c^\mu$ is a null vector, the flux of $\Theta^{\mu\nu}_{{\rm\hskip0.3mm II}}$ through $C_+$ 
vanishes, $d\sigma_\mu\Theta^{\mu\nu}_{{\rm\hskip0.3mm II}}=0$,
implying that $\Theta^{\mu\nu}_{\rm\hskip0.3mm II}$ propagates along rays of $C_+$.
The energy-momentum flux associated with $\Theta^{\mu\nu}_{{\rm\hskip0.3mm II}}$ varies as $\rho^{-2}$, 
which means that the same amount of energy-momentum flows through spheres of different radii.
It is also significant that if the motion is uniform, $a^\mu=0$, then $\Theta^{\mu\nu}_{{\rm\hskip0.3mm II}}=0$.

None of these features is shared by $\Theta^{\mu\nu}_{{\rm\hskip0.3mm I}}$.
Let $d\sigma^\mu$ be the surface element of the future light cone ${C}_{+}$, then 
\begin{equation}
d\sigma_\mu\,\Theta^{\mu\nu}_{{\rm\hskip0.3mm I}}
=\frac{e^2}{8\pi\rho^4}\,c^\nu \rho^2 d\rho\,d\Omega\,.
\label
{c-Theta-I}
\end{equation}
The flux of $\Theta^{\mu\nu}_{{\rm\hskip0.3mm I}}$ through $C_+$ is nonzero, and hence $\Theta^{\mu\nu}_{{\rm\hskip0.3mm I}}$ 
moves slower than light.
One may conclude that $\Theta^{\mu\nu}_{{\rm\hskip0.3mm II}}$ detaches from the source, while 
$\Theta^{\mu\nu}_{{\rm\hskip0.3mm I}}$ remains bound to it.
It is clear from (\ref{Theta-I}) and (\ref{U}) that $\Theta^{\mu\nu}_{{\rm\hskip0.3mm I}}$ falls 
with distance at least as $\rho^{-3}$. 
Therefore, $\Theta^{\mu\nu}_{{\rm\hskip0.3mm I}}$ yields the flux of energy-momentum which dies 
out with distance. 
Furthermore,  $\Theta^{\mu\nu}_{{\rm\hskip0.3mm I}}$ is nonvanishing for any motion of the source.
In other words,  $\Theta^{\mu\nu}_{{\rm\hskip0.3mm I}}$ represents a part of the electromagnetic 
energy-momentum that is dragged by the charge.

The integration of $\Theta^{\mu\nu}_{{\rm\hskip0.3mm I}}$ 
over a three-dimensional surface intersecting the world line results in a divergent expression.
In the language of quantum field theory, such expressions are known as ``ultraviolet divergent''. 
The mathematical reason for ultraviolet divergences is that the product of tempered distributions 
with coincident supports is ill-defined \cite{Bogoliubov1951}.

We will thereafter refer to a symmetric tensor as radiation, and denote it by $\Theta^{\mu\nu}_{{\rm\hskip0.3mm II}}$, 
if 
\begin{equation}
({\rm i})\quad\partial_\mu \Theta_{{\rm\hskip0.3mm II}}^{\mu\nu}=0\,,
\label
{i}
\end{equation}                                       
\begin{equation}
({\rm ii})\quad c_\mu\Theta^{\mu\nu}_{{\rm\hskip0.3mm II}}=0\,,
\label
{ii}
\end{equation}
\begin{equation}
({\rm iii})\quad\Theta^{\mu\nu}_{\rm II}\sim \rho^{-2}\,.
\label{iii}
\end{equation}                                           

It is conceivable that the energy flux produced by $\Theta^{\mu\nu}_{{\rm\hskip0.3mm II}}$ 
is directed inward towards the field source resulting in energy gain rather than energy loss. 
One may regard this as the {absorption} of radiation rather than its emission.  
An alternate view is that the emitted energy is {negative}:
$\Theta^{00}_{{\rm\hskip0.3mm II}}=v_\mu\Theta^{\mu\nu}_{{\rm\hskip0.3mm II}}v_\nu<0$.
An example can be drawn from Sect.~\ref{YMW} where the self-interaction problem in the 
Yang--Mills--Wong theory is analyzed.
There is no universally adopted terminology that distinguishes between $\Theta^{00}_{{\rm\hskip0.3mm II}}>0$ and
$\Theta^{00}_{{\rm\hskip0.3mm II}}<0$.
We normally reserve the term ``radiation'' for the case that the emitted energy is positive.

Making switch from four-dimensional electrodynamics to that in $d$ dimensions, we can apply this 
analysis if we replace a sphere enclosing the source by a $(d-2)$-dimensional sphere.
Then condition (iii) becomes 
\begin{equation}
({\rm iii})\quad \Theta^{\mu\nu}_{\rm II}=O\left(\rho^{2-d}\right),
\quad
\rho\to\infty\,.
\label
{iii-D}
\end{equation}                                           
In addition,  $\Theta^{\mu\nu}_{\rm\hskip0.3mm I}$ should fall more rapidly than 
$\Theta^{\mu\nu}_{\rm\hskip0.3mm II}$ to ensure that $\Theta^{\mu\nu}_{\rm II}$ be distinguished 
asymptotically from $\Theta^{\mu\nu}_{\rm I}$,
\begin{equation}
({\rm iv})\quad\Theta^{\mu\nu}_{\rm\hskip0.3mm I}=o\left(\rho^{2-d}\right),
\quad
\rho\to\infty\,.
\label
{iv}
\end{equation}                                           

The radiated energy-momentum is defined by
\begin{equation}
{\cal P}^{\mu}=
\int_{\Sigma} d\sigma_\nu\,\Theta^{\mu\nu}_{\rm II}\,,
\label
{4D-radiat}
\end{equation}                                       
where $\Sigma$ is a three-dimensional spacelike surface intersecting the world line.
Since $\Theta^{\mu\nu}_{\rm II}$ involves only integrable singularities, and
$\partial_\nu\Theta^{\mu\nu}_{\rm II}=0$, the surface of integration $\Sigma$ in (\ref{4D-radiat}) 
may be chosen arbitrarily.
It is convenient to deform $\Sigma$ to a tubular surface ${T}_\epsilon$ of small invariant radius 
$\rho=\epsilon$ enclosing the world line. 
The surface element on this tube is $d\sigma^\mu=\partial^\mu\!\rho\,\rho^{2}\,d\Omega\,ds=(v^\mu+
\lambda c^\mu)\,\epsilon^{2}\,d\Omega\,ds$.
Inserting (\ref{Theta-II}) into (\ref{4D-radiat}) gives 
\begin{equation}
{\cal P}^\mu=-\frac{e^2}{4\pi}
\int^s_{-\infty}d\tau
\int d\Omega\left[a^2+(a\cdot u)^2\right]  c^\mu\,.
\label
{4D-radiation-rate}
\end{equation}                         
The solid angle integration is simple.
One only need to apply the evident formulas
\begin{equation}
\int d\Omega=4\pi\,,
\quad
\int d\Omega\, u_\mu=0\,,
\quad
\int d\Omega\, u_\mu u_\nu=-\frac{4\pi}{3}\,\stackrel{\scriptstyle v}{\bot}_{\hskip0.5mm\mu\nu}\,,
\label
{solid-int}
\end{equation}                                       
where $\stackrel{\scriptstyle v}{\bot}$ is the projection operator on a hyperplane with normal $v^\mu$,
\begin{equation}
\stackrel{\scriptstyle v}{\bot}_{\mu\nu}=\eta_{\mu\nu}-\frac{v_\mu v_\nu}{v^2}\,.
\label
{projection-operator-df}
\end{equation}                 
The result of integration is
\begin{equation}
{\cal P}^{\mu}=
-{2\over 3}\,e^2\int^{{s}}_{-\infty}d\tau\,a^2(\tau) v^\mu(\tau)\,.
\label
{P-rad}
\end{equation}                        

For this expression to be convergent, the integrand must fall off sufficiently rapidly as  
$s\to-\infty$.
The pertinent asymptotic condition, formulated by Haag \cite{Haag1955}, states: the motion of every charged particle must 
asymptotically approach a uniform regime in the remote past,
\begin{equation}
\lim_{s\to -\infty} a^\mu(s)=0\,.
\label
{a-to-0}
\end{equation}                        
With this asymptotic condition, Eq.~(\ref{P-rad}) represents the four-momentum emitted by the source
over the period from the remote past to the instant $s$.
Differentiating  (\ref{P-rad}) with respect to $s$ we obtain the four-momentum emitted by an accelerated 
charge per unit proper time: 
\begin{equation}
{{\dot{\cal P}}^{\mu}}=-{2\over 3}\,e^2 a^2 v^\mu\,.
\label
{dP-II}
\end{equation}                        
This is the relativistic generalization \cite{Heaviside1902} of the famous {Larmor formula} 
\cite{Larmor1897},
\begin{equation}
\frac{d{\cal E}}{dt}=
{2\over 3}\,e^2\,{\bf a}^2\,,
\label
{dE-Larmor}
\end{equation}                        
describing the rate of radiated energy in an instantaneously comoving Lorentz frame. 

Equation~(\ref{dE-Larmor}) shows that ${d{\cal E}}/{dt}>0$ to evidence that the emission of 
radiation is a dissipative, and hence,  unidirectional process.

The concept of electromagnetic radiation grew up over a long period.
Our interest here is with the definition of radiation developed by Teitelboim \cite{Teitelboim1970}. 
It was argued in \cite{Kosyakov1992}, \cite{Kosyakov1998}, \cite{Kosyakov2007} that only this 
definition can be correctly applied to the Yang--Mills--Wong theory.

\subsubsection{Local balance of energy-momentum}
\label
{balance}
Since $\Theta^{\mu\nu}_{{\rm\hskip0.3mm I}}$ contains $\rho^{-3}$ and $\rho^{-4}$, 
this part of the electromagnetic stress-energy tensor is nonintegrable over three-dimensional 
surfaces intersecting the world line.
An appropriate regularization is called for.
With a Lorentz-invariant cutoff prescription, the result of integration is given by
\begin{equation}
P^\mu_{\rm I}={\rm Reg}_\epsilon\int_{\Sigma}d\sigma_\alpha\,\Theta_{\rm I}^{\alpha\mu}
=
{e^2\over 2\epsilon}\,{v}^{\mu}-{2\over 3}\,e^2 {a}^{\mu}\,,
\label
{P-bound}
\end{equation}                        
where $\epsilon$ is the cutoff parameter which is to go to zero in the end of calculations.
For a thorough derivation of Eq.~(\ref{P-bound}) see, {\it e.~g.}, \cite{Kosyakov2007}.
Observing that a bare particle possesses the four-momentum 
\begin{equation}
p^{\mu}_0
=
\int_{\Sigma}d\sigma_\alpha\,t^{\alpha\mu}
=
{m_0}{v}^{\mu}\,,
\label
{p-0}
\end{equation}                        
one may render $m_0$ a singular function of $\epsilon$, $m_0=m_0(\epsilon)$, add Eqs.~(\ref{P-bound}) and 
(\ref{p-0}) up, and carry out the renormalization of mass, that is, assume that 
\begin{equation}
m=\lim_{\epsilon\to 0}\left[m_0(\epsilon)+ {e^2\over 2\epsilon}\right]
\label
{m-ren}
\end{equation}                        
is finite and positive.
This completes the definition of the measure ${\rm Reg}_\epsilon\,d\sigma_\lambda\left(
\Theta_{\rm I}^{\lambda\mu}+t^{\lambda\mu}\right)$  in the limit $\epsilon\to 0$, and the
regularization-renormalization procedure culminates in the well-defined quantity 
\begin{equation}
p^\mu
=
\lim_{\epsilon\to 0}{\rm Reg}_\epsilon\int_{\Sigma}d\sigma_\alpha\left(
\Theta_{\rm I}^{\alpha\mu}+t^{\alpha\mu}\right)
=
m {v}^{\mu}-{2\over 3}\,e^2 {a}^{\mu}\,,
\label
{p-dressed}
\end{equation}                        
originally deduced in \cite{Teitelboim1970}. 

The regularization-renormalization procedure is a means for completing the definition of the product 
of singular distributions, like ${\Box}^{-1}\delta^4(x)$, as linear continuous functionals on a 
suitable test function space, say, on Schwartz space.

The four-momentum $p^\mu$ defined in Eq.~(\ref{p-dressed}) is attributed to a new entity synthesized 
from mechanical and electromagnetic degrees of freedom.
This entity is reasonable to call a dressed charged particle. 

Turning back to the general solution $F$ of Maxwell's equations (\ref{Eulerian-em-II}) and 
(\ref{Eulerian-em-I}), we recall that $F$ is the sum of the retarded solution $F_{\rm ret}$ 
describing the self field of the delta-function source plus the general solution $F_{\rm ext}$ of 
the homogeneous wave equation describing free (``external'') electromagnetic field, $F=F_{\rm ret}+F_{\rm ext}$.
Accordingly,  $\Theta^{\mu\nu}$ is split into $\Theta=\Theta_{\rm ret}+\Theta_{\rm mix}+\Theta_{\rm ext}$.
Our concern here is with $\Theta_{\rm mix}$ containing mixed contributions of $F_{\rm ret}$ and 
$F_{\rm ext}$, while  $\Theta_{\rm ext}$ is immaterial for the present discussion.
Because the leading singularity of $F_{\rm ret}$ is of the type $\rho^{-2}$, and $F_{\rm ext}$ is 
regular on the world line, the term $\Theta_{\rm mix}$ is integrable.
Besides, taking into account the readily verifiable relationship 
\begin{equation}
\partial_\mu\Theta_{\rm mix}^{\mu\nu}=0\,,
\label
{div-theta-mix}
\end{equation}                        
the four-momentum ${\wp}^{\mu}$ associated with $\Theta^{\mu\nu}_{\rm mix}$ is conveniently evaluated 
by the use of a tube ${T}_\epsilon$ of infinitesimal radius $\epsilon$, enclosing the world line $z^\mu(s)$, 
as the integration surface,
\begin{equation}
{\wp}^{\mu}
=
\int_{T_\epsilon}d\sigma_\alpha \Theta^{\alpha\mu}_{\rm mix}=
-e \int^{{s}}_{-\infty}d\tau\,F_{\rm ext}^{\mu\nu}(z)\, v_\nu(\tau)\,.
\label
{P-mix}
\end{equation}                        
Equation (\ref{P-mix}) represents the four-momentum extracted from an {external} field $F_{\rm ext}$ 
during the whole past history prior to the instant $s$.
The derivative of ${\wp}^{\mu}$ with respect to $s$ equals an {external} Lorentz force exerted on 
the dressed particle at the point $z^\mu(s)$.

Let us integrate (\ref{Total-conserv-law}) over a domain of spacetime bounded by two spacelike 
surfaces ${\Sigma}'$ and ${\Sigma}''$, separated by a short timelike interval, with both normals 
directed towards the future, and a tube ${T}_{R}$ of large radius ${R}$.
With the Gau{ss}--Ostrogradsky theorem, this gives
\begin{equation}
\left(\int_{{\Sigma}''}-\int_{{\Sigma}'}+\int_{{T}_{R}}\right)\,d\sigma_\mu
\left(\Theta^{\lambda\mu}+t^{\lambda\mu}\right)=0\,.
\label
{balance-integ}
\end{equation}                        
We then assume that $F_{\rm ext}$ disappears at spatial infinity.
The only term contributing to the integral over $T_R$ is $\Theta^{\mu\nu}_{\rm II}$.
Taking into account the second equation of  (\ref{cons-Theta-I-II}), the integral of 
$\Theta^{\mu\nu}_{\rm II}$ over $T_R$ can be converted into the integral  
over $T_\epsilon$.
The upshot is
\begin{equation}
\left[m{v}^{\lambda}(s)-{2\over 3}\,e^2 {a}^{\lambda}(s)\right]\Biggr |_{s'}^{s''}
-{2\over 3}\,e^2\int_{s'}^{s''}ds\,{a}^2(s){v}^\lambda(s)=
e\int_{s'}^{s''}ds\,F_{\rm ext}^{\lambda\mu}(z)\,{v}_\mu(s)\,,
\label
{balance-four-momenta}
\end{equation}                        
or, in a concise form,
\begin{equation}
\Delta p^\lambda+\Delta {\cal P}^\lambda+\Delta{\wp}^\lambda=0\,.
\label
{balance-LD}
\end{equation}                        
This is the desired local energy-momentum balance on the world line: the four-momentum 
$\Delta{\wp}^{\lambda}=-eF_{\rm ext}^{\lambda\mu}\,{v}_\mu\Delta s$, extracted from the external 
field $F_{\rm ext}$ during the short period of time $\Delta s$, is expended on  
the increment of four-momentum of the dressed particle, $\Delta p^{\lambda}$, and the four-momentum 
carried away by radiation, ${\Delta{{\cal P}}}^\lambda$.

Intuitively, this local balance is associated with an energy-momentum equilibration of the 
initially unstable system ``a bare particle $+$ electromagnetic field''.
The rearrangement of degrees of freedom in this system may be said to terminate with the formation 
of the dressed particle and radiation. 
If an external field is incorporated in the system along with the self field, the external Lorentz 
force $eF_{\rm ext}^{\mu\nu}\,{v}_\nu$ comes into play in this equilibration.

\subsubsection{The Abraham--Lorentz--Dirac equation}
\label
{ALD}
In an expanded form, Eq.~(\ref{balance-LD}) is an ordinary third-order differential equation for 
$z^\mu(s)$,
\begin{equation} 
m a^\mu-{2\over 3}\,e^2\left({\dot a}^\mu +v^\mu a^2\right)=eF_{\rm ext}^{\mu\nu}\,{v}_\nu\,,
\label
{LD}
\end{equation}                       
originally discovered by Abraham \cite{Abraham1903}, \cite{Abraham1904}, \cite{Abraham1905}, 
Lorentz \cite{Lorentz1904}, \cite{Lorentz1909}, and Dirac \cite{Dirac1938}.
This equation is thus referred to by their names.

With the identities 
\begin{equation} 
v^2=1\,,\quad v\cdot a=0\,,\quad v\cdot{\dot a}=-a^2\,,
\label
{v-2=1,v-cdot-a=0,v-cdot-dot-a=-a-2}
\end{equation}                         
(\ref{LD}) can be rewritten as 
\begin{equation} 
\stackrel{\scriptstyle v}{\bot}({\dot p}-f)=0\,,
\label
{LD-Newton}
\end{equation}                         
where $\stackrel{\scriptstyle v}{\bot}$ stands for the projection operator on a hyperplane with 
normal $v^\mu$, Eq.~(\ref{projection-operator-df}), $p^\mu$ is the four-momentum 
defined in Eq.~(\ref{p-dressed}), and $f^\mu$ is an external four-force.

Equation (\ref{LD-Newton}) is nothing but Newton's second law smoothly embedded in Minkowski 
spacetime.
Dressed particles are therefore dynamical objects governed by Newton's second law.
The dissimilarity of a dressed particle from its ancestor, a bare particle, is that the former has 
the four-momentum 
\begin{equation}
p^\mu=m\left(v^\mu-\tau_0\, a^\mu\right),
\label
{p-mu-dress-charge}
\end{equation}                       
where $\tau_0$ is the characteristic time interval
\begin{equation} 
\tau _{0}={2e^2\over 3m}\,,
\label
{class-radius}
\end{equation}                        
while the four-momentum of the latter depends on kinematical variables as
\begin{equation}
p^{\mu}_0=m_0 v^\mu\,.
\label
{p-mu-bare-charged_particle}
\end{equation}

It follows from (\ref{p-mu-dress-charge}) that
\begin{equation}
p^{2}=m^{2}\left({1+\tau _{0}^{2}\,a^{2}}\right).  
\label
{M-m}
\end{equation}
Suppose that the acceleration of a dressed particle exceeds the critical value,
\begin{equation}
a^2 {\tau_0^{2}}=-1\,,
\label
{a-sqr=-1}
\end{equation}
then the dressed particle becomes a {tachyon}, that is, an object whose four-momentum is spacelike, 
$p^2<0$.
This does not imply superluminal motion.
The potentially tachyonic nature of a dressed particle results from the fact that the curvature of 
its world line can be excessively high.

A central feature of the Abraham--Lorentz--Dirac equation (\ref{LD}) is the lack of invariance under 
time reversal $s\to -s$ because this equation involves both $a^\mu$ whose transformation law 
is $a^\mu\to a^\mu$, and ${\dot a}^\mu$ which transforms according to ${\dot a}^\mu\to-{\dot a}^\mu$.
The rearrangement denudes the Maxwell--Lorentz electrodynamics of the time reversal symmetry 
properties: the emission of radiation is a unidirectional process, and the equation of motion for 
a dressed particle, Eq.~(\ref{LD}), is {irreversible}.

\subsubsection{Another way of looking at the dressed dynamics}
\label
{alternative}
There are other methods of deriving the Abraham--Lorentz--Dirac equation 
without resort to the energy-momentum conservation law, Eq.~(\ref{Total-conserv-law}).
Our interest here is with one of them (see, {\it e.~g.,} \cite{Barut1964}) which is claimed to be 
based on an alternative definition of radiation.

Let us proceed directly from the equation of motion for a bare charged particle (\ref{Eulerian-part}),
\begin{equation}
m_0 a^\mu=ev_\nu F^{\mu\nu}(z)\,,
\label
{eq-motion-bare-part}
\end{equation}                                           
in which $F^{\mu\nu}=F^{\mu\nu}_{\rm ret}+F^{\mu\nu}_{\rm ext}$, 
$F^{\mu\nu}_{\rm ret}$ is the retarded field due to the charge in question, and 
$F^{\mu\nu}_{\rm ext}$ is an external field.
Following Dirac's original 
approach \cite{Dirac1938}, the retarded field $F_{\rm ret}^{\mu\nu}$ is separated into 
regular and singular parts through introducing the corresponding advanced field $F_{\rm adv}^{\mu\nu}$:
\begin{equation}
F_{\rm ret}^{\mu\nu}=\frac12\left(F_{\rm ret}^{\mu\nu}-F_{\rm adv}^{\mu\nu}\right)+
\frac12\left(F_{\rm ret}^{\mu\nu}+F_{\rm adv}^{\mu\nu}\right)={\bar F}^{\mu\nu}+F_{P}^{\mu\nu}\,.
\label
{A-re=A_-+A_+}
\end{equation}                                           

Expressions for the retarded and advanced Green's function, 
\begin{equation}
D_{\rm ret}(t,r)=\frac{1}{r}\,\delta(t-r)\,,
\quad
D_{\rm adv}(t,r)=\frac{1}{r}\,\delta(t+r)\,,
\label
{D-ret,D-adv}
\end{equation}                                           
indicate that the {retarded} field generated by a delta-function source behaves similar 
to the advanced field in the vicinity of the source.
Therefore, ${\bar F}$ is less singular than $F_{\rm ret}$ and $F_{\rm adv}$, while  $F_{P}$ shares 
the singular behavior of $F_{\rm ret}$ and $F_{\rm adv}$.

The resulting regular part of the vector potential is
\begin{equation}
{\bar A}^\mu(x)= \int d^4y\,D(x-y)\, j^\mu(y)\,,
\label
{A-(-)}
\end{equation}                                           
where 
\begin{equation}
D(x)=\frac12\left[D_{\rm ret}(x)-D_{\rm adv}(x)\right]={\rm sgn}(x_0)\,\delta(x^2)\,,
\label
{D-homog}
\end{equation}                                           
\begin{equation}
j^\mu(x)=e\int_{-\infty}^\infty ds\,{v}^\mu(s)\,\delta^{4}\left[x-z(s)\right].
\label
{single-charge-j-4d}
\end{equation}                        
We use (\ref{single-charge-j-4d}) in (\ref{A-(-)}) to give 
\begin{equation}
{\bar A}^\mu(x)=e\int^{\infty}_{-\infty} ds\, v^\mu(s)\,D\left[x-z(s)\right].
\label
{A-(-)-via-v}
\end{equation}                                           

Denoting $R^\mu=x^\mu-z^\mu(s)$, we evaluate the regular part of the field strength: 
\begin{equation}
{\bar F}^{\mu\nu}(x)=e\int^{\infty}_{-\infty} ds\left(\frac{dR^2}{ds}\right)^{-1}
\frac{d}{ds}\,D(R)\left[v^\nu(s)\,\frac{\partial  R^2}{\partial\,x_\mu}-
v^\mu(s)\,
\frac{\partial  R^2}{\partial\,x_\nu}\right].
\label
{F-mu-nu-(-)}
\end{equation}                                           
Since
\begin{equation}
\frac{dR^2}{ds}=-2R\cdot v\,,
\quad
\frac{\partial  R^2}{\partial\,x_\nu}=2 R^\nu\,,
\label
{dR-2/d-tau=}
\end{equation}                                           
we have
\begin{equation}
{\bar F}^{\mu\nu}(x)= 
e\int^{\infty}_{-\infty} ds\,
D(R)\,\frac{d}{ds}\left(\frac{ R^\mu v^\nu-R^\nu v^\mu}{R\cdot v}\right).
\label
{F-(-)}
\end{equation}                                           

Let the observation point $x^\mu$ be on the world line, $x^\mu=z^\mu(\tau)$.
All other points on the world line are separated from $x^{\mu}$ by timelike intervals.
Accordingly, the delta-function in (\ref{D-homog}) should be understood as the limit
\begin{equation}
\delta(R^2)=\lim_{\epsilon\to 0}\delta(R^2-\epsilon^2)\,.
\label
{delta=lim-delta}
\end{equation}                                          
Besides, we can represent the argument of the signum function in (\ref{D-homog}) as $R_0=R\cdot v$.

We now write $s=\tau+\sigma$, and consider the integrand for a small interval $\sigma$.
Using the expansions
\begin{equation}
z^\mu(\tau+\sigma)=z^\mu+\sigma\,v^\mu+{\sigma^2\over2}\,a^\mu+{\sigma^3\over 6}\,{\dot a}^\mu+\ldots,
\label
{z-expansion}
\end{equation}                        
\begin{equation}
{v}^\mu(\tau+\sigma)=v^\mu+\sigma\,a^\mu+{\sigma^2\over2}\,{\dot a}^\mu+\ldots,
\label
{v-expansion}
\end{equation}                        
where the vectors on the right-hand side refer to the instant $\tau$, we find
\begin{equation}
R^\mu=z^\mu(\tau)-z^\mu(\tau+\sigma)=-\sigma\left(v^\mu+{\sigma\over2}\,
a^\mu+{\sigma^2\over 6}\,{\dot a}^\mu\right)+\ldots
\label
{X-expansion}
\end{equation}                                           
It follows that
\begin{equation}
R^\mu v^\nu(\tau+\sigma)-R^\nu v^\mu(\tau+\sigma)=\frac{\sigma^2}{2}
\left(v^\nu a^\mu-v^\mu a^\nu\right)+{\sigma^3\over3}\left(
v^\nu{\dot a}^\mu-v^\mu{\dot a}^\nu\right)+\ldots
\label
{Xv-vX-expansion}
\end{equation}                        
In view of identities (\ref{v-2=1,v-cdot-a=0,v-cdot-dot-a=-a-2}), 
\begin{equation}
R\cdot v(\tau+\sigma)=-\sigma+O(\sigma^3)\,.
\label
{rho-expansion}
\end{equation}                        

Substituting (\ref{Xv-vX-expansion}) and (\ref{rho-expansion}) into (\ref{F-(-)}) and taking into 
account that
\begin{equation}
{\rm sgn}(R\cdot v)\,\delta(R^2-\epsilon^2)={\rm sgn}(-\sigma)\,\delta(\sigma^2-\epsilon^2)=
-\frac{1}{2\epsilon}\left[\delta(\sigma-\epsilon)-\delta(\sigma+\epsilon)\right],
\label
{sgn(R-cdot-v)}
\end{equation}                                           
we obtain
\begin{equation}
{\bar F}^{\mu\nu}(z)={2\over3}\,e^2\left({\dot a}^\mu v^\nu-{\dot a}^\nu v^\mu\right)+O(\epsilon)\,,
\label
{F-(-)-final}
\end{equation}                                           
and 
\begin{equation}
e v_\nu {\bar F}^{\mu\nu}(z)={2\over3}\,e^2\left({\dot a}^\mu +{a}^2 v^\mu\right)+O(\epsilon)\,.
\label
{f-(-)-Abraham-term}
\end{equation}                                           

The term
\begin{equation}
\Gamma^\mu={2\over3}\,e^2\left({\dot a}^\mu +{a}^2 v^\mu\right),
\label
{Abraham-term-df}
\end{equation}                                           
is called the {Abraham term}.
In the literature, $\Gamma^\mu$ is often interpreted as {radiation reaction}, that is, the 
finite effect of the retarded Li\'enard--Wiechert field upon its own source.
This interpretation goes back to Dirac \cite{Dirac1938} who considered $F_{\rm ret}^{\mu\nu}-F_{\rm adv}^{\mu\nu}$ as the radiation field and
$F_{\rm ret}^{\mu\nu}+F_{\rm adv}^{\mu\nu}$ as the bound field.
But this treatment is wrong.
With reference to Eq.~(\ref{A-re=A_-+A_+}), we remark that $\Gamma^\mu$ is derived from
${\bar F}^{\mu\nu}$, rather than from $F_{\rm ret}^{\mu\nu}-F_{\rm adv}^{\mu\nu}$ which is double 
the ${\bar F}^{\mu\nu}$.
We already mentioned that a dressed particle is acted upon by only an external force.
It will transpire in the next section that the concept of radiation reaction causes much 
confusion in understanding the rearranged Maxwell--Lorentz theory.
Furthermore, linear combinations of retarded and advanced fields seem to be of no use for 
nonlinear theories such as the Yang--Mills theory.
Therefore, it is best to think of Eq.~(\ref{A-re=A_-+A_+}) as a mere formal trick for discriminating 
between integrable and nonintegrable singularities of the retarded field. 

Consider the symmetric part $F_P$ of the decomposition (\ref{A-re=A_-+A_+}).
The corresponding Green's function is
\begin{equation}
D_P(R)=\frac12\left[D_{\rm ret}(R)+D_{\rm adv}(R)\right]=\delta(R^2)\,.
\label
{1/2(D-ret+D-adv)}
\end{equation}                                           
We regularize this expression as follows: 
\begin{equation}
\delta(R^2-\epsilon^2)=\delta(\sigma^2-\epsilon^2)=\frac{1}{2\epsilon}\left[\delta(\sigma-\epsilon)+
\delta(\sigma+\epsilon)\right].
\label
{delta(R-2-epsilon-2)}
\end{equation}                                          
Applying this procedure to 
\begin{equation}
F^{\mu\nu}_{P}(x)=e\int^{\infty}_{-\infty} ds\,\delta(R^2-\epsilon^2)\,\frac{d}{ds}
\left(\frac{ R^\mu v^\nu-R^\nu v^\mu}{R\cdot v}\right),
\label
{F-(+)}
\end{equation}                                           
we get
\begin{equation}
F^{\mu\nu}_{P}(z)={e\over2\epsilon}\left(v^\mu {a}^\nu-v^\nu {a}^\mu\right)+O(\epsilon)\,.
\label
{F-(+)-final}
\end{equation}                                           
Therefore,
\begin{equation}
e v_\nu F^{\mu\nu}_{P}(z)=-{e^2\over2\epsilon}\,{a}^\mu+O(\epsilon)\,.
\label
{f-(+)-final}
\end{equation}                                           
Substituting (\ref{f-(-)-Abraham-term}) and (\ref{f-(+)-final}) in (\ref{eq-motion-bare-part}) and 
performing the renormalization of mass, Eq.~(\ref{m-ren}), we come again to the Abraham--Lorentz--Dirac 
equation (\ref{LD}).

For completeness, it is useful to refer to a simple method of regularization proposed in \cite{Barut1974}, 
which obviates the need for advanced fields.
This method can readily be extended to field theories involving scalar and tensor fields, 
and linearized gravitation \cite{BarutVillaroel1975a}, as well as electrodynamics in curved 
spacetime \cite{BarutVillaroel1975b}.
The key idea of this method is that the retarded field $F_{\rm ret}$ can be regularized in the
vicinity of the source  using a kind of analytic continuation.
To be more precise, the field is regarded as a function of two variables $F^{\mu\nu}[x;z(s)]$ and
is continued analytically from null intervals between the observation point $x^\mu$ and the 
retarded point $z^\mu(s)$ to timelike intervals which result from assigning $x^\mu=z^\mu(s+\epsilon)$ 
and keeping the second variable $z^\mu(s)$ fixed.

To summarize, the retarded Li\'enard--Wiechert field can be regularized in different ways.
The regularization scheme is allowed to be arbitrary; the only requirement is that it respects 
the symmetries of the action (\ref{S-Planck-general})--(\ref{S-em-field}). 
Given a regularization characterized by the regularization parameter $\epsilon$, 
the field becomes finite but $\epsilon$-dependent at distances shorter than $\epsilon$. 
This suggests that the mechanical mass should also be a function of regularization, $m_0=m_0(\epsilon)$.
A remarkable fact is that the renormalization of mass (\ref{m-ren}), absorbing the self-energy 
divergence, makes the rearranged Maxwell--Lorentz electrodynamics a finite and unambiguous theory.

\subsubsection{Paradoxes and misconceptions}
\label
{Paradoxes}
The physical validity of the Abraham--Lorentz--Dirac equation
\begin{equation} 
m a^\mu-{2\over 3}\,e^2\left({\dot a}^\mu +v^\mu a^2\right)=f^{\mu}
\label
{LD-interpr}
\end{equation}                        
has been the subject of much controversy.
At present the following view of this equation is of considerable use  \cite{Arzelies1966}, 
\cite{Barut1964}, \cite{Jackson1962},  \cite{LandauLifshitz1971}, \cite{Parrott1987},  
\cite{Rohrlich1965}:
Eq.~(\ref{LD-interpr}) governs a charged radiating particle endowed with the four-momentum
\begin{equation} 
p^\mu=mv^\mu\,.
\label
{four-momentum-wrong}
\end{equation}                        
The particle is assumed to experience both an external force $f^{\mu}$ and the radiation reaction 
\begin{equation} 
\Gamma^\mu={2\over 3}\,e^2\left({\dot a}^\mu +v^\mu a^2\right),
\label
{Abraham-term}
\end{equation}                        
which is also known as the {radiation damping} four-force.

This view leads to many paradoxes and puzzles.
To gain greater insight into why this view is so much persistent, let us take a closer look at the 
notion of radiation field proposed by Dirac \cite{Dirac1938}.
The general solution to Maxwell's equations can be cast in two alternative ways:
\begin{equation}
F=F_{\rm in}+F_{\rm ret}\,,
\label
{ret-in-solution}
\end{equation}
\begin{equation}
F=F_{\rm out}+F_{\rm adv}\,,
\label
{adv-out-solution}
\end{equation}
where $F_{\rm in}$ and $F_{\rm out}$ are respectively incoming and outgoing fields described by 
solutions to the homogeneous wave equation.
Dirac defined the radiation field as
\begin{equation}
F_{\rm rad}=F_{\rm out}-F_{\rm in}\,.
\label
{rad-Dirac-df}
\end{equation}
In view of (\ref{ret-in-solution}) and (\ref{adv-out-solution}), this can also be written as the 
difference between the retarded and advanced solutions: 
\begin{equation}
F_{\rm rad}=F_{\rm ret}-F_{\rm adv}\,.
\label
{rad-Dirac-ret-adv}
\end{equation}
$F_{\rm rad}$ measures the dissimilarity between the field which is going to happen in the far 
future and the field which was formed in the remote past, and hence is advisable to be called the 
radiation field. 
We will see in Sec.~\ref{wheeler-feynman} that $F_{\rm rad}$ plays a crucial role in the Wheeler and 
Feynman absorber theory of radiation.

$F_{\rm rad}$ shares a number of traits with $F_{\rm II}$, the long-range part of the retarded field 
defined by Eq.~(\ref{F-II}).
Indeed, consider a world line composed of two timelike rays connected by a curved fragment, and draw 
two future light cones from the connection points.
In the region enclosed by these cones, $F_{\rm rad}$ approaches $F_{\rm ret}$, and, furthermore, 
$F_{\rm II}$ dominates $F_{\rm ret}$ with distance away from the source.
Meanwhile $F_{\rm II}$ was shown in Sec.~\ref{Radiation} to be completely removable by an appropriate 
local SL$(2,{\mathbb R})$ transformation, so that $F_{\rm rad}$ is removable along with $F_{\rm II}$
from the asymptotic region in which $F_{\rm rad}\to F_{\rm II}$.

It is worthy of note that all manifestations of $F_{\rm rad}$ as a free field are due to linearity 
of Maxwell's equations.
In non-Abelian gauge theories, linear combinations of retarded and advanced solutions are no longer
free fields.
The construction (\ref{rad-Dirac-ret-adv}) may only be of utility in Maxwell's electrodynamics or 
other theories with linear field equations.   
 
Since $\frac12 F_{\rm rad}$ is nonsingular on the world line, one can use it in the Lorentz 
force law, much as was done in Sec.~\ref{alternative}, to yield the Abraham term $\Gamma^\mu$, 
Eq.~(\ref{Abraham-term}), appearing in the Abraham--Lorentz--Dirac equation (\ref{LD}). 
Ignoring the excess factor $2$ in the definition of $F_{\rm rad}$, one commits to recognize 
$\Gamma^\mu$ as the radiation reaction. 

What are the consequences of this recognition? 
The radiating particle feels a recoil, 
\begin{equation}
-{{\dot{\cal P}}^{\mu}}={2\over 3}\,e^2 a^2 v^\mu\,,
\label
{dP-II-}
\end{equation}                        
the negative of the Larmor emission rate (\ref{dP-II}).
However, $-{{\dot{\cal P}}^{\mu}}$ is not a four-force because it is not orthogonal to $v^\mu$.
On the other hand, $\Gamma^\mu$ is orthogonal to $v^\mu$, but differs from the expected recoil by 
the so-called Schott term ${2\over 3}\,e^2 {\dot a}^\mu$  \cite{Schott1915}.
This term makes the energy-momentum balance problematic.
To see this, write the temporal component of (\ref{LD-interpr}) as
\begin{equation} 
\frac{d}{dt}\,\frac{m}{\sqrt{1-{\bf v}^2}}-{2\over 3}\,e^2\,a^2-{2\over 3}\,e^2\,\frac{d}{dt}\,{a}^0={\bf F}\cdot{\bf v}\,.
\label
{0-component-LD}
\end{equation}                        
Following Dirac \cite{Dirac1938}, one may reason: the rate at which the external force ${\bf F}$ 
does work on 
the particle is equal to the increase in the particle's kinetic energy, plus the energy radiated, 
plus the energy stored in the Schott term.
Although the energy stored in the Schott term can in principle be attributed to a ``reversible 
form of emission and absorption of field energy'', its actual role appears mysterious.

In an effort to remedy the situation, we impose the asymptotic condition
\begin{equation} 
\lim_{s\to\pm\infty} a^\mu(s)=0\,,
\label
{Haag-pm-infty}
\end{equation}                        
generalizing the Haag condition (\ref{a-to-0}).
Observing that ${2\over 3}\,e^2 {\dot a}^\mu$ is a perfect differential, and 
integrating (\ref{LD-interpr}) over $s$ in infinite limits, we eliminate the effect of this term and
come to a global energy-momentum balance 
\begin{equation} 
mv^\mu_{\rm out}-mv^\mu_{\rm in}-{2\over 3}\,e^2\int^{\infty}_{-\infty}ds\,a^2v^\mu=\int^{\infty}_{-\infty} ds\,f^{\mu}\,.
\label
{LD-intergr}
\end{equation}                        
It may appear that (\ref{LD-intergr}) is a satisfactory solution to the problem \cite{Rohrlich1974}.
But actually this result is puzzling.
Whatever the history of the particle, $z^\mu(s)$, obeying the asymptotic condition 
(\ref{Haag-pm-infty}), the totality of alternating local emissions and absorptions, controlled by 
the Schott term, is zero, so that the global energy-momentum balance (\ref{LD-intergr}) holds true.
It is as if the particle takes care on that this totality does not become nonvanishing in the end.
The natural question can then be raised: Why is the energy-momentum conserved {globally} rather than 
locally? 
There is nothing in the laws of the Maxwell--Lorentz theory which suggests that the electromagnetic 
interaction is nonlocal to make {local} balance impossible.

The paradigm of rearrangement offers an alternative view of Eq.~(\ref{LD-interpr}) 
as the equation of motion for a dressed particle.
It was established in Sec.~\ref{ALD} that (\ref{LD-interpr}) is equivalent to Newton's second 
law embedded in Minkowski space, Eq.~(\ref{LD-Newton}).
The key point is that a dressed particle possesses the four-momentum $p^\mu$ given by (\ref{p-mu-dress-charge}) rather than by (\ref{four-momentum-wrong}).
The structure of Eq.~(\ref{LD-Newton}) makes it clear that the only force exerting on the
dressed particle is an external force $f^\mu$. 
There is no term in this equation through which the dressed particle interacts with itself.
The notion ``radiation damping force'' and the like are thus to be abandoned as misconceptions.  
The local energy-momentum balance on the world line, Eq.~(\ref{balance-LD}), 
\begin{equation}
{\dot p}^\mu+{\dot{\cal P}}^{\mu}=f^{\mu}\,
\label
{local-balance-dr}
\end{equation}                        
is a mere rewriting of the Abraham--Lorentz--Dirac equation, which tells us that the energy-momentum 
of an external field is converted into the increment of energy-momentum of the dressed particle
and the energy-momentum radiated.

Closely related to the energy-momentum balance problem is the {paradox of uniform acceleration}. 
A covariant condition for uniform acceleration (see, {\it e.~g.},  \cite{Rohrlich1965}) is
\begin{equation}
\left(\stackrel{\scriptstyle v}{\bot}{\dot a}\right)^\mu={\dot a}^\mu+a^2v^\mu=0\,.
\label
{uniforml-accel-motion-gen-}
\end{equation}
This condition implies  $\Gamma^\mu=0$, and the Abraham--Lorentz--Dirac equation (\ref{LD-interpr}) 
becomes
\begin{equation} 
m a^\mu=f^{\mu}\,,
\label
{LD-without-Gamma}
\end{equation}      
which is identical to the equation of motion for a nonradiating particle, say a neutral particle
of mass $m$.
One may see the paradox in the fact that a uniformly accelerated charged particle, 
while emitting electromagnetic radiation, experiences no back reaction.
Besides, it is strange that the case of uniform acceleration is physically distinguished.

No paradox arises for a dressed particle.
As pointed out above, the dressed particle experiences only an external force $f^\mu$.
The Abraham term $\Gamma^\mu$ has nothing to do with radiation reaction. 
$\Gamma^\mu$ says nothing about the emission rate ${{\dot{\cal P}}^{\mu}}$, and $\Gamma^\mu=0$
does not imply that ${{\dot{\cal P}}^{\mu}}=0$.

There is another formulation of this paradox.
Let us compare the behavior of neutral and charged particles, which have identical masses $m$, and
move along a straight line under a constant force $f^\mu$, say fall to the surface of the Earth.
Both are attracted toward the Earth by an approximately constant force ${\bf f}=-mg{\bf n}$, where 
$g$ is the acceleration of gravity, and ${\bf n}$ the normal vector of the surface of the Earth.
With the ansatz 
\begin{equation}
v^\mu=\left(\cosh \nu,{\bf n}\sinh \nu\right),
\label
{v-mu-one-dm}
\end{equation}
\begin{equation}
f^\mu=mg\left(\sinh \nu,-{\bf n}\cosh \nu\right),
\label
{f-mu-one-dm}
\end{equation}
where $\nu=\nu(s)$ is an unknown function, the equation of motion for the charged particle reduces to
\begin{equation}
{\dot\nu}-\tau_0\,{\ddot\nu}=-g\,,  
\label
{LD-one-dm}
\end{equation}
and that for the neutral particle reduces to
\begin{equation}
{\dot\nu}=-g\,.
\label
{neutral-part-one-dim}
\end{equation}
Both equations (\ref{LD-one-dm}) and (\ref{neutral-part-one-dim}) are satisfied by ${\nu}=-gs$.  
Therefore, a given constant force causes both particles move along the same hyperbolic world line
\begin{equation}
z^\mu(s)=z^\mu(0)+{g}^{-1}\left(\sinh gs, -{\bf n}\cosh gs\right),
\label
{hyperbolic-motn-one-dim}
\end{equation}
even if the accelerated charged particle radiates.
Since this radiation carries off energy, the charged particle may be expected to accelerate less
than the neutral one.
 
Note, however, that the energy of a neutral particle is positive definite, while the energy of a 
dressed charged particle is indefinite.
Despite the fact that both particles execute identical motions, the energy associated with these 
motions is different.
Indeed, the energy of a dressed charged particle is 
\begin{equation} 
p^0=mv^0-{2\over 3}\,e^2a^0=m\left[\cosh (gs)-\tau_0 g\sinh (gs)\right].
\label
{p-0=hyperb-motion}
\end{equation}                        
Accordingly, the increment of $p^0$ during a period from $s_1$ to $s_2$ is
\begin{equation} 
\Delta p^0={p}^0({s_2})-{p}^0({s_1})=m\left[\cosh (gs)-\tau_0 g\sinh (gs)\right]|^{s_2}_{s_1}\,.
\label
{Delta-p-0=hyperb-motion}
\end{equation}                        
The energy radiated during this period is
\begin{equation} 
-{2\over 3}\,e^2\int^{s_2}_{s_1} ds\,a^2v^0=m\tau_0 g\sinh (gs) |^{s_2}_{s_1}\,.
\label
{emitted-energy-hyprb-motion}
\end{equation}                        
The sum of (\ref{Delta-p-0=hyperb-motion}) and (\ref{emitted-energy-hyprb-motion}) equals the work 
$W$ of the force $f^\mu$ defined in (\ref{f-mu-one-dm}),
\begin{equation} 
W=\int^{s_2}_{s_1} ds\,f^0=m\cosh (gs) |^{s_2}_{s_1}\,,
\label
{work-hyprb-motion}
\end{equation}                        
as might be expected from the balance equation (\ref{local-balance-dr}).

For the neutral particle,  $p^0=mv^0$, and so
\begin{equation} 
\Delta p^0=m\cosh (gs)|^{s_2}_{s_1}\,,
\label
{Delta-p-0=neutral-hyperb-motion}
\end{equation}                        
which is equal to $W$.

Keeping in mind Eq.~(\ref{M-m}), one may state that, when executing an accelerated motion, the 
dressed particle appears as an object less heavy than the neutral particle.
That is why the increment of energy of the dressed particle caused by a constant force, 
Eq.~(\ref{Delta-p-0=hyperb-motion}), is less than that of the neutral particle,
Eq.~(\ref{Delta-p-0=neutral-hyperb-motion}), by the energy radiated, 
Eq.~(\ref{emitted-energy-hyprb-motion}).

A further concern is with the so-called {counter-acceleration}. 
One normally expects that the smallness of $\Gamma^\mu$ implies small corrections to the 
essentially Newtonian behavior of a charged particle. 
But these expectations are not always realized.

Let a charged particle be moving along a straight line.
Then ${v}^{\mu}=(\cosh\nu,\sinh\nu,0,0)$, $f^{\mu}=f(\sinh\nu,\cosh\nu,0,0)$,
and the Abraham--Lorentz--Dirac equation (\ref{LD}) reduces to
\begin{equation}
m\left({\dot\nu}-\tau_0\,{\ddot\nu}\right)=f\,.
\label
{one-dim_ALD}
\end{equation}                                           
This ordinary differential equation can be readily integrated to give
\begin{equation} 
{\dot\nu}(s)= e^{s/\tau_0}\left[
C-{1\over m\tau_0}\int_0^s d\tau\, e^{-\tau/\tau_0}\,f(\tau)\right],
\label
{general-sol-1D-LD-one-dim}
\end{equation}                        
where $C$ is an arbitrary initial value of ${\dot\nu}$ at $s=0$. 
The comparison of (\ref{LD}) and (\ref{one-dim_ALD}) shows that ${\dot\nu}$ can be identified with
the one-dimensional acceleration in the instantaneously comoving frame of reference. 
Setting $C=0$, one finds that ${\dot\nu}$ and $f$ are oppositely directed.

For a dressed particle, this result presents no difficulty.
Indeed,  Eq.~(\ref{LD-Newton}) shows that the dynamics of a dressed particle is Newtonian.
This, however, in no way implies that acceleration must be aligned with force; $a^\mu$ and $f^\mu$ 
would have the same direction only if one makes the {\it ad hoc} assumption that $p^\mu=mv^\mu$.

We next turn to the problem of {runaways}.
On putting  $f^\mu=0$ in (\ref{LD-interpr}), one can easily check that the general solution to this 
equation is
\begin{equation} 
v^\mu(s)={\rm V}^\mu\cosh(\nu_0+w_0\tau_0e^{s/\tau_0})+{\rm U}^\mu\sinh(\nu_0+w_0\tau_0e^{s/\tau_0})\,,
\label
{runaway-Abraham-Lorentz}
\end{equation}                         
where ${\rm V}^\mu$ and ${\rm U}^\mu$ are constant four-vectors such that ${\rm V}\cdot{\rm U}=0$, 
${\rm V}^2=-{\rm U}^2=1$, and $\nu_0$ and $w_0$ are arbitrary parameters.
This solution of the Abraham--Lorentz--Dirac equation is the most embarrassing feature of the theory: 
a free charged particle moving along the world line defined in (\ref{runaway-Abraham-Lorentz}) continually
accelerates,
\begin{equation}
a^2(s)=-w_0^2\exp\left(2s/\tau_0\right),
\label
{a^2-runaway}
\end{equation}
and, furthermore, continually radiates.
This may seem contrary to energy conservation.
Such particles are said to be self-accelerated, or else executing a runaway motion.

Based on the dynamics of a dressed charged particle, the issue of energy-momentum conservation can 
be solved immediately.
Taking $f^\mu=0$ in (\ref{local-balance-dr}), we have 
\begin{equation}
{\dot p}^\mu=-{\dot{\cal P}}^{\mu}\,,
\label
{local-balance-f=0}
\end{equation}    
which suggests that the rate of change of the energy-momentum of a dressed particle equals the negative of the 
emission rate.
Of crucial importance is the observation that the energy of a dressed charged particle is {indefinite},
\begin{equation}
p^{0}=m\gamma \left[1-\tau _{0}\,\gamma ^{3}\left({\bf a}\cdot {\bf v}\right)\right],
\label
{p-0-em-}
\end{equation}
and hence, increasing $|{\bf v}|$ need not be accomplished by increasing $p^0$.
In fact, the energy of a dressed particle executing a runaway motion (\ref{runaway-Abraham-Lorentz}) 
decreases steadily, which exactly compensates the increase in energy of the electromagnetic field 
emitted,
\begin{equation}
{p}^0({s_2})-{p}^0({s_1})=-\int^{s_2}_{s_1} ds\,{\dot{\cal P}}^{0}\,.
\label
{globocal-balance-f=0}
\end{equation}    

The runaways have long been believed to be unphysical solutions.
The reason for this belief is that a free electron with exponentially increasing acceleration has 
never observed experimentally.
However, by the turn of the 20th century, the idea that free objects can move with 
acceleration became not without appeal (for a review of some objects exemplifying such non-Galilean
motions see  \cite{Kosyakov2003}).
The observational evidence for an accelerating Universe \cite{Riess1998}, \cite{Perlmutter1999}, 
expressed in terms of the scale factor $a$ of the line element for homogeneous and isotropic 
spacetime,
\begin{equation}
a(t)\sim \exp(H_0t)\,,
\label
{universe-accelerated}
\end{equation}    
where $H_0$ is the current Hubble expansion parameter ($H_0^{-1}\sim 10^{10}$ years), is usually 
attributed to the presence of a positive cosmological constant.
An alternative approach to account for this exponentially accelerated motion \cite{Kosyakov2005} 
asserts that the Universe may be regarded as a free massive object, say a brane, which emits 
gravitational radiation and moves in a runaway regime similar to that shown in Eq.~(\ref{runaway-Abraham-Lorentz}),  
with the characteristic time $\tau_0$ being $\sim H_0^{-1}\sim G_{\rm N}m$, where $G_{\rm N}$ is Newton's constant, and $m$ the 
total visible mass of the Universe.

Turning to the lack of observational evidence for self-accelerated motions of electrons, we note
that the critical acceleration $|{\bf a}|=\tau_0^{-1}$ has been attained after a lapse of time
\begin{equation}
\Delta s=-\tau_0\log (\tau_0 w_0)\,,
\label
{lapse-accelerated}
\end{equation}    
and then the four-momentum of a self-accelerated electron becomes spacelike, $p^2<0$.
The period of time over which a self-accelerated subnuclear particle, if any, possesses 
timelike four-momenta is quite tiny. 
From (\ref{class-radius}) and (\ref{lapse-accelerated}), the period $\Delta s$ is estimated at 
$\tau_0\sim 10^{-23}$ s for electrons, and still shorter for more massive charged elementary 
particles. 
All primordial self-accelerated particles with such $\tau_0$'s have long been in the tachyonic state.
But we do not have the slightest notion of how tachyons can be experimentally recorded.

We finally address the problem of {pre-accelerations}.
For simplicity, we consider a charged particle moving along a straight line.
To get rid of runaways, we assume that ${\dot\nu}(s)\to 0$ as $s\to\infty$.
Then the differential equation (\ref{one-dim_ALD}) can be readily converted into an 
integro-differential equation  \cite{IwanenkoSokolow1953},  \cite{Jackson1962}, 
\begin{equation} 
m{\dot\nu}(s)=\int_0^\infty d\xi\,f(s+\tau_0\xi)\, e^{-\xi}\,.
\label
{sol-1D-LD}
\end{equation}                        

It follows from this equation that the particle accelerates before the force is applied.
For example, if a step pulse, $f(t)=f_0\,\theta(t)$, were applied, then the particle would begin its 
acceleration at a time $\tau_0$ before the pulse arrived.
This phenomenon is appraised to be unphysical because the behavior of the particle apparently 
violates causality.   

We should not forget, however, that a dressed particle is an object synthesized from mechanical and 
field degrees of freedom, and hence the third-order differential equation, Eq.~(\ref{LD-interpr}), 
governs its behavior.
If an effort is made to express the Abraham--Lorentz--Dirac equation in terms of the second-order 
equation of motion for a bare particle, then the distinction between these two dynamics shows itself 
as an effective smearing over a small spacetime region of size $\tau_0$.
This imprecise argument can be formulated more neatly: the effective smearing is expressed by 
Eq.~(\ref{sol-1D-LD}).
Of course, the actual interaction of a dressed particle with an external field is local, as 
exemplified by the local energy-momentum balance (\ref{local-balance-dr}).

Most paradoxes in the Maxwell--Lorentz electrodynamics stem from the view that a charged radiating 
particle carries four-momentum $p^\mu=mv^\mu$, and that the Abraham term $\Gamma^\mu$ exerts on 
this particle as the radiation damping force.
The formula $p^\mu=mv^\mu$ has no justification except for keeping an analogy with mechanics.
If a stress-energy tensor ${\mathfrak{T}}^{\mu\nu}$ associated with the integral quantity 
$p^\mu=mv^\mu$ is explored, one can readily verify that the local conservation law
$\partial_\mu{\mathfrak{T}}^{\mu\nu}=0$ does not hold outside the world line. 
Considering ${{T}}^{\mu\nu}- {\mathfrak{T}}^{\mu\nu}$ as the radiation part of 
the stress-energy tensor we come into conflict with the characteristic properties of the 
radiation (\ref{i}) and (\ref{ii}).

Note also that $p^\mu=mv^\mu$ implies that $p^0$ is positive definite. 
But the four-momentum of a dressed charged particle is defined by $p^\mu=P_{\rm I}^\mu+m_0v^\mu$.
The bound four-momentum $P_{\rm I}^\mu$ is a timelike future directed vector, while the four-momentum 
of a bare particle $m_0v^\mu$ is a timelike past directed vector because $m_0(\epsilon)<0$ for 
small $\epsilon$, as Eq.~(\ref{m-ren}) suggests.
Assuming that $P_{\rm I}^\mu+m_0v^\mu$ is a timelike vector, one should recognize that the time component 
of this vector can have any sign. 
This conclusion is certified by Eq.~(\ref{p-0-em-}).

Lack of understanding of the fact that the dynamics of a dressed charged particle attributed to the 
Abraham--Lorentz--Dirac equation (\ref{LD-interpr}) is physically satisfactory has led to numerous 
attempts at developing {\it ad hoc} modifications and approximate versions of this equation adapted 
to applications (see, {\it e.~g.},  \cite{Spohn2004} and references therein).
We omit these developments because they add little to our discussion of the conceptual
aspects of the Maxwell--Lorentz theory.

While the concept of dressed particles is a tenable physical means of disambiguation, the proper 
mathematical treatment of the Abraham--Lorentz--Dirac equation governing a dressed particle is as 
yet imperfectly understood.
When extreme care is not exercised, surprising results may arise  \cite{Eliezer1947}, 
\cite{Plass1961},  
\cite{Parrott1987}, which tempts one to assign the blame to the Abraham--Lorentz--Dirac equation 
itself.

\subsection{Electrodynamics in various dimensions}
\label
{2Q,6D}
It is instructive to see whether the rearrangement of the Maxwell--Lorentz theory remains unchanged 
in other dimension if the field sector of the action is assumed to be valid,
\begin{equation}
{S}=-{1\over 4\Omega_{d-2}}\int d^dx\left(F_{\mu\nu}F^{\mu\nu}+A_\mu j^\mu\right).
\label
{Lagrangian-Maxw-D}
\end{equation}                                          
Here, $\Omega_{d-2}={2{\pi}^{\frac{d-1}{2}}}/{\Gamma\left(\frac{d-1}{2}\right)}$ 
is the area of the unit $(d-2)$-sphere, the field strength $F_{\mu\nu}$ is expressed in terms of 
the vector potentials, 
$F_{\mu\nu}=\partial_\mu A_\nu-\partial_\nu A_\mu$, 
and  
\begin{equation}
j^\mu(x)=e\int_{-\infty}^\infty ds\,{v}^\mu(s)\,\delta^{\hskip0.2mm d}
\left[x-z(s)\right]
\label
{single-charge-j}
\end{equation}                        
represents the current density of a single point charge $e$.
For simplicity, the charge $e$ is put to be unit here and in the next subsubsection.
Greek letters standing for spacetime indices take the values $0,1,\ldots,d-1$.
We adopt the metric $\eta_{\mu\nu}={\rm diag}\,(1,-1,\ldots, -1)$.

The field equation resulting from (\ref{Lagrangian-Maxw-D}), on imposing the Lorenz gauge condition, 
reads 
\begin{equation}
\Box\, A^\mu=\Omega_{d-2}\,j^\mu\,.
\label
{wave-equation-D}
\end{equation}                                          
This equation can be solved with the aid of the Green's function technique \cite{IwanenkoSokolow1953}.
Solutions to the equation of the retarded Green's function 
\begin{equation}
\Box\, G_{\rm ret}(x)=\delta^{d}(x)
\label
{wave-equation-Green-D}
\end{equation}                                          
are given by
\begin{equation}
G_{\rm ret}(x)
=\cases{
\frac{1}{2\pi^n}\, 
\theta(x_0)\,\delta^{(n-2)}(x^2) 
& $d=2n$\,, 
\cr
& \cr
\frac{(-1)^{n-1}}{\pi^{n+\frac12}}\,
\Gamma\left(n-\frac{1}{2}\right)
\theta(x_0)\,\theta(x^2)\left(x^2\right)^{\frac{1}{2}-n}  & $d=2n+1$\,,
\cr}
\label
{Green's-D}
\end{equation}
where $\delta^{(n-2)}(x^2)$ is the Dirac delta-function differentiated $n-2$ times with respect to 
its argument.

A sharp distinction between wave propagation in space of even and odd dimensions can be 
understood from {Huygens' principle}, whereby any retarded signal carries information on the 
state of a point source at the instant of its emission.
This idea is exemplified in the first line of (\ref{Green's-D}): the retarded Green's function 
for ${\mathbb R}_{1,2n-1}$ is concentrated on the forward sheet of the light cone $x^2=0,\,x_0>0$.
By contrast, in ${\mathbb R}_{1,2n}$ the retarded signal measured at a point $x^{\mu}$ derives from 
the entire history of the source which lies on or within the past light-cone of $x^{\mu}$.
If we think of the retarded signal as travelling with the speed of light then it ought to be emitted 
at the instant the source intersects the past light cone of $x^{\mu}$. 
Hence, Huygens' principle fails in odd spacetime dimensions: the second line of (\ref{Green's-D}) 
shows that the support of the retarded Green's function in ${\mathbb R}_{1,2n}$ 
is the interior of the future light cone  $x^2\ge 0$, rather than its surface.

Even-dimensional electromagnetic worlds have little in common with their relatives in odd
dimensions. 
To illustrate this, in ${\mathbb R}_{1,2}$, the action (\ref{Lagrangian-Maxw-D}) can be augmented by 
the addition of the Chern--Simons term 
\begin{equation}
{S}_{\rm CS}=\mu\int d^3x\,\epsilon^{\alpha\beta\gamma}A_\alpha F_{\beta\gamma}\,,
\label
{Chern--Simons-Lagr-EM}
\end{equation}                                          
which is gauge invariant despite the presence of the parameter $\mu$ 
interpreted as the mass of the field $A^\alpha$  \cite{Deser1982}. 
Odd-dimensional versions of the Maxwell--Lorentz electrodynamics are more intricate, both technically 
and conceptually, than its even-dimensional versions, in particular the self-interaction problem in 
${\mathbb R}_{1,2n}$ is less well understood.
Attempts at formal evaluating the effects of self-interaction gained little, if any, insight into this
problem. 
This forces us to focus on the affair in even dimensions.

\subsubsection{${\mathbb R}_{1,2n-1}$}
\label
{d=2n}
The $2n$-dimensional retarded vector potential is given by
\begin{equation}
A^{(2n)}_\mu(x)
=\Omega_{2n-2}\int_{-\infty}^{\infty} ds\,G^{(2n)}_{\rm ret}\left(R\right){v}_\mu(s)\,, 
\label
{vector-potential-Green's-D}
\end{equation}
where $R^\mu=x^\mu-z^\mu(s_{\rm ret})$ is the null $2n$-vector drawn from the retarded point 
$z^\mu(s_{\rm ret})$ on the world line, where the signal is emitted, to the point $x^\mu$, where 
the signal is received.

To simplify our notations as much as possible, we introduce the {net} vector potentials and 
field strengths, ${\cal A}_{\mu}$ and ${\cal F}_{\mu\nu}$ (as opposed to the ordinary vector 
potentials and field strengths, ${A}_{\mu}$ and ${F}_{\mu\nu}$, whose {normalization} is consistent 
with Gauss' law):
\begin{equation}
{A}^{(2p)}_\mu
=N_p^{-1}{\cal A}^{(2p)}_\mu\,,
\quad
{F}^{(2p)}_{\mu\nu}
=N_p^{-1}{\cal F}^{(2p)}_{\mu\nu}\,,
\label
{A-cal-A}
\end{equation}                       
where
\begin{equation}
N_1=1\,,
\quad
N_p=(2p-3)!!\,,
\quad
p\ge 2\,.
\label
{N-p-df}
\end{equation}                       

Consider the vector potentials, calculated through the use of Eq.~(\ref{vector-potential-Green's-D}), 
restricting our attention to $d$ in the range from $d=2$ to $d=10$ (which hold the greatest interest 
in string and brane applications):
\begin{equation}
{\cal A}^{(2)}_\mu=-R_\mu\,,
\label
{cal-A-2}
\end{equation}                       
\begin{equation}
{\cal A}^{(4)}_\mu=\frac{v_\mu}{\rho}\,,
\label
{cal-A-4}
\end{equation}                       
\begin{equation}
{\cal A}^{(6)}_\mu=-\lambda\,\frac{v_\mu}{\rho^3} +\frac{a_\mu}{\rho^2}\,,
\label
{cal-A-6}
\end{equation}                       
\begin{equation}
{\cal A}^{(8)}_\mu
=\left[3\lambda^2
-\rho^2\left({\dot a}\cdot c\right)\right]\frac{v_\mu}{\rho^5} 
-3\lambda\,\frac{a_\mu}{\rho^4}
+\frac{{\dot a}_\mu}{\rho^3}\,,
\label
{cal-A-8}
\end{equation}                       
$$
{\cal A}^{(10)}_\mu
=\left[-15\lambda^3
+10\lambda\rho^2\left({\dot a}\cdot c\right)-\rho^2a^2
-\rho^3\left({\ddot a}\cdot c\right)\right]\frac{v_\mu}{\rho^7} 
$$
\begin{equation}
+\left[15\lambda^2
-4\rho^2\left({\dot a}\cdot c\right)\right]\frac{a_\mu}{\rho^6}
-6\lambda\,\frac{{\dot a}_\mu}{\rho^5}
+\frac{\ddot a_\mu}{\rho^4}\,,
\label
{cal-A-10}
\end{equation}                       
$$
{\cal A}^{(12)}_\mu
=\Biggl\{105\lambda^2\left[\lambda^2
-\rho^2({\dot a}\cdot c)\right]
+ 15\lambda\rho^2[\rho({\ddot a}\cdot c)+a^2]
$$
$$
-\frac52\,\rho^3({a}^2)^{.}
-\rho^4(\stackrel{\ldots}{a}\cdot c)
+10\rho^4({\dot a}\cdot c)^2
\Biggr\}\frac{v_\mu}{\rho^9} 
$$
$$
+5\left\{3\lambda\left[-7\lambda^2
+4\rho^2({\dot a}\cdot c)\right]
-\rho^2[\rho({\ddot a}\cdot c)
+ a^2]\right\}\frac{{a}_\mu}{\rho^8}
$$
\begin{equation}
+5\left[9\lambda^2
-2\rho^2({\dot a}\cdot c)
\right]\frac{{\dot a}_\mu}{\rho^7}
-10\lambda\,\frac{{\ddot a}_\mu}{\rho^6}
+\frac{{\stackrel{\ldots}{a}}_\mu}{\rho^5}\,.
\label
{cal-A-12}
\end{equation}                       

It is possible to show  \cite{Kosyakov2008a} that 
%the corresponding ${\cal F}^{(2n)}_{\mu\nu}$ are 
%expressed in terms of ${\cal A}^{(2m)}_{\mu}$ with the aid of remarkably simple and elegant 
%algebraic relationships
the 
%retarded 
field strength ${\cal F}^{(2n)}_{\mu\nu}$ generated by a point charge living in a
$2n$-dimensional world is expressed in terms of the 
%retarded 
vector potentials 
${\cal A}^{(2m)}_{\mu}$ due to this charge in $2m$-dimensional worlds nearby,
\begin{equation}
{\cal F}^{(2)}
=-{\cal A}^{(2)}\wedge{\cal A}^{(4)}\,,
\label
{F-2}
\end{equation}                       
\begin{equation}
{\cal F}^{(4)}
=-{\cal A}^{(2)}\wedge{\cal A}^{(6)}\,,
\label
{F-4}
\end{equation}                       
\begin{equation}
{\cal F}^{(6)}
=
-{\cal A}^{(2)}\wedge{\cal A}^{(8)} 
-
{\cal A}^{(4)}\wedge{\cal A}^{(6)}\,,
\label
{F-6}
\end{equation}                       
\begin{equation}
{\cal F}^{(8)}
=
-{\cal A}^{(2)}\wedge{\cal A}^{(10)} 
-
2{\cal A}^{(4)}\wedge{\cal A}^{(8)}\,,
\label
{F-8}
\end{equation}                       
\begin{equation}
{\cal F}^{(10)}
=
-{\cal A}^{(2)}\wedge{\cal A}^{(12)} 
-
3{\cal A}^{(4)}\wedge{\cal A}^{(10)}
-
2{\cal A}^{(6)}\wedge{\cal A}^{(8)}\,. 
\label
{F-10}
\end{equation}                       
We mention in passing that these algebraic relationships are not only remarkably simple and 
elegant, but also physically surprising.
The world line $z^\mu(s)$ of the charge generating these field configurations is described by 
various numbers of the principal curvatures $\kappa_j$ for different spacetime dimensions.
To be specific, consider Eq.~(\ref{F-4}).
The world line which gives rise to ${\cal A}^{(2)}_{\mu}$ is a planar curve, specified solely by 
$\kappa_1$, while that giving rise to ${\cal A}^{(6)}_{\mu}$ is a curve characterized by five 
essential parameters $\kappa_1$,  $\kappa_2$, $\kappa_3$,  $\kappa_4$, and $\kappa_5$.  
If we regard the world line $z^\mu(s)$ in ${\mathbb M}_{1,2n-1}$ as the basic object, then both 
projections of this curve onto lower-dimensional spacetimes and its extensions to higher-dimensional 
spacetimes are rather arbitrary.
However, this arbitrariness does not show itself in Eqs.~(\ref{F-2})--(\ref{F-10}) linking  
${\cal F}^{(2n)}_{\mu\nu}$ and ${\cal A}^{(2m)}_{\mu}$.

To reveal the behavior of the retarded electromagnetic field at spatial infinity, we  
segregate in ${\cal A}^{(2n+2)}$ the term scaling as $\rho^{-n}$ by introducing the vectors  
\begin{equation}
{\mathfrak b}^{(2n+2)}_\mu
=
\lim_{\rho\to\infty}\,\rho^{n}{\cal A}^{(2n+2)}_\mu
\label
{infrare-a}
\end{equation}                       
and
\begin{equation}
{\bar{\cal A}}^{(2n+2)}_\mu
=
\frac{1}{\rho^n}\,{\mathfrak b}^{(2n+2)}_\mu]\,.
\label
{infrared-A}
\end{equation}                       
All infrared irrelevant terms are erased by this limiting procedure, so that
\begin{equation}
{{\cal A}}^{(2)}
\wedge 
{\bar{\cal A}}^{(2n+2)}
\label
{infrared-F}
\end{equation}                       
represents the long-distance asymptotics of ${\cal F}^{(2n)}$ 
\cite{Kosyakov1999},  \cite{Kosyakov2007},  \cite{Kosyakov2008a}, \cite{GursesSarioglu2003}.
To see this, it is sufficient to compare the behavior of ${\cal A}^{(2)}\wedge{\cal A}^{(2n+2)}$ 
and ${\cal A}^{(4)}\wedge{\cal A}^{(2n)}$.
Because the least falling terms of ${\cal A}^{(2n+2)}$ and ${\cal A}^{(2n)}$ scale, respectively, 
as $\rho^{-n}$ and $\rho^{1-n}$,  the leading long-distance asymptotics of ${\cal A}^{(2)}\wedge{\cal A}^{(2n+2)}$ 
is given by $\rho^{1-n}$ while that of ${\cal A}^{(4)}\wedge{\cal A}^{(2n)}$ is given by 
$\rho^{-n}$.
The same is true for the comparison of the long-distance behavior of ${\cal A}^{(2)}\wedge{\cal A}^{(2n+2)}$
and that of the remaining two-forms contained in ${\cal F}^{(2n)}$.  

We write explicitly ${\mathfrak b}^{(2n+2)}_\mu$ for $n=1,2,3,4,5$: 
\begin{equation}
{\mathfrak b}^{(4)}_\mu=v_\mu\,,
\label
{mf-4}
\end{equation}                       
\begin{equation}
{\mathfrak b}^{(6)}_\mu=-\left({a}\cdot c\right)v_\mu+{a_\mu}\,,
\label
{mf-6}
\end{equation}                       
\begin{equation}
{\mathfrak b}^{(8)}_\mu=\left[3\left({a}\cdot c\right)^2-\left({\dot a}\cdot c\right)\right]{v_\mu} 
-3\left({a}\cdot c\right){a_\mu}+{{\dot a}_\mu}\,,
\label
{mfr-8}
\end{equation}                       
\begin{eqnarray}
{\mathfrak b}^{(10)}_\mu
=-\left[15\left({a}\cdot c\right)^3
-10\left({a}\cdot c\right)\left({\dot a}\cdot c\right)
+\left({\ddot a}\cdot c\right)\right]{v_\mu} 
\nonumber\\
+\left[15\left({a}\cdot c\right)^2
-4\left({\dot a}\cdot c\right)\right]{a_\mu}
-6\left({a}\cdot c\right){{\dot a}_\mu}
+{\ddot a_\mu}\,,
\label
{mfr-10}
\end{eqnarray}                       
\begin{eqnarray}
{\mathfrak b}^{(12)}_\mu
=\left\{
5\left[
3\cdot 7\left({a}\cdot c\right)^2\left(\left({a}\cdot c\right)^2
-\left({\dot a}\cdot c\right)\right)
+2\left({\dot a}\cdot c\right)^2
+ 
3\left({a}\cdot c\right)\left({\ddot a}\cdot c\right)
\right]
-(\stackrel{\ldots}{a}\cdot c)
\right\}
{v_\mu} 
\nonumber\\
-5\left\{3\left({a}\cdot c\right)
[7\left({a}\cdot c\right)^2
-
4({\dot a}\cdot c)]
+({\ddot a}\cdot c)\right\}{{a}_\mu}
\nonumber\\
+5\left[9\left({a}\cdot c\right)^2
-2({\dot a}\cdot c)\right]{\dot a}_\mu
-10\left({a}\cdot c\right){\ddot a}_\mu
+{\stackrel{\ldots}{a}}_\mu\,.
\label
{mfr-12}
\end{eqnarray}                       

The radiated energy-momentum is defined by
\begin{equation}
{\cal P}^{\mu}=
\int_{\Sigma} d\sigma_\nu\,\Theta^{\mu\nu}_{\rm II}\,,
\label
{six-radiat}
\end{equation}                                       
where $\Sigma$ is a $(2n-1)$-dimensional spacelike hypersurface.
$\Theta^{\mu\nu}_{\rm II}$ involves only integrable singularities, and $\partial_\nu\Theta^{\mu\nu}_{\rm II}=0$.
Therefore, the surface of integration $\Sigma$ in (\ref{six-radiat}) may be chosen arbitrarily.
It is convenient to deform $\Sigma$ to a tubular surface ${T}_\epsilon$ of small invariant radius 
$\rho=\epsilon$ enclosing the world line. 
The surface element on this tube is 
\begin{equation}
d\sigma^\mu=\partial^\mu\!\rho\,\rho^{2n-2}\,d\Omega_{2n-2}\,ds
=(v^\mu+\lambda c^\mu)\,\epsilon^{2n-2}\,d\Omega_{2n-2}\,ds\,.
\label
{tube-measure}
\end{equation}                                       

Since 
the radiation flux through a $(2n-2)$-dimensional sphere enclosing
the source is constant for any radius of the sphere,
%Therefore, 
the terms of $\Theta_{\mu\nu}$ responsible for this flux 
scale as
$\rho^{2-2n}$.
Equation (\ref{six-radiat}) becomes 
\begin{equation}
{{\cal P}}_{\mu}^{(2n)}=
-\frac{1}{N_n^2\Omega_{2n-2}}
\int^s_{-\infty}ds
\int d\Omega_{2n-2}\, c_\mu 
\left({\mathfrak b}^{(2n+2)}\right)^2\,,
\label
{radiation-rate}
\end{equation}                         
so that the radiation rate is given by
\begin{equation}
{\dot{\cal P}}_{\mu}^{(2n)}
=
-\frac{1}{N_n^2\Omega_{2n-2}}
\int d\Omega_{2n-2}\, c_\mu 
\left({\mathfrak b}^{(2n+2)}\right)^2\,.
\label
{radiat-rate}
\end{equation}                         

A large list of generic formulas describing radiation of tensor fields of various ranks from an 
accelerated point source moving in ${\mathbb R}_{1,2n-1}$, which allows to evaluate the total 
intensity and radiated momentum, has been given in \cite{MironovMorozov2008b}). 

The above results may give the impression that self-interaction in any even dimension is qualitatively 
the same.
Indeed, it seems to be imperative that the Maxwell--Lorentz theory, generalized to $2n$ dimensions,
rearranges in a standard way to bring into existence radiation and dressed particles whose momenta 
obey the balance equation (\ref{balance-LD}).
We now take $d=2$ and $d=6$ as examples 
which impeach this impression, namely we mean to show that 
the  $d=2$ and $d=6$ pictures differ drastically from the $d=4$ picture \cite{Kosyakov1999}, 
 \cite{Kosyakov2007}. 
Two-dimensional electrodynamics is immune from rearrangement: there are neither radiation nor
dressing.
Six-dimensional electrodynamics does rearrange, but the upshot is surprising.
The six-momentum of a dressed particle is not defined uniquely; this quantity is given by two 
different expressions $\mathfrak{p}^\mu$ and $p^\mu$.
Each is the six-momentum in a particular context.
If we take the balance equation (\ref{balance-LD}), then the dressed particle is represented by 
$\mathfrak{p}^\mu$, but if Newton's second law (\ref{LD-Newton}) is regarded as the equation of 
motion for the dressed particle, its dynamical state is specified by $p^\mu$.  

\subsubsection{${\mathbb R}_{1,1}$}
\label
{d=2}
The retarded vector potential, generated by a charged point particle, 
\begin{equation}
A_{\mu}=-eR_\mu\,, 
\label
{potential-2D}
\end{equation}                                             
and the associated retarded field strength, 
\begin{equation}
F_{\mu\nu}=e\left(c_\mu v_\nu-c_\nu v_\mu\right), 
\label
{strength-2D}
\end{equation}                                             
are not singular, even if $F_{\mu\nu}$ is discontinuous on the world line.
Besides, $F_{\mu\nu}$ is independent of acceleration.

The stress-energy tensor for the field (\ref{strength-2D}) is 
\begin{equation}
\Theta^{\mu\nu}
={\frac14}\,e^2\left(c^{\mu} v^{\nu}+\,c^{\nu} v^{\mu}-c^{\mu} c^{\nu}\right)=
{\frac14}\,e^2\left(v^{\mu} v^{\nu}-u^{\mu} u^{\nu}\right)={\frac14}\,e^2\eta^{\mu\nu}\,.
\label
{Theta self}
\end{equation}                                           
Here, the completeness relation $v^{\mu} v^{\nu}-u^{\mu}u^{\nu}=\eta^{\mu\nu}$ stemming from the 
fact that $v^{\mu}$ and $u^{\mu}$ form a basis in ${\mathbb R}_{1,1}$ has been used.
It is evident that $\partial_\mu \Theta^{\mu\nu}=0$.
Expression (\ref{Theta self}) contains the term $-\frac14\, e^2 c^\mu c^\nu$. 
Is it possible to interpret it as radiation?
Although this term meets three conditions (\ref{i}), (\ref{ii}), and 
(\ref{iii-D}), the fourth condition (\ref{iv}) is violated because 
$-\frac14\, e^2c^\mu c^\nu$ is similar in its spatial behavior 
to the rest of the stress-energy tensor (\ref{Theta self}), contrary to the 
requirement that the radiation be asymptotically separated from the bound part. 
Hence, the electromagnetic radiation is absent from ${\mathbb R}_{1,1}$.
% two-dimensional spacetime. 

The problem of two particles in ${\mathbb R}_{1,1}$ is readily translated into the problem of two 
parallel plates of a planar immense capacitor in ${\mathbb R}_{1,3}$.
There is only an electric field ${\bf E}$ between the plates, which is constant for any separation 
and velocity of the plates.
The same is true for a system of $N$ charged particles which can be thought of as a system of $N$ 
parallel charged plates of infinite extent.
An important point is that we are dealing with a 
well-defined problem only for systems
whose total charge $Q=\sum e_I$ is vanishing, otherwise infrared divergence of the self-energy 
ensues.
Indeed, restricting our consideration to a single-particle system, $Q=e\ne 0$, we find 
\begin{equation}
P^{\mu}=\int d\sigma_\nu\,\Theta^{\mu\nu}=\frac12\,e^2 v^{\mu}\,{L}\,, \quad {L\to\infty}\,.
\label
{infrared}
\end{equation}                                           
In contrast, if $Q=0$, then the integration range in every integral of the type shown in Eq.~(\ref{infrared})
is finite, as Gauss' law suggests, and the integrals turn out to be convergent.

The resulting dynamics is thus not subject to rearrangement.
The equation of motion for $I$th particle in which the general retarded solution to Maxwell's equations is
used reads
\begin{equation}
m_I{a}_I^\mu=e_I\sum_{J=1}^N e_J\,v_{\lambda}^I\left(v_J^\lambda c_J^\mu-c_J^\lambda v_J^\mu\right).
\label
{equation_of_motion}
\end{equation}                                           
The set of $N$ ordinary differential equations of the form of Eq.~(\ref{equation_of_motion}) is 
integrable.
Exact solutions to (\ref{equation_of_motion}) \cite{Kosyakov2007} show that every particle 
moves along a hyperbolic world line.
Therefore, the Maxwell--Lorentz electrodynamics in ${\mathbb R}_{1,1}$ is locally reversible 
despite the fact that the retarded boundary condition has been switched on. 

Surprisingly, however, the global dynamical picture is irreversible.
Since this subtle feature of the two-dimensional dynamics, underlying the mechanism of 
self-interaction in classical strings, will be reviewed in Sec.~\ref{string}, we will defer its 
consideration until then.

It is interesting to compare the situation in a genuine two-dimensional world with that in an 
effective two-dimensional world which arises when a charged particle is placed in a straight line in 
ambient space. 
The Abraham--Lorentz--Dirac equation (\ref{LD}) then reduces to Eq.~(\ref{one-dim_ALD}), whence 
it follows that the effective dynamics is irreversible.
To explain the discrepancy, let us note that only kinematical aspects of the effective description 
are in fact two-dimensional, whereas self-interaction still remains four-dimensional, because its 
features are attributable to the four-dimensional Lien\'ard--Weichert field developed in ambient 
spacetime.

Why is the Maxwell--Lorentz electrodynamics in ${\mathbb R}_{1,1}$ unaffected by rearrangement?
What is the dissimilarity between ${\mathbb R}_{1,3}$ and ${\mathbb R}_{1,1}$ that may render 
unstable systems stable? 
The automorphism group of ${\mathbb R}_{1,3}$ is the semidirect product of the Lorentz group 
SO$(1, 3)$ and the translation group $T_4$, while that of ${\mathbb R}_{1,1}$ is the semidirect 
product of SO$(1, 1)$ and $T_2$. 
The geometrical dissimilarity stands out:  SO$(1, 3)$ is non-Abeilian, and SO$(1, 1)$ is Abeilian.
It seems plausible that just this distinctive feature resolves the contradiction between the 
manifestations of electromagnetic self-interaction in ${\mathbb R}_{1,3}$ and ${\mathbb R}_{1,1}$.

\subsubsection{${\mathbb R}_{1,5}$}
\label
{d=6}
To grasp the distinctive features of self-interaction in the six-dimensional Maxwell--Lorentz 
theory, let us turn to the ultraviolet behavior of the retarded field due to a point charge, 
\begin{equation}
F=\frac{e}{3\rho^4}\left[c\wedge U +\rho\left({a\wedge v}\right)\right],
\label
{6-strength}
\end{equation}                                             
\begin{equation}
U^\mu =\rho^2{\dot a}^\mu-3\lambda\rho\,{a^\mu}+
\left[3\lambda^2-\rho^2\,({\dot a}\cdot c)\right]{v^\mu}\,.
\label
{6-vector-V}
\end{equation}                                           

The electromagnetic six-momentum $P^\mu$ would result from integrating $\Theta^{\mu\nu}$ over a 
five-dimensional spacelike hypersurface.
But the obstacle to this integration is that the $F_{\mu\nu}$ exhibits  non-integrable
singularities on the world line.
By (\ref{6-strength}) and (\ref{6-vector-V}),
\[
F_{\mu\alpha}F^\alpha_{\hskip1.5mm\nu}
=\frac{e^2}{9}\,\Biggl[-\frac{a_\mu a_\nu+a^2\,v_\mu v_\nu}{\rho^6}+(c\cdot V)\,
(c_\mu V_\nu+c_\nu V_\mu)-c_\mu c_\nu\,V^2+\frac{a_\mu V_\nu+a_\nu V_\mu}{\rho^3}
\]
\begin{equation}
+(a\cdot V)\,\frac{v_\mu c_\nu+v_\nu c_\mu}{\rho^3}
-(v\cdot V)\,\frac{a_\mu c_\nu+a_\nu c_\mu}{\rho^3}
-\frac{\lambda+1}{\rho^4}\,(v_\nu V_\mu+v_\mu V_\nu)\Biggr],
\label
 {FF-6}
\end{equation}                               
so that
\begin{equation}
F_{\alpha\beta}F^{\alpha\beta}=\frac{2e^2}{9}\,\left[\,
\frac{a^2}{\rho^6}-(c\cdot V)^2-\frac{2}{\rho^3}\,(a\cdot V)+
\frac{2}{\rho^4}\,({\lambda+1})\,(v\cdot V)\right].
\label
{F-dot-F-6}
\end{equation}                               
Since the element of measure on a five-dimensional spacelike hyperplane is proportional to 
$\rho^4 d\rho$, the integration of  $\Theta^{\mu\nu}$ results in {cubic} and {linear} divergences.
It is clear from (\ref{FF-6}) and (\ref{F-dot-F-6}) that the cubic divergence occurs even in the 
static case.
In contrast, the linear divergence, which owes its origin to the terms of $F_{\mu\nu}$ scaling as 
$\rho^{-2}$, 
appears only for curved world lines, that is, in the case that either $a^\mu$ or ${\dot a}^\mu$, or 
both are  nonzero.
This implies that the Poincar\'e--Planck action for a bare particle (\ref{S-Planck-general}) 
must be augmented by the addition of terms containing higher derivatives of $z^\mu$ to absorb the 
linear divergence.

The Lagrangian with higher derivatives is said to describe rigid particles. 
The simplest reparametrization invariant Lagrangian for a rigid bare particle is 
\begin{equation}
L=-\sqrt{{v}\cdot{v}}\left(m_0-\nu_0 {a}^2\right),
\label
{rigid-term-action}
\end{equation}                                    
where $m_0$ and $\nu_0$ are real parameters.
The corresponding six-momentum is  
\begin{equation}
p_{\hskip0.5mm 0}^{\mu}=m_0 v^\mu+\nu_0 (2{\dot a}^\mu+3a^2 v^\mu)\,.
\label
{bare-6-momentum}
\end{equation}                                       
On dimensional grounds, it is possible to show that the linearly divergent term arising from the 
integration of $\Theta^{\mu\nu}$ involves ${v}^\mu$ and ${\dot a}^\mu$ in exactly the same way as 
the second term of (\ref{bare-6-momentum}) does.
The cubic and linear divergences are thus eliminated through the respective renormalization of 
$m_0$ and $\nu_0$. 

%The need for using the terms of rigid dynamics to absorb extra ultraviolet divergences resembles the 
%fact that the term $\frac14\lambda({\bar\phi}\phi)^2$ is required in scalar quantum electrodynamics for 
%absorbing extra quadratic divergences of the ``tadpole'' type arising in four-dimensional perturbative 
%calculations, even if this problem is absent from the two-dimensional context. 

The classical dynamics governed by the action (\ref{S-Planck-general})--(\ref{S-em-field})
proves inconsistent for $d>4$ because ultraviolet divergences of the self-energy of a point charge 
proliferate with $d$, while the Poincar\'e--Planck term (\ref{S-Planck-general}) does not involve 
enough free parameters to eliminate all the divergences through the redefinition of these 
parameters.
If we mean to explore ${\mathbb R}_{1,2n-1}$ with $n>2$, preserving the laws of Maxwell's 
electrodynamics encoded in the Schwarzschild and Larmor terms (\ref{S-Schwarzs}) and (\ref{S-em-field}), 
we have to use a rigid particle dynamics. 
This statement has been justified from various perspectives \cite{Kosyakov1999}, \cite{Gal'tsov2002}, 
\cite{Kazinski2002}, \cite{Yaremko2004}, \cite{Galakhov2008}, \cite{BirnholtzHadar2014}).

A regular way for deriving the equation of motion for a dressed charged particle is to evaluate 
regularized expressions for $P^\mu$, renormalize divergent terms, and segregate finite terms.
But we take another route which requires fewer efforts \cite{Kosyakov1999}.
We determine the radiation rate, and then make use of the fact that the equation of motion 
for a dressed particle involves the projector $\stackrel{\scriptstyle v}{\bot }$\,. 
This procedure is independent of a particular regularization prescription because the radiation 
momentum integral is always convergent. 

In order to clarify the origin of the projector $\stackrel{\scriptstyle v}{\bot}$\,, we digress for a while and 
recall the reader the main implication of reparametrization invariance resulting from Noether's 
second theorem  \cite{Noether1918}.
Consider an infinitesimal change of the parameter of a world line,
\begin{equation}
\delta\tau=\epsilon (\tau)\,,
\label
{reparametrization-infinitesimal}
\end{equation}                        
where $\epsilon(\tau)$ is an arbitrary smooth function of $\tau$, close to zero, which becomes vanishing 
at the end points of integration.
Variation of $\tau$ implies the corresponding variation of the world line coordinates
\begin{equation}
\delta z^\mu={\dot z}^\mu\epsilon\,.
\label
{reparam-infsm-coord}
\end{equation}                        
In response to the variations (\ref{reparametrization-infinitesimal}) and (\ref{reparam-infsm-coord}),
the action $S[z]$ varies as 
\begin{equation}
\delta S= \int d\tau\,{\cal E}_\mu{\dot z}^\mu\epsilon\,.
\label
{variation-S-sec2.6}
\end{equation}                        
Let $S$ be reparametrization invariant, $\delta S=0$.
Because $\epsilon$ is an arbitrary function of $\tau$, one concludes that
\begin{equation}
{\dot z}^\mu{\cal E}_\mu=0\,.
\label
{linear-dependence-Eulerians}
\end{equation}                        
This equation expresses Noether's second theorem: if the action is invariant under the group of 
transformations involving an arbitrary function $\epsilon(\tau)$, then the Eulerians ${\cal E}_\mu$ 
are linearly dependent. 
The identity (\ref{linear-dependence-Eulerians}) suggests that ${\cal E}_\mu$ contains the 
projection operator on a hyperplane with normal ${\dot z}^\mu$ defined by Eq.~(\ref{projection-operator-df}).
This operator annihilates identically any vector parallel to ${\dot z}^\mu$.  
Reparametrization invariance bears on the projection structure of the basic dynamical law for a 
bare particle which can be written in the form of Eq.~(\ref{LD-Newton}).
The availability of $\stackrel{\scriptstyle v}{\bot}$\, in the equation of motion for a dressed
particle is therefore the imprint of reparametrization invariance which is to be preserved by the
rearrangement. 

With reference to the general expression for $2n$-dimensional radiated energy-momentum, 
Eq.~(\ref{radiation-rate}), we specify it to $d=6$:  
\[
{\cal P}^{\mu}
=
-\frac{e^2}{9}\int d\tau\int d\Omega_4\Biggl\{\left[(\stackrel{\scriptstyle v}
{\bot}{\dot a})^2+9(a\cdot u)^2a^2+9(a\cdot u)^4+({\dot a}\cdot u)^2\right]
v^\mu
\]
\begin{equation}
-\left[3\,\frac{da^2}{ds}\,(a\cdot u)+6(a\cdot u)^2({\dot a}\cdot u)\right]u^\mu\Biggr\}\,.
\label
{tube-int-6D}
\end{equation}                                       
To make the solid angle integration, we apply the readily derivable formulas 
\begin{equation}
\int d\Omega_{4}\, u_\mu u_\nu=-\frac{\Omega_{4}}{5}\,
\stackrel{\scriptstyle v}{\bot}_{\hskip0.5mm\mu\nu}\,,
\label
{solid-int-6D-2}
\end{equation}                                       
\begin{equation}                                       
\int d\Omega_{4}\,u_\alpha u_\beta u_\mu u_\nu=\frac{\Omega_{4}}{5\cdot
7}\,\left(
\stackrel{\scriptstyle v}{\bot}_{\hskip0.5mm\mu\nu}\,
\stackrel{\scriptstyle v}{\bot}_{\hskip0.5mm\alpha\beta}+
\stackrel{\scriptstyle v}{\bot}_{\hskip0.5mm\alpha\mu}\,
\stackrel{\scriptstyle v}{\bot}_{\hskip0.5mm\beta\nu}+
\stackrel{\scriptstyle v}{\bot}_{\hskip0.5mm\alpha\nu}\,
\stackrel{\scriptstyle v}{\bot}_{\hskip0.5mm\beta\mu}\right),
\label
{solid-int-6D-4}
\end{equation}                                       
which gives 
\begin{equation}                                       
{\cal P}^{\mu}=
\frac{e^2}{9}
\int_{-\infty}^s d\tau\left\{-\frac{4}{5}\left[{\dot a}^2-\frac{16}{7}\,
(a^2)^2 \right]v^\mu-\frac{3}{7}\, \frac{da^2}{ds}\,a^\mu
+\frac{6}{5\cdot 7}\,a^2
(\stackrel{\scriptstyle v}{\bot}{\dot a})^\mu\right\},
\label
{6D-rad}
\end{equation}                                       
and
\begin{equation}
{\dot {\cal P}}^{\mu}=\frac{e^2}{9}\left\{-\frac{4}{5}\,\left[{\dot a}^2
-\frac{16}{7}\,(a^2)^2 \right]v^\mu-\frac{3}{7}\,\frac{da^2}{ds}\,a^\mu 
+\frac{6}{5\cdot 7}\,a^2(\stackrel{\scriptstyle v}{\bot}{\dot a})^\mu\right\}.
\label
{P-rad-6D}
\end{equation}                               
One can easily verify the inequality 
\begin{equation}
v\cdot{\dot{\cal P}}=-\frac{4e^2}{45}\left[
{\dot a}^2-\frac{16}{7}\,(a^2)^2 \right]>0\,, 
\label
{E-rad-6D}
\end{equation}                               
which evidences that ${\cal P}^{0}$ represents positive field energy flowing outward from the source.
 
The bound momentum contains divergent and finite parts, ${\mathfrak{p}}^{\mu}=p_{\hskip0.5mm\rm div}^{\mu}+p_{\hskip0.5mm\rm fin}^{\mu}$.
The finite part $p_{\hskip0.5mm\rm fin}^{\mu}$ is free of dimensional parameters other than the 
overall factor $e^2$, and inherits the appropriate dimension from kinematical variables:
\begin{equation}
p_{\hskip0.5mm\rm fin}^{\mu}=c_1\,{\ddot a}^{\mu}+
c_2\,a^2a^\mu+c_3\,\frac{da^2}{ds}\,v^\mu\,,
\label
{p-mu-f-6D}
\end{equation}                               
where $c_1$, $c_2$, and $c_3$ are numerical coefficients.
The presence of $\stackrel{\scriptstyle v}{\bot }$\, in the equation of motion for a dressed 
particle implies the transversality condition 
\begin{equation}
v\cdot\left({\dot p}_{\hskip0.5mm\rm fin}+{\dot{\cal P}}\right)=0\,.
\label
{p-f+cal-P-perp-v}
\end{equation}                               
With the identities 
\begin{equation}
(a\cdot v)=0\,,
\quad 
(\dot{a}\cdot v)=-a^2\,,
\quad 
(\ddot{a}\cdot v)=-\frac{3}{2}\,\frac{da^2}{ds}\,,
\quad 
(\stackrel{\ldots}{a}\cdot\,v)=-2\,\frac{d^2a^2}{ds^2}+\dot{a}^2\,,
\label
{identities-extnd-6D}
\end{equation}                               
this gives
\begin{equation}
p_{\hskip0.5mm\rm fin}^{\mu}
=\frac{4}{45}\,e^2\left({\ddot a}^{\mu}+\frac{16}{7}\,a^2\,a^\mu+2\,\frac{da^2}{ds}\,v^\mu\right).
\label
{05}
\end{equation}                        

The kinematical structure of the divergent part $p_{\hskip0.5mm\rm div}^{\mu}$ is similar to that of 
the bare particle momentum (\ref{bare-6-momentum}). 
We therefore should renormalize $m_0$ and $\nu_0$ in $p_{\hskip0.5mm0}^{\mu}$, and combine it with 
$p_{\hskip0.5mm\rm fin}^{\mu}$, 
Eq.~(\ref{05}), to yield
\begin{equation}
{\mathfrak{p}}^{\mu}
=
m v^\mu+\nu\,(2{\dot a}^\mu+3a^2\,v^\mu)+\frac{4}{45}\,e^2\left({\ddot a}^{\mu}
+\frac{16}{7}\,a^2a^\mu+2v^\mu\,\frac{da^2}{ds}\right).
\label
{pi}
\end{equation}                            

The six-momentum ${\wp}^{\mu}$ extracted from an external field $F^{\mu\nu}_{\hskip0.3mm \rm ex}$ is 
found by integrating the mixed term of the stress-energy tensor $\Theta^{\mu\nu}_{\hskip0.3mm\rm mix}$ 
over a tube ${T}_\epsilon$ of small radius $\epsilon$  enclosing the world line,
\begin{equation}
{\wp}^{\mu}
=\int_{{T}_\epsilon}d\sigma_\nu\,\Theta^{\mu\nu}_{\rm mix}
=-\int^{{s}}_{-\infty}d\tau\,e F^{\mu\nu}_{\hskip0.3mm \rm ex} v_\nu\,,
\label
{P-mix-Lorentz-6D}
\end{equation}                        
whence 
\begin{equation}
{\dot{\wp}}^{\mu}=-f^\mu=-e F^{\mu\nu}_{\hskip0.3mm \rm ex} v_\nu\,.
\label
{Lorentz-6D}
\end{equation}                        

The local energy-momentum balance reads
\begin{equation}
\dot{\mathfrak{p}}^{\mu}+{\dot{\cal P}}^{\mu}+{\dot {\wp}}^{\mu}=0\,.
\label
{balance-p-6D}
\end{equation}                      
Substituting (\ref{pi}), (\ref{P-rad-6D}), and (\ref{Lorentz-6D}) in 
(\ref{balance-p-6D}), we obtain the equation of motion for a dressed charged particle 
\begin{equation}
\stackrel{\scriptstyle v}{\bot}\!(\,{\dot p}-f)=0\,,
\label
{eq-motion-dress-6d}
\end{equation}                            
where 
\begin{equation}
p^\mu=m v^\mu+\nu\,(2{\dot a}^\mu+3a^2 v^\mu)+\frac{1}{9}\,e^2\left(\frac{4}{5}\,\ddot{a}^\mu+2a^2a^\mu+
\frac{da^2}{ds}\,v^\mu\right).
\label
{p-mu-6D}
\end{equation}                      
In an explicit form, this equation is written as
\begin{equation}
m a^\mu+\nu\!\left(2{\ddot a}^\mu+3a^2 a^\mu+3\,\frac{da^2}{ds}\,v^\mu\right)+\Gamma^\mu=f^\mu\,,
\label
{prbl10.2.7-2}
\end{equation}                                     
\begin{equation}
\Gamma^\mu=
\frac{e^2}{9}\left[\frac{4}{5}\,{\stackrel{\ldots}a}^{\hskip0.3mm\mu}+
2a^2{\dot a}^\mu+3\,\frac{da^2}{ds}\,a^\mu+\left(\frac{8}{5}\,
\frac{d^2a^2}{ds^2}+
2(a^2)^2-\frac{4}{5}\,{\dot a}^2\right)v^\mu\right].
\label
{Prbl10.2.7-1}
\end{equation}                                     

A surprising result is the occurrence of two different six-momenta $\mathfrak{p}^\mu$ and $p^\mu$,
defined by Eqs.~(\ref{pi}) and (\ref{p-mu-6D}).
Each can be recognized as the energy-momentum of the dressed particle: the $\mathfrak{p}^\mu$ plays 
this role in the balance equation (\ref{balance-p-6D}), whereas Newton's second law (\ref{eq-motion-dress-6d}) 
tells us that the dressed particle is endowed with the six-momentum $p^\mu$.  

The rearrangement outcome, Eqs.~(\ref{balance-p-6D}) and (\ref{eq-motion-dress-6d}), has been 
obtained with the aid of the condition of transversality (\ref{p-f+cal-P-perp-v}), which greatly 
alleviates the problem.
However, this trick is inadequate for $d\ge 8$  \cite{MironovMorozov2008a}. 
In higher dimensions, it seems impossible to avoid tedious calculations with explicit regularizations 
and choosing counterterms, similar to those proposed in \cite{Kazinski2002}, \cite{Galakhov2008}. 

When comparing the symmetries behind the rearrangement in four and six dimensions, one further 
comment is in order.
It has been shown in Sec.~\ref{Radiation} that the retarded field $F$ generated by a point charge 
in ${\mathbb R}_{1,3}$, Eq.~(\ref{F-LW-III}), is a decomposable 2-form invariant under local 
SL$(2,{\mathbb R})$ transformations.
Recall that sl$(2,{\mathbb R})$ is equivalent to sp$(1)$, see, {\it e.~g.}, \cite{BarutRaczka1977}.
In contrast, the retarded field $F$ in ${\mathbb R}_{1,5}$, Eqs.~(\ref{6-strength}) and (\ref{6-vector-V}), 
contains two exterior products. 
The 2-form $F$ of this structure is invariant under similar transformations forming the Sp$(2)$ 
group.
A key geometrical distinction between these groups is that Sp$(2)$ is non-Abeilian, and Sp$(1)$ is 
Abeilian.
It is conceivable that this distinction may explain the fact that the six-momentum of a dressed
particle appears in two guises, $\mathfrak{p}^\mu$ and $p^\mu$, while its four-dimensional 
counterpart  $p^\mu$ is uniquely defined. 
In Sec.~\ref{GR}, some evidence in support of this suggestion will arise in a wider mathematical 
context of the Banach--Tarski theorem.

\subsection{Massless charged particles}
\label
{MASSLESS}
The idea of zero-mass particles has several essential aspects.
We begin with the very possibility to give a consistent Lagrangian formulation of a modified
Maxwell--Lorentz electrodynamics involving massless charged particles.
Imagine a particle which is moving along a smooth null world line,
\begin{equation}
{\dot z}^2=0\,.
\label
{dot-z-sqr=0}
\end{equation}                                            
Here, $z_\mu(\tau)$ stands for a curve parametrized by a monotonically increasing variable $\tau$
associated with the evolution in time. 
We differentiate Eq.~(\ref{dot-z-sqr=0}) to give
\begin{equation}
{\dot z}\cdot{\ddot z}=0\,.
\label
{dot-cdot-ddot=0}
\end{equation}                                            
Because ${\dot z}_\mu$ is null, ${\ddot z}_\mu$ may be either spacelike or lightlike, aligned 
with ${\dot z}_\mu$.
If ${\ddot z}^2<0$, then the trajectory is bent.
As an illustration, we refer to a particle that orbits the origin in a circle of radius $r_0$ at a
constant angular velocity of $1/r_0$.
The world line is a helical null curve of radius $r_0$ wound around the time axis.
On a large scale, the particle traverses timelike intervals.

If ${\ddot z}^2=0$, then ${\ddot z}_\mu$  and ${\dot z}_\mu$ are parallel, and the trajectory is 
straight.
Although ${\ddot z}_\mu$ has nonzero components, the motion is uniform.
Indeed, whatever the evolution parameter $\tau$, the history is depicted by a straight world line 
aligned with the null vector ${\dot z}_\mu$. 
Therefore, ${\ddot z}_\mu$ is an artefact concerning the choice of $\tau$ for parametrizing the 
world line rather than a quantity related to actual acceleration.
To put it otherwise, planar world lines, other than straight lines, are unrelated to the history of 
particles moving with the speed of light.
The properties of null curves are discussed at greater length in \cite{Bonnor1969}. 

It is reasonable to suppose that massless particles move along null world lines.
The Poincar\'e--Planck action (\ref{S-Planck-general}) is unsuited for such particles 
because the four-momentum $p^\mu$ derived from it vanishes as  $m_0\to 0$, and the 
dynamics proves to be trivial.

In contrast, the action
\begin{equation}
{S}=-\frac{1}{2}\int d\tau\left(\eta\,{\dot z}^{2}+\frac{m_0^{2}}{\eta}\right),
\label
{Brink-action}
\end{equation}
proposed in \cite{Brink1976}, is sound for both massive and massless particles.
Here, $\eta$ is an auxiliary dynamical variable, called ``einbein''.
We assume that $\eta$ transforms as
\begin{equation}
{\eta}
\to
{\bar\eta}
=\frac{d{\bar\tau}}{d\tau}\,\eta
\label
{eta-eta'}
\end{equation}                                            
in response to reparametrizations ${\tau}\to{\bar\tau}$.
With this transformation law for $\eta$, the action (\ref{Brink-action}) is reparametrization 
invariant.

Varying the action (\ref{Brink-action}) with respect to $\eta$ gives 
\begin{equation}
{\dot z}^2 -\eta^{-2}\,m_0^2=0\,.
\label
{eq-eta}
\end{equation}
For $m_0\ne 0$, the solution to this equation is 
\begin{equation}
\eta=m_0\,({\dot z}\cdot{\dot z})^{-1/2}\,.  
\label
{solution}
\end{equation}
Substitution of (\ref{solution}) in (\ref{Brink-action}) regains (\ref{S-Planck-general}).
Therefore, the action (\ref{Brink-action}) is equivalent to the Poincar\'e--Planck action 
(\ref{S-Planck-general}) provided that the Euler--Lagrange equation (\ref{eq-eta}) is taken into 
account.
 
We combine (\ref{Brink-action}) with (\ref{S-Schwarzs}) and put $m_0=0$ to obtain the action which
encodes the dynamics of a massless charged particle,
\begin{equation}
{S}=-\int d\tau\left(\frac{1}{2}\,\eta\,{\dot z}^{2}+e{\dot z}\cdot A\right).  
\label
{Brink-massless-action}
\end{equation}

Variation of (\ref{Brink-massless-action}) with  respect to $\eta$ results in the basic constraint, 
Eq.~(\ref{dot-z-sqr=0}), which shows that {the massless particle governed by this action does move 
along null world lines}.
Note also that the resulting constraint, Eq.~(\ref{dot-z-sqr=0}), is independent of $\eta$.
 
Variation of (\ref{Brink-massless-action}) with respect to $z^\mu$ gives the equation of motion for 
a massless particle
\begin{equation}
\eta\,{\ddot z}^\mu+{\dot\eta}\,{\dot z}^\mu=e {\dot z}_\nu F^{\mu\nu}\,.
\label
{eq-motion-massive&massless-charged-p}
\end{equation}                        

Since $\eta$ does not contribute to the other Euler-Lagrange equation, Eq.~(\ref{dot-z-sqr=0}), this 
quantity is undetermined.
However, we are entitled to handle the reparametrization freedom making the dynamical equations 
as simple as possible.
For some choice of the evolution variable, the einbein can be converted to a constant,
$\eta=\eta_0$.
Then (\ref{eq-motion-massive&massless-charged-p}) becomes 
\begin{equation}
\eta_0{\ddot z}^\mu=e {\dot z}_\nu F^{\mu\nu}\,,
\label
{eq-motion-massless-charged-p}
\end{equation}                        
which closely resembles the equation of motion for a massive particle (\ref{Eulerian-part}).

The next important issue, concerning the high energy phenomenology, is the fact that zero-mass 
leptons do not appear to exist.
The existence of massless charged particles is ``clearly permitted by Maxwell's equations'' \cite{Bonnor1970}.
The same is also true for the interaction of massless particles with the Yang--Mills field.
Take, for example, neutrinos which interact with the SU$(2)\times$U$(1)$ Yang--Mills field in the 
Standard Model.
These particles were long assumed to be massless, but recent experimental data suggest that neutrinos 
are likely to be endowed with a finite, albeit small, mass. 
On the other hand, it is commonly supposed that quarks in quark--gluon plasma may reveal themselves 
as zero-mass particles because deconfinement triggers the chiral-symmetry-restoring phase 
transition, whereby quarks attain their masslessness.
Were such indeed the case, the disparity between realizable zero-mass quarks and unfeasible 
zero-mass leptons would be of even great concern.

However, the most important issue of the present discussion is {conformal} invariance.
The Maxwell--Lorentz electrodynamics of massless charged particles, as will soon become clear, does not 
experience rearrangement. 
Both {dressing} and {radiation} are {absent} from this theory \cite{Kosyakov2008b}.
If conformal invariance is overlooked, then one may be under wrong impression that this theory 
is amenable to the rearrangement  \cite{KazinskiSharapov2003}.
It is perhaps no wonder that a ``massless dressed particle'' is impossible to synthesize from a 
massless bare 
particle and electromagnetic field: for lack of $m_0$ in a scale invariant theory, the 
renormalization of mass is forbidden, 
whence it follows that the self-energy term must of necessity be vanishing.
However, the fact that a massless charged particle moving along a curved world line does not radiate 
may seem surprising if one remembers that the radiation is inevitable for an accelerated charged 
particle having arbitrarily small mass.
To explain the difference, we remark that a conformal theory need not be conceived as a continuous 
limit of fading away terms that violate conformal invariance.
As discussed above, the set of allowable histories of particles moving with the speed of light
is free from planar world lines, which, contrastingly, are well suited for the histories of subluminal particles.
In general, the set of physically allowable smooth timelike world lines does not asymptotically 
approach the set of physically allowable smooth null world lines.

To establish the above assertion on the absence of the rearrangement, we follow the previous line of 
reasoning, taking, as the starting point, the Noether identity 
\begin{equation}
\left(\partial_\mu T^{\lambda\mu}\right)(x)={1\over 8\pi}\left({\cal E}^{\lambda\mu\nu}F_{\mu\nu}\right)(x)
+{1\over 4\pi}\left({\cal E}_\mu F^{\lambda\mu}\right)(x)
+\int^\infty_{-\infty}d\tau\,\varepsilon^\lambda(z)\,\delta^4\left[x-z(\tau)\right],
\label
{Noether-1-id-em-zero}
\end{equation}                         
in which some terms are slightly modified as against Eq.~(\ref{Noether-1-id-em}) to accommodate the 
fact that the bare particle under examination moves along null world lines.
The proper time $s$ is unusable as a parameter of such curves, and we should look at 
another variable $\tau$, for example, the laboratory time $t$ in a particular Lorentz frame.
Accordingly, the modified ${\cal E}_\mu$, $\varepsilon^\lambda$, and $t^{\lambda\mu}$ are 
\begin{equation}
{\cal E}_\mu=\partial^\nu F_{\mu\nu}+4\pi e\int^\infty_{-\infty}\!d\tau\,{\dot z}_\mu(\tau)\,
\delta^4\left[x-z(\tau)\right]\,, 
\label
{Eulerian-em-II-zero-mass}
\end{equation}                        
\begin{equation}
\varepsilon^\lambda=\eta\,{\ddot z}^\lambda+{\dot\eta}\,{\dot z}^\lambda-e {\dot z}_\mu F^{\lambda\mu}\,,
\label
{varepsilon-massless-charged-p}
\end{equation}                        
\begin{equation}
t^{\lambda\mu}(x)=
\int_{{-\infty}}^{{\infty}} d\tau\,\eta(\tau)\,
{\dot z}^\lambda(\tau) {\dot z}^\mu(\tau)\,\delta^4\left[x-z(\tau)\right],
\label
{stress-energy-part}
\end{equation}                                            
while ${\cal E}^{\lambda\mu\nu}$ and $\Theta^{\lambda\mu}$ are given, as before, by
Eqs.~(\ref{Eulerian-em-I}) and (\ref{Theta-mu-nu}).

Evidently
\begin{equation}
T^\mu_{~\mu} =\Theta^\mu_{~\mu}+t^\mu_{~\mu}=0\,.
\label
{stress-energy-tr}
\end{equation}                                            
This implies that the system is invariant under the group of conformal transformations  C$(1,3)$
 \cite{Bessel-Hagen1921},  \cite{Fulton1962}.

The retarded electromagnetic field due to a charge moving along a null world line is 
\begin{equation}
F_{\mu\nu}=F^{\rm r}_{\mu\nu}+F^{\rm ir}_{\mu\nu}\,.
\label
{field=reg-irreg}
\end{equation}                                            
The first term $F^{\rm r}_{\mu\nu}$ (r for regular) is  
\begin{equation}
F^{\rm r}_{\mu\nu}=R_\mu V_\nu-R_\nu V_\mu\,,
\label
{field}
\end{equation}                                            
\begin{equation}
V_\mu=\frac{e}{\rho^2}\left(-{\dot z}_\mu\,\frac{{\ddot z}\cdot
R}{\rho}+{\ddot z}_\mu\right).
\label
{V-mu}
\end{equation}                                            
Here, all kinematical variables refer to the point $\tau_{\rm ret}$.
The scalar
\begin{equation}
\rho =R\cdot{\dot z}
\label
{rho-df-zero}
\end{equation}                                            
measures the separation between $z^\mu(\tau_{\rm ret})$ and $x^\mu$.
To see this, let us choose a particular Lorentz frame in which 
\begin{equation}
{\dot z}^\mu=\left(1,0,0,1\right), 
\quad
R^\mu=r\left(1,{\bf n}\right)
= r\left(1,\sin\vartheta\cos\varphi,\sin\vartheta\sin\varphi,\cos\vartheta\right).
\label
{lorentz-frame}
\end{equation}                                            
From $R\cdot{\dot z}=r\left(1-\cos\vartheta\right)$ follows that $\rho$ varies smoothly from 0 
to $\infty$ as $x^\mu$ moves away from $z^\mu(\tau_{\rm ret})$, except for the case that $R^\mu$ 
points in the direction of ${\dot z}^\mu$.
The surface swept out by the singular ray ${R}_\mu$  aligned with the tangent vectors ${\dot z}_\mu$
forms a two-dimensional warped manifold ${\cal M}_2$.

The second term $F^{\rm ir}_{\mu\nu}$ (ir for irregular) in Eq.~(\ref{field=reg-irreg}) is  
\begin{equation}
F^{\rm ir}_{\mu\nu}=e\,\frac{{\dot z}^2}{\rho^3}\left(R_\mu {\dot z}_\nu-R_\nu {\dot z}_\mu\right).
\label
{irreg-field}
\end{equation}  
Since ${\dot z}^2=0$, the irregular term $F^{\rm ir}_{\mu\nu}$ is everywhere zero except for the 
manifold  ${\cal M}_2$.

Note that the Gauss' law integral is saturated with the ${\bf E}^{\rm ir}$ alone. 
Indeed, integrating ${\bf E}^{\rm ir}$ over a sphere $r=\ell$ in a Lorentz frame, in which
${\dot z}^{\hskip0.3mm\mu}$ and $R^\mu$ take the form shown in Eq.~(\ref{lorentz-frame}), we obtain
\begin{equation}
\int d{\bf S}\cdot {\bf E}^{\rm ir}=e{\dot z}^2 \int_0^{2\pi}d\varphi \int_0^{\pi} d\vartheta\,
\frac{\sin\vartheta}{\left({\dot z}_0-|{\dot{\bf z}}|\cos\vartheta\right)^2}=4\pi e\,,
\label
{Gauss-law}
\end{equation}                                            
and furthermore, the same surface integration of ${\bf E}^{\rm r}$ can be shown to be zero.

It may be worth pointing out that the factor ${\dot z}^2$ disappears from Eq.~(\ref{Gauss-law}) 
because it is cancelled by the identical factor arising in the denominator owing to the solid angle 
integration of $\rho^{-2}$.
If we would have $\rho^{-s}$ with $s$ other than ${2}$, then this mechanism would fall short of
the required cancellation. 
Consider for example the term of the stress-energy tensor $\Theta^{\rm ir}_{\mu\nu}$ built from
$F^{\rm ir}_{\mu\nu}$.
By  (\ref{irreg-field}), 
\begin{equation}
F^{{\rm ir}\hskip0.5mm\alpha}_{\mu}F^{\rm ir}_{\alpha\nu}+\frac{\eta_{\mu\nu}}{4}\,
F^{\rm ir}_{\alpha\beta}F^{{\rm ir}\hskip0.5mm\alpha\beta}=e^2\,\frac{({\dot z}^2)^2}{\rho^4}
\left(\frac{R_\mu {\dot z}_\nu+R_\nu {\dot z}_\mu}{\rho}-\frac{{\dot z}^2 R_\mu R_\nu}{\rho^2}-
\frac{\eta_{\mu\nu}}{2}\right).
\label
{Theta-irr}
\end{equation}                                            
To define the corresponding part of four-momentum $P^{\rm ir}_\mu$, we integrate 
$\Theta^{\rm ir}_{\mu\nu}$ over the future light cone $C_+$, drawn from $z^\mu(\tau_{\rm ret})$, 
with the use of the surface element $d\sigma^\mu=R^{\mu}\rho\,d\rho\,d\Omega$, after prior
regularizing the integral over $\rho$ to make it convergent. 
In response to the solid angle integration of $\rho^{-2}$, the denominator gains the factor 
${\dot z}^2$.
However, this factor does not kill the factor $({\dot z}^2)^2$ in the numerator, and hence 
the regularized integrand, involving the overall factor ${\dot z}^2$, is vanishing.
In the limit of regularization removal, we have $P^{\rm ir}_\mu=0$.

As to the term of stress-energy tensor containing mixed contribution of $F^{\rm r}_{\mu\nu}$ and 
$F^{\rm ir}_{\mu\nu}$, it is not unduly difficult to show that this term, being contracted with the 
surface element of the future light cone $C_+$, is annihilated by ${R}^2=0$ and ${\dot z}^2=0$.

This completes the proof of our assertion that the self-energy term is vanishing. 

Turning to  $F^{\rm r}_{\mu\nu}$, we first note that both invariants 
${\cal P}=\frac12\,{}^\ast\!F^{\rm r}_{\mu\nu}F^{{\rm r}\hskip0.5mm\mu\nu}$ and 
${\cal S}=\frac12\,F^{\rm r}_{\mu\nu}F^{{\rm r}\hskip0.5mm\mu\nu}$ are zero.
In other words, $F^{\rm r}_{\mu\nu}$ is a null field over all spacetime minus the manifold 
${\cal M}_2$.

The term of the stress-energy tensor $\Theta_{\rm r}^{\mu\nu}$ built from $F^{\rm r}_{\mu\nu}$,
\begin{equation}
\Theta_{\rm r}^{\mu\nu}=-\frac{e^2}{4\pi}\,\frac{{\ddot z}^2}{\rho^4}\,R^\mu R^\nu\,,
\label
{Theta-self}
\end{equation}                        
has an integrable singularity on the world line.
To evaluate the four-momentum associated with $\Theta_{\rm r}^{\mu\nu}$, we take, as before, the 
surface of integration to be a tubular surface ${T}_\ell$ of small radius $\ell$ enclosing the world 
line.
It remains to manage the singularity on ${\cal M}_2$.
A pertinent regularization prescription is to perforate a hole in the intersection of ${T}_\ell$
with ${\cal M}_2$.
Let the normal of ${T}_\ell$ be $n^\mu=\left(0,{\bf n}\right)$, where the unit vector ${\bf n}$ is 
defined in Eq.~(\ref{lorentz-frame}), then 
\begin{equation}
 \Theta_{\rm r}^{\mu\nu} d\sigma_\nu
=-\frac{e^2{\ddot z}^2}{4\pi}\left(\frac{1}{1-\cos\vartheta}\right)^4 
\left(1,\sin\vartheta\cos\varphi,\sin\vartheta\sin\varphi,\cos\vartheta\right)
\sin\vartheta\, d\vartheta\, d\varphi\,d\tau\,,
\label
{d-sigma-Theta}
\end{equation}                         
so that the part of the four-momentum associated with $\Theta_{\rm r}^{\mu\nu}$ is
\begin{equation}
P^\mu_{\rm r}=\int_{{T}_\ell} d\sigma_\nu\Theta_{\rm r}^{\mu\nu}
=-\frac23 e^2\Lambda\int^\tau_{-\infty}d\tau\,{\dot z}^\mu{\ddot z}^2\,.
\label
{P-mu}
\end{equation}                         
Here, $\Lambda=4\,{\delta}^{-6}$, and $\delta$ is a small regularization parameter, the lower limit 
of integration over $\vartheta$, required for smearing the ray singularity.
In Eq.~(\ref{P-mu}), we have  omitted finite terms which are negligibly small in comparison with 
the term proportional to ${\delta}^{-6}$.

At first glance $P^\mu_{\rm r}$ is the four-momentum radiated by a charge moving along a null world line.
But closer inspection shows that the contribution of $P^\mu_{\rm r}$ to the energy-momentum 
balance equation is absorbed by some reparametrization of the null curve.
To see this, we reiterate {\it mutatis mutandis} the argument used for the derivation of Eq.~(\ref{balance-integ})
from the Noether identity (\ref{Noether-1-id-em}) to attain
\begin{equation}
\left(\int_{\Sigma''}-\int_{\Sigma'}+\int_{{T}_{R}}\right)d\sigma_\mu\,\Theta^{\lambda\mu}
+\int_{{\tau}'}^{{\tau}''}d\tau\left({\dot\eta}{\dot z}^\lambda+{\eta}{\ddot z}^\lambda\right)
=0\,,
\label
{Noether-integr}
\end{equation}              
where $\Sigma'$ and $\Sigma''$ are spacelike surfaces separated by a short timelike interval.
If we impose the Haag asymptotic condition 
\begin{equation}
\lim_{\tau\to -\infty} {\ddot z}^\mu(\tau)=0\,,
\label
{a-to-0-null}
\end{equation}                        
then the integration over the tube $T_R$ approaches zero as $R\to\infty$.
Only $F^{\rm r}_{\mu\nu}$ contributes to the integrations over $\Sigma'$ and $\Sigma''$, and the 
result of such integrations is typically expressed by Eq.~(\ref{P-mu}). 
Using (\ref{P-mu}) in (\ref{Noether-integr}) gives
\begin{equation}
\int_{{{\tau}}'}^{{{\tau}}''}d\tau\left({\dot\eta}{\dot z}^\lambda+{\eta}{\ddot z}^\lambda
-\frac{2}{3}\,e^2\Lambda\,{\ddot z}^2{\dot z}^\lambda\right)=0\,.
\label
{Noether-integr-}
\end{equation}              
The first and the last terms have similar kinematical structures. 
This suggests that there is a particular parametrization ${\bar\tau}$ such that these terms cancel.
To verify this suggestion, we go from $\tau$ to ${\bar\tau}$ through the reparametrization
\begin{equation}
d{{\tau}}
=d{{\bar\tau}}\left[1
+\frac{1}{{\bar\eta}({\bar\tau})}\,\frac{2}{3}\,e^2\Lambda
\int_{-\infty}^{{\tau}}d\sigma\,{\ddot z}^2(\sigma)\right].
\label
{reparam}
\end{equation}                                            
In fact, Eqs.~(\ref{reparam}) and  (\ref{eta-eta'}) constitute a set of two functional differential 
equations in which ${\bar\tau}={X}\left(\tau\right)$ and ${\bar\eta}=Y\left[\eta(\tau);{\bar\tau}(\tau)\right]$
are the unknown quantities, and appropriate solutions to these equations can hopefully be found.
By (\ref{eta-eta'}), 
\begin{equation}
\eta(\tau)=
{\bar\eta}({\bar{\tau}})
+
\frac{2}{3}\,e^2\Lambda
\int_{-\infty}^{{\tau}}d\sigma\,{\ddot z}^2(\sigma)\,,
\label
{reparam-einb}
\end{equation}                                            
and 
\begin{equation}
\frac{d\eta}{d\tau}=
\frac{d{\bar\eta}}{d\tau}
+
\frac{2}{3}\,e^2\Lambda
{\ddot z}^2({\tau})\,,
\label
{reparam-einb-dot}
\end{equation}                                            
whence it follows that the term $-\frac{2}{3}\,e^2\Lambda\,{\ddot z}^2{\dot z}^\lambda$
disappears from Eq.~(\ref{Noether-integr-}).

We thus come to recognize that the net effect of the putative radiated four-momentum is actually 
removable through an appropriate reparametrization of the null world line.

We finally compare these results with those obtained in the Maxwell--Lorentz theory of massive charged 
particles.
It is reasonable to begin with the Noether identity (\ref{Noether-1-id-em-zero}) which is universally
suitable for both massive and massless cases.
If $m_0\ne 0$, then the usual way to explore this identity further is to consider $\eta$ to be 
the solution (\ref{solution}) of the constraint equation (\ref{eq-eta}), which implies 
that the world line is parametrized by the proper time $ds=d\tau\,\sqrt{{\dot z}\cdot{\dot z}}$.
However, there is nothing to prevent us from following the above route.
In doing so, we come to Eq.~(\ref{Noether-integr}).
A closer look at the integrals of $\Theta^{\lambda\mu}$ over $\Sigma'$ and $\Sigma''$, which 
represents the increment of the four-momentum of electromagnetic field for the period $\tau''-\tau'$, 
shows a dramatic change of the affair.
For a particle moving along a timelike world line, 
\begin{equation}
\left(\int_{\Sigma''}-\int_{\Sigma'}\right)d\sigma_\mu\,\Theta^{\lambda\mu}=
\int^{\tau''}_{\tau'}d\tau\left({e^2\over 2\epsilon}\,{\ddot z}^{\lambda}
+{\dot\eta}{\dot z}^\lambda+{\eta}{\ddot z}^\lambda
-{2\over 3}\,e^2\, {{{\stackrel{\ldots}{z}}}}^{\hskip0.3mm\lambda}
-{2\over 3}\,e^2\,{\ddot z}^{\hskip0.3mm 2} {\dot z}^\lambda\right).
\label
{P-field-def}
\end{equation}                        
Evidently the term $-{2\over 3}\,e^2\, {{\stackrel{\ldots}{z}}}^{\hskip0.3mm\lambda}$ cannot be 
cancelled by other terms of Eq.~(\ref{P-field-def}), no matter what is the parametrization 
of the world line.
Furthermore, $\left({e^2/2\epsilon}\right){\ddot z}^{\lambda}$ is divergent.
For this divergence to be absorbed by the mass renormalization, the gauge must be fixed, 
$\eta=m_0/\sqrt{{\dot z}\cdot{\dot z}}$, which implies that the world line is parametrized by the 
proper time. 
Accordingly, the term $-{2\over 3}\,e^2\,{\ddot z}^{\hskip0.3mm 2} {\dot z}^\lambda$
survives in the energy-momentum balance. 

It was emphasized in Sec.~\ref{heart} that the responsibility for rearranging the initial degrees 
of freedom  rests with instabilities peculiar to the system.
For lack of the rearrangement, the system with the action (\ref{Brink-massless-action}) 
is stable, which is likely to be due to 
the conformal symmetry group C$(1,3)$.
Recall that C$(1,3)$ is the {lowest dimensional} group containing the Poincar\'e group.
Furthermore, C$(1,3)$ is {semisimple}, even though the Poincar\'e group is the semidirect product of 
the Lorentz and translation groups. 
Admittedly, however, why the features of this fascinating symmetry provide a way for the acquisition 
of stability has been something of mystery. 

Of special note is that the Yang--Mills--Wong theory of zero-mass particles carrying the 
pertinent non-Abelian charges also enjoys the property of conformal invariance.
The absence of the rearrangement from this theory can lead to far-reaching speculations.
As stated above, quarks in quark--gluon plasma are most likely to be massless.
Such quarks do not emit electromagnetic and Yang--Mills radiation, and hence do not lose their 
energy in collisions.
This might help to illuminate the fact that the quark--gluon plasma is the most perfect fluid ever 
observed, see, {\it e.~g.,} \cite{Shuryak2005}.
 
\subsection{Action at a distance}
\label
{wheeler-feynman}
The Maxwell--Lorentz electrodynamics of $N$ charged particles can be recast in terms 
of the direct interparticle action, proposed by Fokker \cite{Fokker1929}, 
\begin{equation}
S_{\rm F}=-\sum_{I}\int\!d\tau_I\!\left\{m_I\sqrt{{\dot z}_I^2}+\frac12\int\! d\tau_J\!
\sum_{(J\ne I)}e_I e_J{\dot z}_I^{\hskip0.3mm\mu}(\tau_I){\dot z}_{J \mu}(\tau_J)\,\delta\!\left[(z_I-z_J)^2
\right]\right\},
\label
{S-Wheeler-Feynman}
\end{equation}                          
where capital Latin letters are used to label the particles.
The distinctive feature of this formulation of electrodynamics is the presence of retarded and 
advanced interactions on an equal footing.  
Owing to the delta-function, the typical points $z_I$ and $z_J$ on $I$th and $J$th world lines 
can be thought of as ``interacting'' if they are connected by a null interval, which is a 
relativistic generalization of interactions by contact occurring at zero distance.
The Fokker action (\ref{S-Wheeler-Feynman}) does not contain unconstrained field degrees of freedom.
It is as if particle $I$ were affected by particle $J$ {directly}, that is, without mediation of 
the electromagnetic field.
Accordingly, self-interaction seems to be absent from (\ref{S-Wheeler-Feynman}). 

Wheeler and Feynman assumed \cite{WheelerFeynman1945}, \cite{WheelerFeynman1949} that radiation is 
completely absorbed.
To clarify the precise meaning of this assumption, we first decompose the retarded Green's function 
$D_{\rm ret}$ into even and odd parts by introducing the advanced Green's function $D_{\rm adv}$:
\begin{equation}
D_{\rm ret}=\frac12\left(D_{\rm ret}+D_{\rm adv}\right)+\frac12\left(D_{\rm ret}-D_{\rm adv}\right) 
=D_{P}+D\,.
\label
{GreensWF-rad}
\end{equation}                          
The even part $D_{P}(x)=\frac12\left[D_{\rm ret}(x)+D_{\rm adv}(x)\right]=\delta(x^2)$ is a solution to 
the inhomogeneous wave equation with the delta-function source,
\begin{equation}
\Box\,D_{P}(x)=4\pi{\delta}^4(x)\,,
\label
{Green-Maxwell-Wheeler-Feynman}
\end{equation}                          
while the odd part 
$D(x)=\frac12\left[D_{\rm ret}(x)-D_{\rm adv}(x)\right]={\rm sgn}\,(x_0)\,\delta(x^2)$ obeys the 
homogeneous wave equation
\begin{equation}
\Box\,D(x)=0\,.
\label
{Box-D=0-}
\end{equation}                          
Turning to the dynamics underlying the Fokker action (\ref{S-Wheeler-Feynman}), we note that the 
interactions between particles are such that they simulate electromagnetic field between them: the 
vector potential and the field strength adjunct to particle  $I$ are given by half the retarded and 
half the advanced solutions to Maxwell's equations,
\begin{equation}
A^\mu_{I}(x)=\int d^4y\,D_{P}(x-y)\, {j}_{I}^\mu(y)
=
e_I\int_{-\infty}^\infty d\tau_I\,{\dot z}_I^\mu(\tau_I)\,\delta\left[(x-z_I)^2\right],
\label
{vector-potent-Wheeler-Feynman}
\end{equation}                          
\begin{equation}
F^{\lambda\mu}_{I}={\partial^\lambda A_{I}^\mu}-{\partial^\mu A_{I}^\lambda}\,,
\label
{field-Wheeler-Feynman}
\end{equation}                          
where
\begin{equation}
j^\mu_{I}(x)=e_I\int_{-\infty}^\infty d\tau_I\,{\dot z}_I^\mu(\tau_I)\,\delta^4\left[x-z_I(\tau_I)\right]
\label
{current-Wheeler-Feynman}
\end{equation}                          
is the current density of $I$th charged point particle.
These quantities satisfy the wave equation and the Lorenz gauge condition:
\begin{equation}
\Box\,A^\mu_{I}=4\pi\,{j}_{I}^\mu\,,
\label
{Maxwell-Wheeler-Feynman}
\end{equation}                          
\begin{equation}
\partial_\mu A^\mu_{I}=0\,.
\label
{Lorenz-Wheeler-Feynman}
\end{equation}                          

In contrast, 
\begin{equation}
{\bar A}^\mu_{I}(x)= \int d^4y\,D(x-y)\,j_{I}^\mu(y) 
\label
{A(-)=4-pi-int-D(x-y)j(y)}
\end{equation}                          
is the vector potential of a free field obeying the homogeneous wave equation 
\begin{equation}
\Box\,{\bar A}^\mu_{I}=0\,,
\label
{Box-F(-)=0}
\end{equation}                          
because $\Box_x D(x-y)=0$, as indicated by Eq.~(\ref{Box-D=0-}).

With zero initial data on a spacelike hyperplane $\Sigma$, ${\bar A}^\mu_{I}|_{\Sigma}=0$ and
$\left(n\cdot\partial\right){\bar A}^\mu_{I}|_{\Sigma}=0$, the solution to the
Cauchy problem for the wave equation (\ref{Box-F(-)=0}) is trivial
${\bar A}^\mu_{I}(x)=0$. 

Imagine for a while that only $N$ charged particles are in the Universe, and ${\bar A}^\mu(x)$ is 
the total free field adjunct to these particles, ${\bar A}^\mu=\sum {\bar A}^\mu_{I}$. 
If ${\bar A}^\mu(x)$ vanishes at one time, then it is zero at all times.  
Wheeler and Feynman \cite{WheelerFeynman1945}, \cite{WheelerFeynman1949} adopted 
\begin{equation}
{\bar A}^\mu(x)=0\, 
\label
{total-absorbt}
\end{equation}                          
as a supplementary constraint to the action (\ref{S-Wheeler-Feynman}), and interpreted it as the 
condition of {total absorption} (for an extended discussion see  \cite{Pegg1975}, \cite{HoyleNarlikar1995}).
Ever since this approach is often referred to as the absorber theory of radiation. 

However, the fact that  ${\bar A}^\mu(x)$ is vanishing does not amount to the lack of radiation in 
the sense of the definition  (\ref{i})--(\ref{iii}).
Rather, this fact suggests that the Fokker action (\ref{S-Wheeler-Feynman}) is inadequate to 
describe the system completely.
The system under examination is actually the union of $N$ particles undergoing the direct mutual 
interactions and the fabric of spacetime which is integrated in the system by Eq.~(\ref{total-absorbt}).

Let us express the action (\ref{S-Wheeler-Feynman}) in terms of $A^\mu_{I}(x)$, and vary the $I$th 
world line, 
\begin{equation}
\delta S=\sum_{I}\,\int d\tau_I
\left[ m_I\,\frac{d}{d\tau_I}\left(\frac{{\dot z}_I^\lambda}{\sqrt{\,{\dot z}_I\cdot {\dot z}_I}}\right)
-e_I\sum_{J(\ne I)}\left(\frac{\partial A^\mu_{J}}{\partial z_{I\lambda}} 
-\frac{\partial A_{J}^\lambda}{\partial z_{I\mu}}\right){\dot z}_{I\mu}\right]
\delta z_{I\lambda}\, ,
\label
{delta-S-Wheeler-Feynman}
\end{equation}                          
to obtain
\begin{equation}
m_I{a}_I^\lambda 
=e_I v_{I\mu}\sum_{J(\ne I)}F^{\lambda\mu}_{J}(z_I)\,.
\label
{eq-motion-Wheeler-Feynman}
\end{equation}                          

This equation differs from the equation of motion for a bare charged particle in that the Lorentz 
force exerted on particle $I$ involves the symmetric combination (half-retarded plus half-advanced) 
of fields due to all particles, except that for particle $I$ itself:
\begin{equation}
\frac12\,\sum_{J(\ne I)}\left[F^{J}_{\rm ret}(z_I)+F^{J}_{\rm adv}(z_I)\right]. 
\label
{F-(+)-to-rad-reac}
\end{equation}                          
For lack of the self field, the usual infinities associated with it do not occur, and so there is
no need to renormalize $m_I$.

On the other hand, Eq.~(\ref{eq-motion-Wheeler-Feynman}) looks quite different from the 
Abraham--Lorentz--Dirac equation governing the behavior of a dressed charged particle.
Note, however, that the Wheeler--Feynman condition (\ref{total-absorbt}) still remains untapped.
One can recast (\ref{F-(+)-to-rad-reac}) as 
\begin{equation}
\sum_{J(\ne I)}
F^{J}_{\rm ret}(z_I) +\frac12\left[F^{I}_{\rm ret}(z_I)-F^{I}_{\rm adv}(z_I)\right] 
-
\frac12\,\sum_{J}\left[F^{J}_{\rm ret}(z_I)-F^{J}_{\rm adv}(z_I)\right], 
\label
{WF-(+)-to-rad-reac}
\end{equation}                          
where the last term is the sum over all particles.
By 
(\ref{total-absorbt}),
\begin{equation}
\sum_{J}\left[F^{J}_{\rm ret}(z_I)-F^{J}_{\rm adv}(z_I)\right]=0
\label
{total-absorption}
\end{equation}                          
at every point on the world line of particle $I$, and so (\ref{F-(+)-to-rad-reac}) takes the form
\begin{equation}
\sum_{J(\ne I)}
F^{J}_{\rm ret}(z_I) 
+\frac12\left[F^{I}_{\rm ret}(z_I)-F^{I}_{\rm adv}(z_I)\right]. 
\label
{WF-rad-reac}
\end{equation}                          
The expression in the square bracket can be elaborated further, as in Sec.~\ref{alternative}, to give 
the Abraham term $\Gamma^\mu$, and Eq.~(\ref{eq-motion-Wheeler-Feynman}) becomes  
\begin{equation}
m_I{a}_I^\lambda-\frac23\,e_I^2\left({\dot a}_I^\lambda+a^2_I v_I^\lambda\right) 
=e_I v_{I\mu}\sum_{J(\ne I)} F^{\lambda\mu}_{J\,{\rm ret}}(z_I)\,.
\label
{eq-mot-Wheeler-Feynman}
\end{equation}                          

In summary, combining Eq.~(\ref{eq-motion-Wheeler-Feynman}) with the Wheeler--Feynman constraint 
(\ref{total-absorption}) results in the conventional equation of motion for a dressed particle 
in the retarded field of all other particles.
Furthermore, Eq~(\ref{eq-mot-Wheeler-Feynman}) is equivalent to the local energy-momentum balance,
Eq.~(\ref{balance-LD}),
\begin{equation}
\dot{{p}}^{\mu}_I+{\dot{\cal P}}^{\mu}_I+{\dot {\wp}}^{\mu}_I=0\,,
\label
{balance-energy-momp}
\end{equation}                      
implying that radiation effects have been properly incorporated in this description. 

Wheeler and Feynman assumed that the total matter in the Universe behaves as a perfect absorber, 
and proposed Eq.~(\ref{total-absorption}) as a cosmological {absorber} condition.
If we keep track of particle $I$, then the radiation of this particle is to be completely absorbed 
by other particles.  
The absorber exerts on particle $I$ a force which is the sum of {retarded} forces due to other 
particles, and endows it with the four-momentum $p^\mu_I=m_Iv^\mu-\frac23\,e^2_Ia^\mu_I$. 
The rearrangement of the initial degrees of freedom appearing in the action (\ref{S-Wheeler-Feynman})
has thus attained in a somewhat meandering way.
The system disguises the radiation, but the effect of dressing can be discerned in the local 
energy-momentum balance, Eq.~(\ref{balance-energy-momp}).

A similar procedure can be readily developed for any theory with linear field equations to convert 
it to a theory of direct interparticle action.
However, this approach is unsound for system with nonlinear field equations, such as the Yang--Mills
and Einstein equations.
The reason for this is that the Green's function method, versatile enough to get rid of reference to
field degrees of freedom, is unfit for use in nonlinear theories. 

\subsection{Nonlinear electrodynamics}
\label
{born-infeld}
The term ``nonlinear electrodynamics'' is usually taken to mean a generalization of the Larmor 
action (\ref{S-em-field}) in which the Lagrangian ${\cal L}$ is nonlinear in the invariants 
${\cal S}=\frac12\,F_{\mu\nu}F^{\mu\nu}$ and ${\cal P}=\frac12\,F_{\mu\nu}{}^\ast\!F^{\mu\nu}$, 
where the {field strength} is expressed in terms of vector potentials, $F_{\mu\nu}=
\partial_\mu A_\nu-\partial_\nu A_\mu$, and the dual field tensor is given by ${}^\ast\!F_{\mu\nu}=
\frac12\,\epsilon_{\mu\nu\alpha\beta}F^{\alpha\beta}$.
The best known theory of this type, proposed by {Born and Infeld} \cite{BornInfeld1934}, is 
characterized by 
\begin{equation}
{\cal L}_{\rm BI}=
{b^2}\left(1-\sqrt{1+\frac{1}{b^{2}}\,{\cal S}- \frac{1}{4b^{4}}\,{\cal P}^2}\,\right).
\label
{BI-1}
\end{equation}                          
Here and throughout this section, Heaviside units are adopted, so that the factor of $1/4\pi$ is 
absent from the field Lagrangians; ${b}$ stands for a constant having dimension of the {field strength}. 
For weak fields, ${\cal L}_{\rm BI}$ approaches $-\frac12\,{\cal S}$ which is the Larmor Lagrangian, 
Eq.~(\ref{S-em-field}).

An unpleasant novelty related to nonlinear  electrodynamics is that it allows for the creation of
electromagnetic shock waves.
Since coefficients of higher derivatives in the field equation are functions of $F_{\mu\nu}$ and
${}^\ast\!F_{\mu\nu}$, the equation for determining the normals to the characteristic surfaces may 
have degenerate real solutions, which is a prerequisite to shock waves.
The presence of the shock waves introduces a further topological aspect associated with 
uncontrollable irreversibility.
The nonlinear set of hyperbolic equations in the $(1+1)$-dimensional Born--Infeld theory is unique 
in that their characteristics cannot intersect, and hence no electromagnetic shock wave occurs \cite{BlokhintsevOrlov1953}.
Besides, the Born--Infeld theory is the only version of nonlinear 
electrodynamics with a sensible weak field limit which is free of the birefringence, that is, only 
this theory describes signal propagation along a single characteristic cone, regardless of their 
polarization  \cite{Boillat1970}.  

The major concern over the blowup of local interactions that was voiced in the early stages of the
development of the modern field theory can be settled in the framework of classical nonlinear 
electrodynamics. 
To verify this, we begin with 
\begin{equation}
{S}=- \int d\tau\left(m_0\sqrt{{\dot z}\cdot{\dot z}}+ e A_\mu {\dot z}^\mu\right)
+\int d^4x\,{\cal L}\left({\cal S},{\cal P}\right),
\label
{action-nonlin-ED}
\end{equation}                          
assuming that the Lagrangian ${\cal L}\left({\cal S},{\cal P}\right)$ reduces to $-\frac12\,{\cal S}$ in the weak 
field limit.
Let us define the field {excitation} 
\begin{equation}
E_{\mu\nu}=\frac{\partial{\cal L}}{\partial F^{\mu\nu}}
=2\left(\frac{\partial{\cal L}}{\partial{\cal S}}\,F^{\mu\nu}+
\frac{\partial{\cal L}}{\partial{\cal P}}\,{}^\ast\!F^{\mu\nu} \right).
\label
{ast-G-df}
\end{equation}                          
Then the Euler--Lagrange equations resulting from (\ref{action-nonlin-ED}) read
\begin{equation}
m_0 a^\mu=e v_\nu F^{\mu\nu}\,,
\label
{Newton-Lorentz-nonlin}
\end{equation}                          
\begin{equation}
\left(\partial_\mu E^{\mu\nu}\right)(x)=-e\int_{-\infty}^\infty ds\,v^\nu(s)\,\delta^4\left[x-z(s)\right].
\label
{partial-ast-G=j}
\end{equation}                          
This set of equations should be augmented by the addition of the Bianchi identity 
\begin{equation}
\partial_\mu{}^\ast\!F^{\mu\nu}=0\,
\label
{Bianchi-partial-ast-F}
\end{equation}                          
(which is a mere restatement of the initial assumption that the {field strength} is expressed in 
terms of vector potentials), and the {constitutive equations}
\begin{equation}
E^{\mu\nu}=E^{\mu\nu}(F)\,,
\label
{constitutive}
\end{equation}                          
following from (\ref{ast-G-df}).
The constitutive equations of Maxwell's  electrodynamics are linear, $E_{\mu\nu}=-F_{\mu\nu}$. 
In the general case, however, Eq.~(\ref{constitutive}) need not be linear in $F^{\mu\nu}$, hence the 
name {nonlinear electrodynamics}.
For example, (\ref{BI-1}) implies the following constitutive equations
\begin{equation}
F^{\mu\nu}=-
\left({1-\frac{1}{b^{2}}\,{\Sigma}-\frac{1}{4b^{4}}\,{\Pi}^2}\right)^{-\frac12}\!\left(E^{\mu\nu}+
\frac{{\Pi}}{2b^{2}}\,\,{}^\ast\! E^{\mu\nu} \right),
\label
{constitutive-BI-inverse}
\end{equation}                          
where 
\begin{equation}
{\Sigma}=\frac12\,E_{\mu\nu}\,E^{\mu\nu},
\quad
{\Pi}=\frac12\,{}^\ast\! E_{\mu\nu}\,E^{\mu\nu}.
\label
{Sigma-Pi-df}
\end{equation}                          

Equations (\ref{Newton-Lorentz-nonlin})--(\ref{constitutive}) form the entire set of equations of 
nonlinear electrodynamics. 
It is clear from (\ref{partial-ast-G=j}) and (\ref{Newton-Lorentz-nonlin}) that a point charge generates $E^{\mu\nu}$ 
but evokes response through $F^{\mu\nu}$. 

The symmetric stress-energy tensor of electromagnetic field
\begin{equation}
\Theta^{\mu\nu}=-F^{\mu}_{~\alpha}\,E^{\alpha\nu}-\eta^{\mu\nu}{\cal L}
\label
{energy-tensor-nonlin-gen}
\end{equation}                          
obeys the equation 
\begin{equation}
\partial_\nu\Theta^{\mu\nu}=- F^{\mu\nu}j_{\nu}\,,
\label
{div-energy-tensor-nonlin}
\end{equation}                          
so that, in a region free of electric charges, 
\begin{equation}
\partial_\nu\Theta^{\mu\nu}=0\,.
\label
{energy-tensor-nonlin-cons}
\end{equation}                          
  
Just as $F_{\mu\nu}$ is associated with the electric field intensity ${\bf E}$ and the magnetic 
induction ${\bf B}$ in a particular Lorentz frame, so $E_{\mu\nu}$ can be related to the {electric 
displacement}, ${\bf D}_i=E_{i0}$, and the {magnetic field intensity}, 
${\bf H}_k=\frac12\epsilon_{0klm}E^{lm}$.
Let us take a look at the static case ${\bf j}=0,\, \partial{\bf D}/\partial t=0$.
Equation (\ref{partial-ast-G=j}) reduces to
\begin{equation}
\nabla\cdot{\bf D}({r})= e\delta^3({\bf r})\,.
\label
{div-D=4-pi-delta}
\end{equation}                          
To be specific, we turn to the Born--Infeld theory. 
The constitutive equation (\ref{constitutive-BI-inverse}) becomes
\begin{equation}
{\bf E}=\frac{b{\bf D}}{\sqrt{b^{2}+{\bf D}^2}}\,.
\label
{E-via-D-BI}
\end{equation}                         

The Poisson equation (\ref{div-D=4-pi-delta}) is obeyed by the Coulomb solution
\begin{equation}
{\bf D}({r})=\frac{e}{4\pi r^2}\,{\bf n}\,,
\label
{D-Coulomb}
\end{equation}                          
which is singular at $r=0$.
However, the field strength ${\bf E}$ derived from (\ref{D-Coulomb}) with the 
help of (\ref{E-via-D-BI}) is regular,
\begin{equation}
{\bf E}({r})=\frac{e}{4\pi\sqrt{r^4+\ell^4}}\,{\bf n}\,.
\label
{E-BI-solution}
\end{equation}                          
Here, $\ell$ is a characteristic length related to the critical field $b$ as
\begin{equation}
b=\frac{e}{4\pi \ell^2}\,.
\label
{charact-length-BI}
\end{equation}                          
At large $r$, ${\bf E}(r)$ approaches the Coulomb field.
Note also that  ${\bf E}(r)\to{\bf D}(r)$ as $\ell\to 0$.

The energy density results from (\ref{energy-tensor-nonlin-gen}):
\begin{equation}
{\Theta}_{00}=-F_{0i}\,E_{0i}-{\cal L}={\bf E}\cdot{\bf D}-{\cal L}\,.
\label
{Teta-00-BI}
\end{equation}                          
By (\ref{E-via-D-BI}), 
\begin{equation}
{\Theta}_{00}=b\left(\sqrt{b^2-{\bf E}^2}-b\right)+
{\bf E}\cdot{\bf D}=b\left(\sqrt{b^2+{\bf D}^2}-b\right).
\label
{energy-BI}
\end{equation}                         
Using (\ref{D-Coulomb}) in (\ref{energy-BI}) gives ${\Theta}_{00}\sim 1/r^2$ near $r=0$, but this 
singularity is integrable, and hence the self-energy is finite,
\begin{equation}
\delta m=\int d^3x\,\Theta_{00}=\frac{e^{2}}{4\pi\ell}\int_{0}^\infty dy\left(\sqrt{1+y^4}-y^2\right).
\label
{delta-m-BI}
\end{equation}                  

In addition, the Born--Infeld theory shows clear evidence that a static point charge is stable
because it is free of tearing strains. 
Consider the force exerting on the charge within an infinitesimal solid angle: 
\begin{equation}
d{\bf F}= e {\bf E}(r) d\Omega |_{r=0}
=\frac{e^2}{4\pi \ell^2}\,{\bf n}\,d\Omega\,.
\label
{explosive-force-BI}
\end{equation}                  
This $d{\bf F}$ is balanced by the infinitesimal force equal in magnitude and opposite in direction.
Both are put to the same point, so that the net effect is zero.
One can then explain the stability of a point charge in the Maxwell--Lorentz theory taking 
(\ref{explosive-force-BI}) as a regularized expression for the infinitesimal force, integrating  
this expression over solid angle, and taking the limit $\ell\to 0$.

The Born--Infeld electrodynamics may thus be understood as a modification of the Maxwell--Lorentz 
theory such that the product of singular distributions turns out to be well defined. 

The solution (\ref{E-BI-solution}) for a charge at rest can be readily
generalized for a uniformly moving charge by a Lorentz boost.
However, solutions for an arbitrarily moving charge, similar to the retarded Li\'enard--Wiechert 
solution of Maxwell's theory, are still not found.
We have no inkling what is the mechanism of rearrangement in nonlinear electrodynamics. 

For an extended discussion of the Born--Infeld electrodynamics we refer the reader to 
\cite{Sommerfeld1949}, and  \cite{Bialynicki-Birula1975}.
The history and some remarkable features of this theory can be learned from the essay
\cite{Bialynicki-Birula1983}.
One passage of it reads: ``There is no evidence that their theory has any direct connection 
with the physical reality''.  
However, as soon as two years later, the Born--Infeld Lagrangian was found to be the low energy 
effective Lagrangian of gauge fields on open strings \cite{FradkinTseytlin1985}.
 
\subsection{Nonlocal interactions}
\label
{efimov}
Another way for avoiding the blowup inherent in local field theories is to ``smear out'' the 
interaction over a small spacetime region. 
Early attempts to render the interaction nonlocal were focussed on the search of suitable {form factors} 
in the interaction,
\begin{equation}
\int d^4 x\int d^4 y\, A_\mu(x)\,F(x-y)\,j^\mu(y)\,.
\label
{S-int-nonloc}
\end{equation}                          
The form factor $F$ was conceived as a smooth function of $(x-y)^2$ which looks like a sharp pulse normalized 
to unit area, for instance 
\begin{equation}
F(x-y)=F_0\exp\left\{-\left[\frac{(x-y)^2}{\ell^2}\right]^2\right\}.
\label
{formfactor-sampl}
\end{equation}                          

Two closely related problems of nonlocal theories of this type are causality violations and angular 
divergences \cite{Kirzhnits1966}.
In fact, there is much evidence that nonlocal interactions can be mediated by superluminal signals. 
The reason for occurrence of the angular divergences is the necessity of integration over an infinite 
range of hyperbolic angles parametrizing the pseudoeuclidean momentum space because the form factor
like that shown in Eq.~(\ref{formfactor-sampl}) 
prevents the Wick rotation making the description euclideanized. 
To meet the challenge, the support of $F(x-y)$ must be {compact}, and its characteristic size 
$\ell$ small.
Note, however, that the topology of four-dimensional real Euclidean space is distinctly different 
from that of Minkowski space  \cite{Zeeman1967}.

Let $F$ be a function of $(x-y)^2$.
Then the invariant region where the acausal effects are confined,
\begin{equation}
(x-y)^2<\ell^2\,,
\label
{(x-y)-2<l-2}
\end{equation}                          
is noncompact: near the light cone $(x-y)^2=0$, the extension of this region in spatial and temporal directions 
is arbitrarily large.
Alternatively, it is possible to use a unit vector $q^\mu$ for constructing a positive definite quadratic 
form 
\begin{equation}
d(x,y)=[q\cdot(x-y)]^2-(x-y)^2\,,
\label
{d(x,y)=n-cdot-(x-y)-2}
\end{equation}                          
and take $F$ to be a function of $d(x,y)$ to limit the acausal effects to a compact, invariant 
region $d(x,y)\le \ell^2$.
This brings up the question: Where can the $q^\mu$ come from?
We may think of $q^\mu$ as the four-velocity $v^\mu$ of some particle.
However, in the absence of particles, we are forced to use a fixed unit vector $q^\mu$, which would
distinguish a privileged frame of reference, and violate explicit Lorentz invariance.
We may also regard $q^\mu$ as an auxiliary unit vector, and average $F$ over directions of $q^\mu$,
but this procedure appears too arbitrary.
 
Another line of attack for overcoming these difficulties is as follows.
Let a form factor be obtained by acting a function $K$ of the d'Alembertian ${\Box}$ on the Dirac 
delta-function,
\begin{equation}
F(x-y)=K({\Box})\,\delta^4(x-y)=\sum_{n=0}^\infty c_n {\Box}^n \delta^4(x-y)\,.
\label
{formfactor-E}
\end{equation}                          
The relativistic invariance of $F(x-y)$ is apparent: $\delta^4(x-y)$ is invariant under Poincar\'e 
transformations, and ${\Box}\,\delta^4(x-y)$ shares this property.
With the Fourier transform 
\begin{equation}
F(x)=\frac{1}{\left(2\pi\right)^4}\int d^4x\,e^{-ik\cdot x}\,{\widetilde F}(k)\,,
\label
{formfactor-F(x)-Fourier}
\end{equation}                          
(\ref{formfactor-E}) becomes  
\begin{equation}
{\widetilde F}(k)=
K({-k^2})=\sum_{n=0}^\infty\,c_n k^{2n}\,.
\label
{formfactor-Fourier}
\end{equation}                          
The radius of convergence of this series depends on $c_n$.                                          
We will discuss only those power series which are convergent in the whole complex $k^2$-plane.
In other words, $K({-k^2})$  is an {entire} function. 

If $c_n=0$ for $n\ge N$, then $K({-k^2})$ is a polynomial, and we are led to an ordinary higher-derivative Lagrangian. 
This suggests that if the coefficients $c_n$ decrease too much rapidly, even though their sequence 
does not terminate, then the interaction is not smeared out, but actually remains local.
Such interactions are called {localizable}.
The line of demarcation between localizable and nonlocal interactions  \cite{Meiman1964}, 
\cite{Jaffe1967}, \cite{Efimov1968} separates entire functions $K({-k^2})$ into two classes: 
\begin{eqnarray}
({\rm L})
\qquad
\lim_{n\to\infty} n\,|c_n|^{1/n}=0\,,
\nonumber\\ 
({\rm N})
\qquad
\lim_{n\to\infty} n\,|c_n|^{1/n}=A\,.
\label
{case3}
\end{eqnarray}                          
Condition (L) was shown to be equivalent to the following bound of asymptotic growth of $K({-k^2})$ 
as $k^2$ approaches infinity in the complex plane: 
\begin{equation}
({\rm L})
\qquad
|K({-k^2})|<C\exp\left(\epsilon\,\sqrt{|k^2|}\right),
\label
{localizable-Fourier-ff}\end{equation}
where $\epsilon$ is arbitrarily small.
As for nonlocal interactions, the class of {entire} functions $K({-k^2})$ showing considerable 
promise as appropriate form factors satisfies the asymptotic condition
\begin{equation}
({\rm N})
\qquad
|K({-k^2})|< C\exp(\sigma \sqrt{|k^2|})
\label
{localizable-Fourier-f}
\end{equation}                          
for a fixed, positive $\sigma$.
With such form factors, finite nonlocal theories of scalar fields obey all general conditions
of quantum field theory: unitarity, covariance, and macroscopic causality  \cite{Efimov1968}, 
 \cite{Efimov1970}.
Furthermore, most results of the axiomatic quantum field theory, in particular $PCT$-invariance and 
connection between spin and statistics, can be extended to the case that the Wightman vacuum 
expectation values reveal exponential energy growth \cite{IofaFainberg1969}, which is 
characteristic of nonlocal interactions. 

Since our interest here is with the rearrangement of classical electrodynamics, we restrict our 
discussion to the following modification of the action  \cite{Kosyakov1976}:
\begin{equation}
{S}=-m_0 \int d\tau\,\sqrt{{\dot z}\cdot{\dot z}} -\int d^4x\left[A_\mu K({\Box})j^\mu
+\frac{1}{16\pi}\, F^{\mu\nu} F_{\mu\nu}\right],
\label
{Maxwell-Lorentz-nonloc}
\end{equation}                          
where $j^\mu$ is the usual four-current of a delta-function source defined in 
(\ref{single-charge-j-4d}), and $K({\Box})$ is given by (\ref{formfactor-E}).
To maintain the link with the Maxwell--Lorentz theory, we require that
\begin{equation}
K(0)=1\,.
\label
{K(0)=1}
\end{equation}                         

Variation of $z^\mu$ gives the equation of motion for a bare particle  
\begin{equation}
m_0 a^\lambda=e v_\mu K({\Box}) F^{\lambda\mu}\,,
\label
{Newton-Lorentz-nonloc}
\end{equation}                          
and varying $A^\mu$, we obtain the equation of motion for the electromagnetic field 
\begin{equation}
\partial_\mu F^{\mu\nu}=4\pi K({\Box})j^\nu\,.
\label
{Maxwell-nonloc}
\end{equation}                          
It is instructive to begin with the static case ${\bf E}=-\nabla\phi$.
Equation (\ref{Maxwell-nonloc}) becomes 
\begin{equation}
\nabla^2\phi({\bf r})=-4\pi e K(-{\nabla^2})\,\delta^3({\bf r})\,.
\label
{Poisson-sampl}
\end{equation}                          
Using  
\begin{equation}
\phi({\bf r})=\frac{1}{(2\pi)^3}\int d^3k\, e^{i{\bf k}\cdot {\bf r}}\, 
{\widetilde\phi}({\bf k})
\label
{phi(br r)=Fourier1}
\end{equation}                         
in (\ref{Poisson-sampl}) gives
\[
\phi({\bf r})=\frac{e}{(2\pi)^3}\int\! d^3k\, e^{i{\bf k}\cdot {\bf r}}\, 
\frac{4\pi K({\bf k}^2)}{{\bf k}^2}
=\frac{2e}{\pi r}\int_0^\infty dk\, \frac{\sin kr}{{k}}\, K({ k}^2)
\]
\[
=\frac{e}{\pi r}\,{\rm P}\!\int_{-\infty}^\infty\! dk\,\frac{K({ k}^2)}{{k}}
\,{\sin}\,{kr}
=\frac{e}{\pi r}\,{\rm Im}\left\{\int_{-\infty}^\infty\! dk\left[
\frac{1}{{k}+i\epsilon}+i\pi\delta(k)\right]\! K({k}^2)\,e^{ikr}\right\}
\]
\begin{equation}
=\frac{e}{\pi r}\,{\rm Im}\left[\int_{-\infty}^\infty dk\,
\frac{K({k}^2)}{{k}+i\epsilon}\,e^{ikr}\right]+\frac{e}{r}\,.
\label
{long-alg-exprsn)=1}
\end{equation}                         
In the next to last equation of Eq.~(\ref{long-alg-exprsn)=1}), the Sokhotski relation has been used, 
\begin{equation}
\frac{1}{x+i\epsilon}={\rm P}\!\left(\frac{1}{x}\right)-i\pi\delta(x)\,,
\label
{Sokhotsky}
\end{equation}                         
in which ${\rm P}$ stands for the Cauchy principal value.
The last equation of Eq.~(\ref{long-alg-exprsn)=1}) has been obtained with regard to Eq.~(\ref{K(0)=1}).

Let us verify that the interaction associated with entire functions $K({k}^2)$ 
obeying (\ref{localizable-Fourier-ff}) is indeed ``localizable''. 
To this end we examine the behavior of 
\begin{equation}
h_R(r)=\frac{1}{\pi }\int_{-R}^R dk\,\frac{K({ k}^2)}{{k}+i\epsilon} \,e^{ikr}\,.
\label
{alpha-R-df}
\end{equation}                          
By Cauchy's theorem, integration over the real axis can be replaced by integration over a semicircle  
$\Gamma_R$ of large radius $R$ at the upper half-plane ${\rm Im}\,k>0$.
Letting $k=Re^{i\vartheta}$, 
\begin{equation}
\Bigl|\int_{\Gamma_R} \frac{dk}{{k}} \,K({ k}^2)\,e^{ikr}\Bigr|\le
\int_{\Gamma_R}\Bigl| \frac{dk}{{k}} \,K({ k}^2)\,e^{ikr}\Bigr|<
\int_{0}^\pi d\vartheta\,\exp(\epsilon R)\,\exp(-rR\sin\vartheta)\,,
\label
{inequality alpha-R-df}
\end{equation}                          
where we have taken into account (\ref{localizable-Fourier-ff}), and
\begin{equation}
|e^{ikr}|=\exp|iR\,(\cos\vartheta+i\sin\vartheta) r|=\exp(-rR\sin\vartheta)\,.
\label
{mod(exp-ikr)}
\end{equation}                          
In the sector $0<\vartheta\le\pi/2$, it is helpful to use the inequality $\sin\vartheta\ge{2\vartheta}/{\pi}$,
\[
\int_{0}^\pi d\vartheta\,e^{\epsilon R}\,e^{-rR\sin\vartheta}=
2e^{\epsilon R}\int_{0}^{\pi/2} d\vartheta\,e^{-rR\sin\vartheta}
<2e^{\epsilon R}\int_{\delta}^{\pi/2}d\vartheta\,e^{-\frac{2}{\pi}\vartheta rR}
\]
\begin{equation}
=\frac{\pi}{rR}\left[e^{-R\left(\frac{2}{\pi}r\delta-\epsilon\right)}-
e^{-R\left(r-\epsilon\right)}\right].
\label
{frac(pi)(rR)alpha-R-df}
\end{equation}                          
For finite $r$ and $\epsilon<2r\delta/\pi$, this expression vanishes in the limit $R\to\infty$.

To summarize, if $K({k}^2)$ obeys (\ref{localizable-Fourier-ff}), then $h_R(r)\to 0$, and 
the potential (\ref{long-alg-exprsn)=1}) takes the form of the Coulomb potential $e/r$.
The singularity of $\phi({\bf r})$ remains unaffected when $K(-{\nabla^2})$ acts on $\delta^3({\bf r})$.

We next turn to nonlocal interactions. 
Supposing that 
\begin{equation}
|K(R^2e^{2\vartheta})|\le C\exp(\ell R\sin\vartheta)
\quad
{\rm as}
\quad
R\to\infty\,,
\label
{K-nonlocal}
\end{equation}                          
we find
\[
\Bigl|\int_{\Gamma_R} \frac{dk}{{k}} \,K({ k}^2)\,e^{ikr}\Bigr|\le
2C\int_{0}^{\frac{\pi}{2}} d\vartheta\,e^{R\sin\vartheta(\ell-r)}\le
2C\int_{0}^{\frac{\pi}{2}} d\vartheta\,e^{\frac{2R\vartheta}{\pi}(\ell-r)}
\]
\begin{equation}
=
\frac{\pi C}{R(\ell-r)}\left[e^{R\left(\ell-r\right)}-1\right].
\label
{K-nonlocal-nu-calc}
\end{equation}                          
For $r>\ell$, this expression vanishes in the limit $R\to\infty$.
To enquire into the situation in the region $r<\ell$, write the solution to 
(\ref{Poisson-sampl}) in the form
\begin{equation}
\phi(r)=e\,\theta(r)\frac{\alpha(r)}{r}\,,
\label
{phi=alpha-by-r}
\end{equation}                          
where $\theta(r)$ is the Heaviside step function indicating that the range of $r$ is reduced to the 
semiaxis ${\mathbb R}_+$, and $\alpha(r)$ is a differentiable function satisfying two conditions
\begin{equation}
\alpha(r)=\alpha_0r+O(r^3)\,, \quad r\to 0\,,
\label
{alpha-near-origin}
\end{equation}                          
\begin{equation}
\alpha(r)=1, \quad |r|\ge\ell\,.
\label
{alpha-beyond-nonloc}
\end{equation}                          
Combining (\ref{phi=alpha-by-r}) with (\ref{long-alg-exprsn)=1}), we obtain
\begin{equation}
\alpha'(r)=\frac{1}{\pi}\int_{-\infty}^{\infty} dk\,K({ k}^2)\,e^{ikr}\,,
\label
{alpha-via-K}
\end{equation}                          
where the prime stands for differentiation with respect to $r$.
The inverse of (\ref{alpha-via-K}) is
\begin{equation}
K({k}^2)=\frac{1}{2}\int_{-\ell}^{\ell} dr\,\alpha'(r)\,e^{-ikr}\,.
\label
{K-via-alpha}
\end{equation}                          
This formula is convenient for constructing $K(k^2)$ with the required properties 
(\ref{K(0)=1}) and (\ref{K-nonlocal}).
One can to show \cite{PaleyWiener1934} that if $\alpha(r)$ is a differentiable function obeying (\ref{alpha-near-origin}) 
and (\ref{alpha-beyond-nonloc}), then (\ref{K-via-alpha}) represents an entire function $K(k^2)$ of 
order $\frac12$, which is square integrable on ${\mathbb R}$, normalized to $K(0)=1$, and whose
indicatrix is $H(\vartheta)=\ell\sin\vartheta$.
Conversely, let $K(k^2)$ be a square integrable on ${\mathbb R}$ entire function of order $\frac12$ 
and type $\ell$.
Then $\alpha'(r)$, defined in (\ref{alpha-via-K}) is zero for $|r|\ge\ell$.

Therefore, to formulate the nonlocal electrodynamics (\ref{Maxwell-Lorentz-nonloc}), we may proceed 
from the static potential (\ref{phi=alpha-by-r}) with an arbitrarily chosen $\alpha(r)$, satisfying 
the conditions (\ref{alpha-near-origin}) and (\ref{alpha-beyond-nonloc}), and use Eq.~(\ref{K-via-alpha})
to obtain the explicit form of $K({\Box})$.

As a simple example of $K(k^2)$ and $\alpha'(s)$ we take
\begin{equation}
K(k^2)=\frac{\sin \left(k\ell\right)}{k\ell}\,,
\quad
\alpha'(r)=\cases{{\ell}^{-1} & $|r|<\ell$\,\cr 0 & $|r|\ge \ell$\,.\cr}
\label
{K-extrem}
\end{equation}                          
The corresponding potential $\phi(r)$ is a truncated Coulomb potential.
A similar static solution for a charged sphere of radius $\ell$ is offered by Maxwell's electrodynamics.
However, the similarity is deceptive.
While the charged sphere is subject to the repulsive static forces, and cannot be stable without 
resort to Poincar\'e cohesive forces, the delta-function source generating the potential 
(\ref{phi=alpha-by-r}) with $\alpha(r)$ given by (\ref{K-extrem}) is free 
of tearing strains, 
\begin{equation}
d {\bf F}=d\Omega\int_0^\infty dr\,r^2 \delta(r)\,{\bf E}(r)=0\,.
\label
{delta-f}
\end{equation}                         

The square integrability of $K(k^2)$ implies that the self-energy is finite:
\begin{equation}
\delta m=\int d^3x\,\frac{{\bf E}^2}{8\pi}=\frac{e^2}{2\pi}\int_{-\infty}^\infty dk\,K^2({ k}^2)<\infty.
\label
{delta-m-efimov}
\end{equation}                         

We are now in position to consider the general case that a point charge is moving along an arbitrary 
timelike smooth world line.
To solve the field equation (\ref{Maxwell-nonloc}), we adopt the retarded boundary condition, and
impose Lorenz gauge.
The solution is given by 
\begin{equation}
A^\mu(x)=-\frac{1}{4\pi^3}\int d^4k\,e^{-ik\cdot x}\,\frac{K(-{k}^2)}{{k}^2+2ik_0\epsilon}
\,{\tilde{\jmath}}^{\hskip0.3mm\mu}(k)\,,
\label
{solution-nonlocal-ED}
\end{equation}                          
where
\begin{equation}
{\tilde{\jmath}}^{\hskip0.3mm\mu}(k)=\int d^4x\, e^{ik\cdot x}\,j^\mu(x)
=
e\int_{-\infty}^{\infty} ds\,e^{ik\cdot z(s)}\,{v}^\mu(s)\,.
\label
{Fourier-nonlocal-current}
\end{equation}                          

Looking at (\ref{K-nonlocal}), we observe that $K(-k^2)$ grows exponentially when $k^2\to\infty$.
This implies that the integral (\ref{solution-nonlocal-ED}) fails to converge unless 
${\tilde{\jmath}}^{\hskip0.3mm\mu}(k)$ decrease appropriately in timelike directions.
It is possible to demonstrate 
 \cite{Kosyakov1976}, \cite{Kosyakov2007} the existence of world lines such that  
\begin{equation}
|{\tilde{\jmath}}^{\hskip0.3mm\mu}(k)|<C e^{-(\ell+\delta)\sqrt{k^2}}\,,
\quad
k^2\to\infty\,,
\label
{bahavior-FT-current}
\end{equation}                          
which is to say that the integral in Eq.~(\ref{solution-nonlocal-ED}) is convergent. 
If the nonlocal electrodynamics is to be consistent, the allowable class of world lines must be
narrowed in the following way.
Let $t$ be laboratory time in a particular Lorentz frame, so that $dz^\mu/dt=(1,{\bf v})$.
Consider a set of timelike smooth curves $z^\mu(t)$ which are capable of being parametrized by a
complex variable $t+iu$, and assume that these curves lend themselves to the requirements of 
analyticity in the strip
\begin{equation}
-\infty<t<\infty\,,\quad |u|<\ell+\Delta\,,\quad 0<\delta<\Delta\,,
\label
{strip}
\end{equation}                          
and integrability 
\begin{equation}
\int_{-\infty}^\infty dt\,|{\bf v}_+(t+iu)|<\infty\,,
\label
{integrability-strip}
\end{equation}                          
where ${\bf v}_+(t)=\frac12[{\bf v}(t)+{\bf v}(-t)]-\frac12({\bf v}_{\rm in}+{\bf v}_{\rm out})$.
Then this set of curves represents the class of world lines for which the asymptotic
estimate (\ref{bahavior-FT-current}) holds. 

The retarded solutions $A^\mu(x)$ prove to be identical to the 
Li\'enard--Wiechert vector potentials everywhere outside 
a thin tube $T_{\varrho}$ of radius ${\varrho}\sim{\ell}$ enclosing the world line. 
All acausal phenomena are confined to a spacetime region ${\cal M}$ bounded by $T_{\varrho}$. 

We thus come to the following picture of self-interaction.
The initial degrees of freedom appearing in the action (\ref{Maxwell-Lorentz-nonloc}) rearrange 
just as they do in the Maxwell--Lorentz theory with few exceptions concerning the region ${\cal M}$.

Substitution of the retarded field (\ref{solution-nonlocal-ED}) in $\Theta^{\mu\nu}$ gives a finite
stress-energy tensor.
The integrability is therefore not the feature unique to $\Theta^{\mu\nu}_{\rm II}$.
In fact, the segregation between $\Theta^{\mu\nu}_{\rm I}$ and $\Theta^{\mu\nu}_{\rm II}$ is  
confidently made only at a distance well away from the region ${\cal M}$.
To define $\Theta^{\mu\nu}_{\rm II}$, the part of the retarded solution $F^{\mu\nu}_{\rm II}$ that 
scales as $\rho^{-1}$ in the limit $\rho\gg\ell$ should be substituted in $\Theta^{\mu\nu}$.
This makes possible to reproduce most of the properties of radiation in the Maxwell--Lorentz theory, 
the generalized Larmor formula (\ref{dP-II}) included, for sufficiently smooth world lines, that is,
for curves whose local acceleration and higher derivatives are always small, 
$\ell^2|a^2|\ll 1$, $\ell^4|{\dot a}^2|\ll 1$,...  

We remark parenthetically that the product of distributions ${\Box}^{-1}K({\Box})\,\delta^4(x)$, 
whose smearing functions $K({-k^2})$ satisfy the asymptotic condition (\ref{localizable-Fourier-ff}),
is well defined if the appropriate basic space ${\cal Z}$ consists of slowly
increasing entire functions $\chi(k^2)$ of the complex variable $k^2=\xi+i\zeta$ subject to the
conditions 
\begin{equation}
|(k^2)^n\chi(k^2)|\le C_n\exp\left(-a|\xi|+b|\zeta|\right),\quad n=0,1,\ldots, N\,,
\label
{test-functions}
\end{equation}                        
where $a$, $b$, $C_n$, and $N$ are constants (dependent on $\chi$). 
For more details of ${\cal Z}$ see \cite{Efimov1968}.
 
Equation (\ref{Newton-Lorentz-nonloc}) is free of the ultraviolet disease and hence is the 
well-defined equation of motion for a dressed charged particle.
It is possible to bring this equation to the form of Newton's second law, Eq.~(\ref{LD-Newton}).  
Indeed, the general solution to the field equation (\ref{Maxwell-nonloc}) is the sum of $F_{\rm ret}$, 
the retarded field due to the delta-function source smeared by $K({\Box})$, plus $F_{\rm ext}$, 
a solution to the homogeneous wave equation describing an external field.
Therefore, the right side of Eq.~(\ref{Newton-Lorentz-nonloc}) is decomposed into two terms, 
$I^\lambda+f^\lambda$, where $I^\lambda$ is attributed to the effect of the retarded field, and 
$f^\lambda$ stands for the external force.
Note also that the terms $m_0a^\lambda$, $I^\lambda$, and $f^\lambda$ are orthogonal to $v^\lambda$ 
each. 
This explains the reason for the occurrence of the overall projector $\stackrel{\scriptstyle v}{\bot}$\,,
and the separation of $f^\lambda$ from the remainder, which, together with $m_0a^\lambda$, is 
associated with ${\dot p}^\lambda$. 
For sufficiently smooth world lines, 
Eq.~(\ref{Newton-Lorentz-nonloc}) is closely approximated by the Abraham--Lorentz--Dirac equation, 
Eq.~(\ref{LD}), in which $m=m_0+\delta m$, with $\delta m$ being given by (\ref{delta-m-efimov}).
Let $\kappa$ be a characteristic curvature of such world lines, and $\epsilon=\ell \kappa$ a small
dimensionless parameter.
Then Eq.~(\ref{p-mu-dress-charge}) represents the four-momentum of a dressed particle $p^\mu$  
accurate to $O(\epsilon^5)$. 

It is notable that the self-energy $\delta m$ is finite as (\ref{delta-m-efimov}) suggests.
The renormalization of mass, $m=m_0+\delta m$, is therefore not a means for rendering the product of 
singular distributions well-defined.
The reason for fusing two quantities of different nature, $m_0$ and $\delta m$, into a single 
quantity, $m$, is again the fact that the initial degrees of freedom appearing in the action 
(\ref{Maxwell-Lorentz-nonloc}) are unstable, which begets their gathering into a dressed particle 
whose inert property is expressed in terms of $m$.

\subsection{Particles with spin}
\label
{spin}
By now, several approaches to the dynamics of classical charged particles with spin have been 
developed.
All can be separated into two main groups: those characterized by the use of ordinary commuting 
variables describing spin degrees of freedom, and those marked by the use of anticommuting 
Grassmannian variables (for a review see, {\it e.~g.},  \cite{Rivas2001}).
Frenkel \cite{Frenkel1926} pioneered in a relativistic Lagrangian description of classical spinning 
particles in electromagnetic field.
Grassmannian variables were first applied to describe spin at the classical level in \cite{Martin1959},
\cite{BerezinMarinov1975}, \cite{BerezinMarinov1977}, and \cite{Casalbuoni1976}.

The Maxwell--Lorentz electrodynamics can be extended to cover a particle possessing a charge $e$ and 
a dipole moment $m^{\mu\nu}=-m^{\nu\mu}$ by introducing the current density
\begin{equation}
j^\mu(x)=\int_{-\infty}^\infty ds\left[e v^\mu(s)+m^{\mu\nu}(s)\frac{\partial}{\partial x^\nu}\right]
\delta^4\!\left[x-z(s)\right].
\label
{currenr-dipole-moment}
\end{equation}                        
To specialize our discussion to the case of a magnetic dipole, we connect $m^{\mu\nu}$ with spin 
variables $S^{\mu\nu}$ of the particle, 
\begin{equation}
m^{\mu\nu}=\mu S^{\mu\nu}\,,
\label
{spin-magnetic-moment}
\end{equation}                        
satisfying the Frenkel condition
\begin{equation}
 S^{\mu\nu} v_\nu=0\,.
\label
{Frenkel_condition}
\end{equation}                        
The particle is assumed to be an object whose nature is preserved under time evolution.
In particular, the magnitude of the tensor $S^{\mu\nu}$ must be unchanged,
\begin{equation}
 S^{\mu\nu} S_{\mu\nu}=2S^2={\rm const}\,.
\label
{spin-unchanged}
\end{equation}                        
The equation of motion for a dipole must be consistent with conditions (\ref{Frenkel_condition}) 
and (\ref{spin-unchanged}).

In a particular Lorentz frame, the antisymmetric tensor $S_{\mu\nu}$ is written as $S_{\mu\nu}=
({\bf N},{\bf S})$, where $\sigma_{0i}={\rm N}_i$ and $\sigma_{ij}=-\epsilon_{ijk}\,{\rm S}_k$.
In the rest frame, Frenkel's constraint (\ref{Frenkel_condition}) implies that only the components 
of ${\bf S}$ are nonzero, while ${\bf N}={\bf 0}$. 
Equation (\ref{spin-unchanged}) can be recast as $S^{\mu\nu} S_{\mu\nu}=2{\bf S}^2$.
Therefore,  ${\bf S}$ may be used to mean spin as viewed by a comoving observer.

The symmetry properties of the extended theory have been subjected to dramatic changes.
Since $m^{\mu\nu}$ has dimension of length, Maxwell's equations cease to be invariant under the
group of conformal transformations C$(1,3)$.
Furthermore, in view of (\ref{currenr-dipole-moment}), it seems difficult to conceive of a simple 
law of transformation for $m^{\mu\nu}$ in 
response to reparametrizations of world lines so as to preserve reparametrization invariance of the 
action. 

The retarded solution to Maxwell's equations with the source $j^\mu$ shown in (\ref{currenr-dipole-moment})
is 
\begin{equation}
A^\mu(x)=e\,\frac{v^\mu}{\rho}+\mu\,\frac{\partial}{\partial x^\nu}\left(\frac{S^{\mu\nu}}{\rho}\right).
\label
{vector_potential-dipole-moment}
\end{equation}                        
All the quantities on the right-hand side are taken at the retarded instant $s_{\rm ret}$.

It is clear from (\ref{vector_potential-dipole-moment}) that the retarded field $F^{\mu\nu}$ 
contains terms proportional to $\rho^{-3}$, $\rho^{-2}$, and $\rho^{-1}$.
Accordingly, the four-momentum $P^\mu$ of this field involves terms that diverge as $\epsilon^{-3}$, $\epsilon^{-2}$, and $\epsilon^{-1}$ in the limit of regularization removal 
$\epsilon\to 0$.
To be more specific, we refer to \cite{BhabhaCorben1941} where it was found that the only singular 
term of $P^\mu$ proportional to $e^2$ looks like
\begin{equation}
\frac{1}{2\epsilon}\,v^\mu\,,
\label
{four-momentum_singular_e-2}
\end{equation}                        
singular terms proportional to $e\mu$ appear as 
\begin{equation}
-\frac{1}{3\epsilon}\left(2\,{\dot S}^{\mu\nu}v_\nu+S^{\mu\nu}a_\nu\right),
\label
{four-momentum_singular_e_mu}
\end{equation}                        
and singular terms proportional to ${\mu}^2$ are of the form
\begin{eqnarray}
\frac{1}{6\epsilon^3}\,v^\mu S^2
+\frac{1}{15\epsilon^2}\left(3a^\mu S^2+5S^{\mu\sigma}{\dot S}_{\sigma\nu}v^\nu
+4S^{\mu\sigma}\,{S}_{\sigma\nu}a^\nu\right)-\frac{1}{15\epsilon}\Bigl[v^\mu\Bigl(
3a^2S^2-5{\dot S}^2
\nonumber\\
+{20}v_\sigma{\dot S}^{\mu\sigma}{\dot S}_{\mu\nu}v^\nu
-4v_\sigma{\dot S}^{\mu\sigma}{S}_{\mu\nu}a^\nu
-a_\mu{S}^{\mu\sigma}{S}_{\sigma\nu}a^\nu\Bigr)
+4S^{\mu\sigma}{\dot S}_{\sigma\nu}a^\nu-6{\dot S}^{\mu\kappa}S_{\kappa\sigma}a^\sigma\Bigr].
\label
{four-momentum_singular_mu-2}
\end{eqnarray}

To absorb these divergences, we need to redefine three free parameters which are held in the 
particle sector of the action.
But only two such parameters, namely $m_0$ and $S$, are available.
This bears some resemblance to the Maxwell--Lorentz electrodynamics in ${\mathbb R}_{1,5}$, 
discussed in Sec.~\ref{d=6}, where $P^\mu$ contains cubic and linear divergences.
We augmented the Poincar\'e--Planck action by the addition of terms containing higher derivatives of 
$z^\mu$, in other words, we adopted the rigid dynamics with a simple Lagrangian (\ref{rigid-term-action}), 
which offered a means to eliminate the divergences and make the description consistent.

However the situation is different now.
As indicated above, we are dealing with a system devoid of conformal and reparametrization 
invariances. 
It is unlikely to contrive a simple Lagrangian for a spinning particle containing, in addition to 
$m_0$ and $S$, a further parameter, so as to be able to cope with elimination of the divergences.
A regular approach to deriving a consistent dynamics of a spinning particle interacting with 
electromagnetic field remains to be developed.
Yet, at times, researchers propose {\it ad hoc} versions of the desired dynamics taking into account 
self-interaction effects  \cite{Bhabha1940},  \cite{BhabhaCorben1941}, \cite{Teitelboim1980}, \cite{RoweRowe1987},
 \cite{BarutUnal1989},  \cite{Gralla2009}.
We do not delve into the essence of these heuristic approaches.
We only note that the eventual outcome is given by two comparatively simple equations,
the equation of motion for a dressed spinning particle, and the equation governing the precession 
of spin of this particle.
For example, the following dynamical equations were obtained in \cite{Teitelboim1980}: 
\begin{equation}
ma^\lambda=e\left({\hat F}^{\lambda\mu}v_\mu+\frac{g}{4m}\,{\hat S}_{\mu\nu}\,\partial^{\lambda}
{\hat F}^{\mu\nu}\right)
+\frac{d}{ds}\left[{\hat S}^{\lambda\mu}\left(g\,\frac{e}{2m}\,{\hat F}_{\mu\nu}v^\nu-a_\mu\right)
+g\,\frac{e}{2m}\,{\hat S}^{\mu\nu}{\hat F}_{\mu\nu} v^\lambda\right],
\label
{equation-of-motion-four-momentum}
\end{equation}                        
and 
\begin{equation}
({\stackrel{\scriptstyle v}{\bot}})_{\lambda\nu}\,({\stackrel{\scriptstyle v}{\bot}})_{\mu\sigma}\,
\frac{d}{ds}\,{\hat S}^{\nu\sigma}=g\,\frac{e}{2m}\left[{\hat S}_{\lambda\nu}\,
({\stackrel{\scriptstyle v}{\bot}})_{\mu\sigma}-{\hat S}_{\mu\nu}\,({\stackrel{\scriptstyle v}{\bot}})_{\lambda\sigma}\right]\!
{\hat F}^{\nu\sigma}\,,
\label
{equation-of-motion-spin}
\end{equation}                        
where $g$ is the gyromagnetic ratio defined by
\begin{equation}
\mu=g\,\frac{e}{2m}\,.
\label
{gyromagnetic ratio}
\end{equation}                        
The rearranged dynamics, Eqs.~(\ref{equation-of-motion-four-momentum}) and (\ref{equation-of-motion-spin}), 
is expressed in terms of two auxiliary quantities ${\hat F}^{\mu\nu}$ and ${\hat S}^{\mu\nu}$ 
originating in the following way.
Let ${F}^{\mu\nu}_{(-)}$ denote half the difference of the retarded and advanced fields 
generated by the particle.
Then ${\hat F}^{\mu\nu}$ is defined by
\begin{equation}
{\hat F}^{\mu\nu}={F}^{\mu\nu}_{\rm ext}+{F}^{\mu\nu}_{(-)}\,.
\label
{hat-F-def}
\end{equation} 
The definition of ${\hat S}^{\mu\nu}$ is more complicated. 
The derivative of this quantity is defined by 
\begin{equation}
\frac{d}{ds}\,{\hat S}_{\mu\nu}={\hat P}_{\mu}v_\nu-{\hat P}_{\nu}v_\mu
+m_{\mu\sigma}{\hat F}_\nu^{~\sigma}-m_{\nu\sigma}{\hat F}_\mu^{~\sigma}\,,
\label
{hat-S-def}
\end{equation}                        
where the auxiliary momentum  variable ${\hat P}^{\mu}$ satisfies the equation
\begin{equation}
\frac{d}{ds}\,{\hat P}^{\mu}=e{\hat F}^{\mu\nu}v_\nu
+m_{\kappa\sigma}\partial^\mu{\hat F}^{\kappa\sigma}\,.
\label
{hat-P-def}
\end{equation}                        
The constraints (\ref{spin-magnetic-moment}), (\ref{Frenkel_condition}), and (\ref{spin-unchanged}) 
therewith change, respectively, as
\begin{eqnarray}
m^{\mu\nu}=\mu{\hat S}^{\mu\nu}\,,
\\
{\hat  S}^{\mu\nu} v_\nu=0\,,
\\
{\hat  S}^{\mu\nu}\,\frac{d}{ds}\,{\hat S}_{\mu\nu}=0\,.
\label
{spin-unchanged-}
\end{eqnarray}                        
It is not unduly difficult to verify that the dynamical equations (\ref{equation-of-motion-four-momentum}) 
and (\ref{equation-of-motion-spin}) are consistent with these modified constraints.

To sum up, the classical dynamics of charged spinning particles formulated in terms of
ordinary commuting variables is a challenging task which still remains to be solved.

The surprising thing is that the Lagrangian description of charged spinning particles with the aid 
of anticommuting Grassmannian variables turns out to be quite simple.
As an illustration, we refer to the model of \cite{GalvaoTeitelboim1980} characterized by the use of
real-valued odd elements of a Grassmann algebra $\theta^\mu$ and $\theta_5$ to describe spin degrees 
of freedom.
An appropriate reparametrization invariant action is
\begin{eqnarray}
S=\int_{\tau_1}^{\tau_2} d\tau\Bigl[-P_\mu{\dot z}^\mu-\frac{i}{2}\left({\dot\theta}^\mu
\theta_\mu+{\dot\theta}_5\theta_5\right)-eA^\mu {\dot z}_\mu
\nonumber\\
+iM\left(P^2-m^2-ieF_{\mu\nu}\theta^\mu\theta^\nu\right)-N\left(\theta^\mu P_\mu+m\theta_5\right)\Bigr]
\nonumber\\
-\frac{i}{2}\left[\theta^\mu(\tau_1)\theta_\mu(\tau_2)+\theta_5(\tau_1)
\theta_5(\tau_2)\right],
\label{W-Grassmann}
\end{eqnarray}
and the endpoint variation conditions are
\begin{equation}
\Delta z^\mu(\tau_1)=\Delta z^\mu(\tau_2)=0\,,
\quad
\Delta\theta^\mu(\tau_1)+\Delta\theta^\mu(\tau_2)=0\,, 
\quad
\Delta\theta_5(\tau_1)+\Delta\theta_5(\tau_2)=0\,.  
\label
{bound-Grassmann}
\end{equation}
The variables $N$ and $M$ are Lagrange multipliers of the constraints 
$P^2-m^2-ieF_{\mu\nu}\theta^\mu\theta^\nu\approx 0$ and $\theta^\mu P_\mu+m\theta_5\approx 0$.
The even Grassmannian construction $i\theta^\mu\theta^\nu$ is similar to the spin tensor 
$S^{\mu\nu}$. 
 
A stumbling block for the study of such models lies in the following fact.
Although even elements of a Grassmann algebra contain usual numbers (Dirac called them $c$-numbers), 
the subset of $c$-numbers does not exhaust all the possibilities. 
There are even elements different from $c$-numbers. 
Solutions to the equation of motion for a spinning particle, $z^\mu(\tau)$, may well be built from 
even Grassmannian variables appearing in the action (\ref{W-Grassmann}) which are not $c$-numbers, 
say, from $N\theta^\mu$ and $i\theta^\mu\theta^\nu$.
The complete collection of world lines $z^\mu(\tau)$ may involve curves which need not be mappings 
${\mathbb R}\to{\mathbb R}_{1,3}$.
Therefore, this spinning particle lives in a realm which is not identical to Minkowski spacetime or
its warpings.
This realm is described in terms of arbitrary even Grassmannian variables. 
However, in this realm, 
most physical quantities are deprived of operational definition, and 
we do not have the slightest idea of how they can be experimentally recorded.

\section{SELF-INTERACTING GAUGE FIELD SYSTEMS}
\label
{YM}
Armed with the insight gained from the self-interaction problem in the Maxwell--Lorentz 
electrodynamics and simplified nonlinear field models possessing nontrivial topological structures, 
we begin to consider this problem in non-Abelian gauge field systems.
Both of the studied manifestations of self-interaction can occur there.
If a system does not involve particles, that is, delta-function sources of the field are missing 
from this system, solutions associated with phases of distinct topological structures are the 
sole manifestation of self-interaction.
The Yang--Mills--Higgs model is an example.
One phase of this system can be related to the Coleman plane waves, and the other phase can be 
attributed to the 't~Hooft--Polyakov magnetic monopole field.
The former enjoys the property of the internal SU$(2)$ symmetry, which is spontaneously broken to 
U$(1)$ in the latter.
These configurations differ not only in their topological setups, but also in physical properties.
On the other hand, if a system contains particles interacting with a non-Abelian gauge field, 
the phase manifestation of self-interaction is supplemented with rearranging initial mechanical 
and field degrees of freedom.
This phenomenon shows a general resemblance to that in the Maxwell--Lorentz theory, but occurs 
variously for different phases.
To illustrate, in one phase of the Yang--Mills--Wong theory, accelerated dressed particles emit 
Yang--Mills field waves of positive energy, while in the other phase, accelerated dressed particles 
emit waves of negative energy.
The gauge groups of these phases are different even though none of them is a subgroup of the other.
They are the compact and a noncompact real forms of the complexification of the gauge symmetry group.
This trait of self-interaction in the Yang--Mills--Wong theory is called ``spontaneous deformation of 
symmetry''.   

\subsection{Self-interaction in the Yang--Mills--Higgs theory}
\label
{pure}
A regular way for obtaining nontrivial static solutions of the pure SU$(2)$ gauge theory is to use 
the so-called Wu and Yang ansatz \cite{WuYang1969} in which space coordinates are interwoven with gauge field
coordinates,
\begin{equation} 
A_0^a({\bf x})=i\,\frac{x_a\, a(r)}{g r^2}\,,
\quad
A_j^a({\bf x})=\frac{\epsilon_{ajk}\, x_k\left[1-b(r)\right]}{gr^2}\,.
\label
{Wu-Yang-ansatz}
\end{equation}                                 

't~Hooft \cite{tHooft1974} and Polyakov \cite{Polyakov1974} pioneered the use of the Wu--Yang ansatz in the Yang--Mills--Higgs 
theory, and discovered a field configuration possessing properties of the magnetic monopole.
This configuration is a finite-energy smooth solution to the SO$(3)$ gauge theory with a Higgs 
triplet.
The Lagrangian is: 
\begin{equation}
{\cal L}=-{\frac14}\,G_{\mu\nu}^aG^{\mu\nu}_a+{\frac12}\,D_\mu\phi^a D^\mu\phi_a-V(\phi)\,,
\label
{YM-Higgs-Lagrn}
\end{equation}                                        
where 
\begin{equation}
G_{\mu\nu}^a=\partial_\mu A^a_\nu-\partial_\nu A^a_\mu+g\epsilon^{abc}A_\mu^bA_\nu^c\,,
\label
{YM-f-strength-GG}
\end{equation}                                        
\begin{equation}
D_\mu\phi_a=\partial_\mu\phi_a+g\epsilon_{abc}A_\mu^b\phi^c\,,
\label
{covar-der-Higgs-GG}
\end{equation}                                        
\begin{equation}
V(\phi)=
\frac{\lambda^2}{4}\left(\frac{\mu^2}{\lambda^2}-\phi^2\right)^2\,,
\quad
\phi^2=\phi_a\phi^a\,,
\quad
a=1,2,3\,.
\label
{Higgs-potentialGG}
\end{equation}                                        

The field equations read
\begin{equation}
\partial^\mu G_{\mu\nu}^a=g\epsilon^{abc}\left[\left(D_\nu\phi\right)_b\phi_c+A^\mu_bG_{\mu\nu}^c\right],
\label
{YM-eq-Higgs-GG}
\end{equation}                                        
\begin{equation}
\partial^\mu D_\mu\phi_a=g\epsilon_{abc}\left(D^\mu\phi^b\right)A_\mu^c+\mu^2\phi_a-\lambda^2\phi_a\phi^2\,.
\label
{eq-mot-Higgs-GG}
\end{equation}                                        

The Higgs potential $V(\phi)$ must approach zero as $r\to\infty$, which means that the Higgs 
field has a nonzero limit at spatial infinity                                  
\begin{equation}
\phi_a\to \frac{\mu}{\lambda}\,{\hat\phi}_a({\bf n})\,,
\quad
{\hat\phi}^a{\hat\phi}_a=1\,,
\quad
{\bf n}^2=1\,,
\quad
r\to\infty\,.
\label
{Higgs-asympt-GG}
\end{equation}                                        
The boundary condition (\ref{Higgs-asympt-GG}) singles out a particular axis ${\hat\phi}_a$ in the 
parameter space of the SO$(3)$ group of internal symmetry for each spatial direction ${\bf n}$, thus 
breaking this symmetry.
Solutions subject to this condition are invariant under the group of rotations about 
${\hat\phi}_a$.
Therefore, the unbroken symmetry is SO$(2)$ or, equivalently, the U$(1)$ subgroup of SO$(3)$. 
The resulting U$(1)$ gauge theory can be identified with Maxwell's theory of charged vector 
and scalar fields if generators $T_a$ of the initial SO$(3)$ group are projected on ${\hat\phi}^a$. 
Then the Abelian vector potential $A^\mu$ associated with the local U$(1)$ gauge group is
\begin{equation}
A^\mu=\left({\phi}^a A^\mu_a\right)\frac{\lambda}{\mu}\,,
\label
{A=g-hat-phi-lambda-by-mu}
\end{equation}                                        
and the electric charge is
\begin{equation}
e=g\left({\phi}^a T_a\right)\frac{\lambda}{\mu}\,.
\label
{e=g-hat-phi-lambda-by-mu}
\end{equation}                                        

Our concern here is with static spherically symmetric solution to the field equations (\ref{YM-eq-Higgs-GG})
and (\ref{eq-mot-Higgs-GG}). 
Let us take the gauge condition $A^a_0=0$.
This implies  $D_0\phi_a=0$ and $G^a_{0j}=0$.
With the ansatz
\begin{equation} 
\phi_a({\bf x})=\frac{x_a\, a(r)}{g r^2}\,,
\label
{tHooft-ansatz}
\end{equation}                                 
\begin{equation} 
A_j^a({\bf x})=\frac{\epsilon_{ajk}\, x_k\left[1-b(r)\right]}{gr^2}\,,
\label
{tHooft-ansatz-}
\end{equation}                                 
(\ref{YM-eq-Higgs-GG}) and (\ref{eq-mot-Higgs-GG}) become 
\begin{equation} 
r^2a''=a\left(2b^2-\mu^2 r^2+\frac{\lambda^2}{g^2}\,a^2\right),
\label
{tHooft-reduction-YM}
\end{equation}                                 
\begin{equation} 
r^2b''=b\left(b^2-1+a^2\right),
\label
{tHooft-reduction-Higgs}
\end{equation}                                 
where the prime stands for differentiation with respect to $r$.

Combining (\ref{Higgs-asympt-GG}) with (\ref{tHooft-ansatz}) gives 
\begin{equation} 
\lim_{r\to\infty}\phi({\bf r})=\frac{\mu}{\lambda}\,{\bf n}\,,
\label
{Higgs-at-ifty}
\end{equation}                                 
that is, ${\hat\phi}^a=n^a$.
The ansatz (\ref{tHooft-ansatz-}) is consistent with Eqs.~(\ref{tHooft-reduction-YM}) and 
(\ref{tHooft-reduction-Higgs}) if $b(r)=O(1)$ as $r\to\infty$.
The ``electromagnetic'' field strength $F_{\mu\nu}$ is to be identified with the component of the 
Yang--Mills strength $G^a_{\mu\nu}$ in the direction of ${\hat\phi}^a$, which corresponds to the 
unbroken U$(1)$ symmetry.
Taking into account that $G^a_{0j}=0$, one can show that, far from the origin,
\begin{equation} 
G^a_{ij}=-\epsilon_{ijk}\,\frac{n_k n_a}{4\pi er^2}\,, 
\label
{tHooft-monopole-G}
\end{equation}                                 
or
\begin{equation} 
F_{ij}={\hat\phi}_a\,G^a_{ij}=-\frac{1}{4\pi er^2}\,\epsilon_{ij a}n_a\,,
\label
{tHooft-monopole-F}
\end{equation}                                 
which represents a radial static magnetic field
\begin{equation} 
B_k=-\frac{n_k}{4\pi er^2}\,,
\label
{tHooft-monopole-B}
\end{equation}                                 
generated by the total magnetic charge $e^\star=1/e$.
This solution, called the 't~Hooft--Polyakov monopole, describes the Yang--Mills field of
magnetic type because  ${}^\ast{G}^a_{\mu\nu}{G}_a^{\mu\nu}=0$ and ${G}^a_{\mu\nu}{G}_a^{\mu\nu}>0$.

The 't~Hooft--Polyakov monopole and the Dirac monopole \cite{Dirac1931}, \cite{Dirac1948} differ 
in their structure within a 
core of size $\sim \lambda/\mu$.
The Dirac monopole solution (\ref{tHooft-monopole-B}) has a singularity for which a point source has to be
introduced explicitly in the action, while the 't~Hooft--Polyakov monopole is smooth everywhere and 
satisfies the field equations (\ref{YM-eq-Higgs-GG}) and (\ref{eq-mot-Higgs-GG}) without external 
sources.
Outside the core these configurations are similar.

There is another class of solutions in the Yang--Mills--Higgs theory, the non-Abelian analogues of 
electromagnetic plane waves, originally discovered by Coleman \cite{Coleman1977}.
To describe these solutions of the field equations (\ref{YM-eq-Higgs-GG}) and (\ref{eq-mot-Higgs-GG}),
it is convenient to adopt light-cone coordinates $x^\mu=\left(x^+, x^-,x^2,x^3\right)$, $x^\pm=x^0\pm x^1$.
The metric, expressed in terms of these coordinates, is $ds^2=dx^+dx^--(dx^2)^2-(dx^3)^2$.
The Coleman solutions for plane waves moving in the negative $x^1$-direction are given by
\begin{equation} 
\phi^a=0\,,
\quad
G^a_{2+}=-G^a_{+2}=f^a(x^+)\,,
\quad
G^a_{3+}=-G^a_{+3}=g^a(x^+)\,,
\label
{electrictype}
\end{equation}                                 
where the $f$'s and $g$'s are arbitrary bounded functions of $x^+$, with all other components of 
the field strength being vanishing. 
For these solutions, the energy density is bounded throughout spacetime, the direction of the
Poynting vector is constant, the magnitude of the Poynting vector is equal to the energy density, 
and ${}^\ast{G}^a_{\mu\nu}{G}_a^{\mu\nu}=0$,  ${G}^a_{\mu\nu}{G}_a^{\mu\nu}=0$, which allows to 
classify these ${G}^a_{\mu\nu}$ as null-field configurations.
The Coleman plane waves represent a phase distinct from the phase associated with the 
't~Hooft--Polyakov magnetic monopoles.
There is no bijective continuous mapping between these two configurations.

The 't~Hooft--Polyakov monopole is electrically neutral due to the combination of the gauge 
condition $A_0^a = 0$ and the requirement that the field configurations should be static.
Julia and Zee \cite{JuliaZee1975}  have abandoned this gauge condition and used the ansatz
\begin{equation} 
A^a_0({\bf x})=\frac{c(r)\,n_a}{4\pi gr^2}\,,
\label
{Julia-Zee-ansatz}
\end{equation}                                 
where $c(r)$ is an unknown function.
The resulting static solution exhibits nonzero $G^a_{0j}$. 
This configuration with electric and magnetic charges, presently known as the Julia--Zee dyon,
offers a further phase of the Yang--Mills--Higgs system.

For arbitrary $\mu$ and $\lambda$, the set of ordinary differential equations (\ref{tHooft-reduction-YM}) 
and (\ref{tHooft-reduction-Higgs}) has never been solved analytically.
However, in the limit $\mu\to 0$, $\lambda\to 0$, with $\mu/\lambda<\infty$,
an exact solution
\begin{equation} 
a(r)=\beta r\,{\coth(\beta r)}-1\,, 
\quad 
b(r)=\frac{\beta r}{\sinh(\beta r)}\,,
\label
{Bogomol'ny-Prasad-Sommerfeld}
\end{equation}                                 
where $\beta$ is an arbitrary constant, was obtained by Bogomol'nyi \cite{Bogomol'nyi1976}, and 
Prasad and Sommerfield \cite{PrasadSommerfield1975}.
The Bogomol'nyi--Prasad--Sommerfield configuration is self-dual,
\begin{equation} 
{}^\ast G_{\mu\nu}^a= iG_{\mu\nu}^a\,,
\label
{self-duality-df-gen}
\end{equation}                                 
where the Hodge dual field ${}^\ast G_{\mu\nu}^a$ is defined by ${}^\ast G_{\mu\nu}^a=\frac12
\epsilon_{\mu\nu\alpha\beta}G^{a\hskip0.5mm\alpha\beta}$.
This implies that the Yang--Mills term of the stress-energy tensor $\Theta_{\mu\nu}$ is vanishing.
Indeed, 
\begin{eqnarray}
\Theta_{\mu\nu}
=
\frac{1}{4\pi} 
\left(G_{a\mu}^{\hskip2.5mm\lambda} G^a_{\lambda\nu}
+\frac{\eta_{\mu\nu}}{4}\,G_a^{\alpha\beta}G^a_{\alpha\beta}\right)
=
\frac{1}{8\pi}\left(
G_{a\mu}^{\hskip2.5mm\lambda}G^a_{\lambda\nu}+
{}^\ast G_{a\mu}^{\hskip2.5mm\lambda}\,{}^\ast G^a_{\lambda\nu}
\right)
\nonumber\\ 
=
\frac{1}{8\pi}
\left(G_{a\mu}^{\hskip2.5mm\lambda}+i\,{}^\ast G_{a\mu}^{\hskip2.5mm\lambda}\right)
\!
\left( G^a_{\lambda\nu}-i\,{}^\ast G^a_{\lambda\nu}\right),
\label
{energy-tens-self-dualit}
\end{eqnarray}                                 
which shows that $\Theta_{\mu\nu}=0$ for 
${}^\ast G_{\mu\nu}=\pm i\, G_{\mu\nu}$.
Thus, the Bogomol'nyi--Prasad--Sommerfield configuration carries zero energy and momentum.

This cursory glance at nontrivial solutions of the Yang--Mills--Higgs system is suffice for present 
purposes.
The availability of these solutions makes it clear that self-interaction in this system gives
rise to several phases with different geometric and physical properties.

The literature on exact solutions of classical Yang--Mills theories is rather extensive.
The reader interested in this topics may consult the books \cite{Rajaraman1982}, \cite{Schwarz1991}, 
and \cite{Rubakov2002}.
A detailed treatment of monopole solutions is given in \cite{GoddardOlive1978}, \cite{JaffeTaubes1980}, 
\cite{Coleman1983}, \cite{AtiyahHitchin1988}.
A largely complete review of classical solutions to SU$(2)$ gauge theories that were known by the 
end of the 1970s is represented in the survey \cite{Actor1979}.

\subsection{Self-interaction in the Yang--Mills--Wong theory}
\label
{YMW}
The Yang--Mills--Wong theory has many features in common with the Maxwell--Lorentz theory. 
This is a classical gauge field theory which describes the interaction of particles carrying 
non-Abelian charges with the Yang--Mills field.
A closed system of $K$ particles interacting with the SU$({\cal N})$ Yang--Mills field is governed 
 by the action \cite{Balachandran1978}
\begin{eqnarray}
S=-\sum_{I=1}^K\,\int ds_I\left\{m^I_0\,\sqrt{{v}_I\cdot{v}_I}
+\sum_{a=1}^{{\cal N}^2-1}\sum_{i,j=1}^{{\cal N}}\, q_I^a\,\eta_{Ii}^{\ast}\left
[\delta^i_{\hskip0.5mm j}\frac{d}{ds_I}
+
{v}_I^\mu\!\left( A^a_{~\mu} T_a\right)^{i}_{\hskip0.5mm j}\right]
\eta_{I}^{\hskip0.3mm j}\right\}
\nonumber\\
-\frac{1}{16\pi}\int d^4x\,G_a^{\mu\nu}G^a_{\mu\nu}\,,
\label
{Balachandr-action}
\end{eqnarray}                                           
where $T_a$ are generators of SU$({\cal N})$, 
$G^a_{\mu\nu}=\partial_\mu A^a_{\nu}-\partial_\nu A^a_{\mu}+if^{a}_{~bs}A^b_{\mu} A^c_{\nu}$ is the 
field strength, $f_{abc}$ are the structure constants of SU$({\cal N})$.
The SU$({\cal N})$ gauge group is thereafter called the ``color'' gauge group, and the classical 
particles go under the name of ``quarks'', with the understanding, however, that all such terms 
refer to the Yang--Mills--Wong theory, which may in some respects stretch the truth of subnuclear 
realm. 

Quarks, labelled with $I$, carry color charges $Q_I$ in the adjoint representation of SU$({\cal N})$, 
$Q_I=Q^a_I\,T_a$, which can be written in terms of the basic variables $\eta_{Ij}$ in the 
fundamental representation,
\begin{equation}
Q_I=\sum_{a=1}^{{\cal N}^2-1}\sum_{i,j=1}^{{\cal N}}\, q_I^a\,\eta_{Ii}^{\ast}\left(T_a\right)^{i}_{\hskip0.5mm j}
\eta_{I}^{\hskip0.3mm j} 
\,.
\label
{Q-in-terms-eta}
\end{equation}                                           

The Euler--Lagrange equations for $\eta$ and $\eta^{\ast}$, in which the label $I$ is omitted, 
\begin{eqnarray}
{\dot\eta}^i=-({v}\cdot A^a)\left(T_a\right)^i_{\hskip0.5mm j}\eta^j\,\,,
\nonumber\\ 
{\dot\eta}^{\ast}_j=\eta^{\ast}_i\,({v}\cdot A^a)\left(T_a\right)^i_{\hskip0.5mm j}\,\,,
\label
{Euler-Lagrange-eta-YMW}
\end{eqnarray}                                          
can be combined into 
\begin{equation}
{\dot Q}_a=-if_{abc}\,Q^b\left({v}\cdot A^c\right).
\label
{8}
\end{equation}                                           
Equation (\ref{8}) was originally derived by Wong  \cite{Wong1970}.
It shows that the color charge $Q_a$ shares with a top the property of precessing. 
Indeed, $Q_a$ precesses around the axis ${v}\cdot A^a$ in the color space.

In contrast to the electric charge $e$, which is a constant, the color charge $Q_a$ is a dynamical 
variable governed by Eq.~(\ref{8}).
But the color charge magnitude is a constant of motion,
\begin{equation}
\frac{d}{ds}\,{Q}^2=2{\dot Q}^aQ_a=0\,,
\label
{Q-square}
\end{equation}                                          
which follows from (\ref{8}), written in the Cartan basis, because in this basis, $f_{abc}=-f_{bac}$.
Furthermore, there is good reason to look for solutions of the Yang--Mills--Wong theory 
satisfying the condition
\begin{equation}
Q_a(s)= {\rm const}.
\label
{Q-a=const}
\end{equation}                                          
Abandoning this condition would pose the problem of an infinitely rapid precession of $Q_a$ 
because the retarded field $A^a_\mu$ is singular on the world line.

Varying $A^\mu$ in the action (\ref{Balachandr-action}) gives the Yang--Mills equations:
\begin{equation}
{\cal E}^a_\mu(x)
=\left(D^{\lambda} G_{\lambda\mu}\right)^a(x)-4\pi g\sum_{I=1}^K\,\int_{-\infty}^\infty ds_I\,Q^a_I(s_I)\,
{v}^I_\mu(s_I)\,\delta^4\left[x-z_I(s_I)\right]=0\,.
\label
{YM-eq-}
\end{equation}                                          

Before proceeding further let us recall that knowing the retarded solution to Maxwell's equations 
with the source involving a single point charge, 
\begin{equation}
A^\mu =e\,\frac{v^\mu}{\rho}\,, 
\label
{A_LW}
\end{equation}                                          
its extension to the case that the source is composed of $K$ charges follows immediately: 
\begin{equation}
A^\mu=\sum_{I=1}^K e_I\,\frac{v^\mu_I}{\rho_I}\,.
\label
{A_LW_K}
\end{equation}                                          
Allowing for linear combinations of solutions with arbitrary real coefficients, Eq.~(\ref{A_LW_K}), 
is tantamount to stating that electric charges $e_I$ take arbitrary real values.
By contrast, the superposition principle does not apply to the Yang--Mills equations unless they 
become Abelian, and hence linearize.  
Non-Abelian solutions of the Yang--Mills equations with the single-quark source are prevented from 
superposing, and we are forced to solve Eq.~(\ref{YM-eq-}) for each $K$ individually.

A systematic method for finding exact retarded solutions of Eq.~(\ref{YM-eq-}) with the source 
composed of $K$ quarks moving along arbitrary timelike smooth world lines was proposed and developed 
in a series of papers \cite{Kosyakov1991}, \cite{Kosyakov1993}, \cite{Kosyakov1994}, 
\cite{Kosyakov1998}, \cite{Kosyakov1999}, \cite{Kosyakov2007}, and rediscovered in part in
\cite{Sarioglu2002}.
Without going into detail of this method we simply present typical solutions and discuss their 
properties bearing on self-interaction.

There are {two kinds of retarded solutions} to Eq.~(\ref{YM-eq-}), Abelian and non-Abelian.
Abelian solutions are defined on a set of ${\cal N}-1$ commuting matrices $H_a$ which form the 
Cartan subalgebra of the Lie algebra su$({\cal N})$.
To build non-Abelian solutions, we need extended subalgebras of su$({\cal N})$, containing the
Cartan subalgebra.

Consider the simplest case that the SU$(2)$ Yang--Mills field is generated by a 
single quark moving along a timelike smooth world line.
The retarded Abelian solution is
\begin{equation}
A^\mu=q T_3\,{{v}^\mu\over\rho}\,,
\label
{Abelian-1-Pauli-}
\end{equation}                                       
where $q$ is an arbitrary real parameter whereby the color charge of the quark is measured,
$Q=q T_3$.
This solution has much in common with the Li\'enard--Wiechert vector potential (\ref{A_LW}), 
in particular the field $G^{\mu\nu}$ evaluated from (\ref{Abelian-1-Pauli-}) is of electric type.

The corresponding retarded non-Abelian solution is 
\begin{equation}
A^\mu=\mp\,{2i\over g}\,T_3\,{{v}^\mu\over\rho}+i\kappa\left(T_1\pm iT_2\right)R^\mu\,.
\label
{Nabelian-1-Pauli-}
\end{equation}                                       
Here $T_a$ $(a=1,2,3)$, are three generators of SU$(2)$, which can be expressed in terms of the Pauli
matrices, $T_a=\frac12\sigma_a$, and $\kappa$ is an arbitrary real nonzero parameter.
The field strength associated with this vector potential is
\begin{equation}
{G}^{\mu\nu}=c^\mu {W}^\nu-c^\nu {W}^\mu\,,
\label
{F-single-quark}
\end{equation}                                         
\begin{equation}
{W}^\mu =\mp\,{2i\over g}\,{T}_3\,{U^{\mu}\over\rho^2}+i\kappa\left(T_1\pm iT_2\right)v^\mu\,,
\label
{W-df}
\end{equation}                                         
where  $U^\mu=-\lambda v^\mu+\rho a^{\mu}$ which is the same as that given by Eq.~(\ref{U}).

With the prescription that observable color singlets must involve either ``$+$'' or 
``$-$'' representatives of the non-Abelian solutions, not their mix, one can deduce that the 
configuration defined by Eqs.~(\ref{F-single-quark})--(\ref{W-df}) qualifies as a Yang--Mills field of 
magnetic type,
\begin{equation}
{\cal P}=\frac12\,{}^\ast{G}^a_{\mu\nu}{G}_a^{\mu\nu}=0\,,
\quad
{\cal S}=\frac12\,{G}^a_{\mu\nu}{G}_a^{\mu\nu}={4\over{g^2\rho^4}}>0\,.
\label
{invariants_YM_cold}
\end{equation}                                          

The Yang--Mills equations determine not only the retarded non-Abelian field (\ref{Nabelian-1-Pauli-}), 
but also the magnitude of the color charge that generates this field.
Indeed, the quantity
\begin{equation}
{Q}^2=-\frac{4}{g^2}\,
\label
{Q-square=2-g}
\end{equation}                                          
appearing in (\ref{Nabelian-1-Pauli-}) is ordered by the structure of the Yang--Mills equations~(\ref{YM-eq-}).

The solution (\ref{Nabelian-1-Pauli-}) acquires the form $A_\mu={\cal A}_\mu^a\,{\cal T}_a$
where all coefficients ${\cal A}_\mu^a$ are pure imaginary with the use of the matrix 
basis
\begin{equation}
{\cal T}_1=\tau_1\,,
\quad{\cal T}_2=i\tau_2\,,
\quad{\cal T}_3=\tau_3\,.
\label
{sl(2,R) basis}
\end{equation}                                       
Elements of this basis obey the commutation relations
\begin{equation}
[{\cal T}_1,{\cal T}_2]=-{\cal T}_3\,,
\quad
[{\cal T}_2,{\cal T}_3]=-{\cal T}_1\,,
\quad
[{\cal T}_3,{\cal T}_1]={\cal T}_2\,,
\label
{sl(2,R) cr}
\end{equation}                                       
which underlie the sl$(2,{\mathbb R})$ Lie algebra. 
The color space becomes a pseudoeuclidean space with the metric $\eta_{ab}={\rm diag}\,(-1,1,-1)$.
The automorphism group of this space is SO$(2,1)$. 
On the other hand, the gauge group of the solution (\ref{Abelian-1-Pauli-}) 
is the initially chosen SU$(2)$.

Should we adopt the initial gauge group Sp$(1)$, rather than SU$(2)$ or SO$(3)$, we would come to 
identical results owing to the equivalence of three complex Lie algebras 
\begin{equation}
{\rm sp}(1,{\mathbb C})\sim{\rm  sl}(2,{\mathbb C})\sim {\rm so}(3,{\mathbb C})\,, 
\label
{equiv-compl}
\end{equation}                                       
their real compact forms
\begin{equation}
{\rm sp}(1)\sim{\rm  su}(2)\sim{\rm  so}(3)\,, 
\label
{equiv-com}
\end{equation}                                       
and their real noncompact forms
\begin{equation}
{\rm sp}(1,{\mathbb R})\sim{\rm  sl}(2,{\mathbb R})\sim {\rm su}(1,1)\sim {\rm so}(2,1)\,, 
\label
{equiv-n-com}
\end{equation}                                       
see, {\it e.~g.}, \cite{BarutRaczka1977}.

Imagine for a little that a single quark is in the Universe. 
The system ``the quark plus its own Yang--Mills field'' exists in two phases which are distinguished 
by their groups of gauge symmetry: SU$(2)$ and SL$(2,{\mathbb R})$.
These phases will be conventionally referred to as {``hot''} and {``cold''}.

A closer look at SU$(2)$ and SL$(2,{\mathbb R})$ shows that none of them is a subgroup of the other.
The origin of SL$(2,{\mathbb R})$ bears no relation to spontaneous symmetry breakdown.
SU$(2)$ and SL$(2,{\mathbb R})$ are the {compact and a noncompact real forms} of the complex group 
SL$(2,{\mathbb C})$. 
Invariance of the action (\ref{Balachandr-action}) under SU$(2)$ automatically entails its 
invariance under the complexification of this group, SL$(2,{\mathbb C})$.
However, a complex-valued Yang--Mills field may seem problematic in the classical context, 
particularly where observable quantities, such as energy, were involved.
Only real forms of SL$(2,{\mathbb C})$ appear to be satisfactory as gauge groups. 
The emergence of a solution invariant under a real form of SL$(2,{\mathbb C})$ different from the 
initial SU$(2)$ is a phenomenon specific to the Yang--Mills--Wong theory.
We call it the {``spontaneous symmetry deformation''}.
The cold and hot phases differ from each other not only in their symmetry; a cold quark generates 
the Yang--Mills field of magnetic type, while a hot quark generates the Yang--Mills field of 
electric type.

While on the subject of systems composed of $K$ quarks and their field evolving in the non-Abelian 
regime, we note that the Yang--Mills--Wong theory of such systems can be consistently formulated 
for the color gauge group SU$({\cal N})$ with ${\cal N}\ge K+1$.
To illustrate, we refer to a system of two quarks whose initial gauge group is assumed to be 
SU$(3)$. 
A retarded non-Abelian solution that describes the field due to two quarks moving along arbitrary 
timelike smooth world lines is 
\begin{equation}
A^\mu =\mp{2i\over g}\left(H_1\,{{v}^\mu_1\over\rho_1}+g\kappa E_{13}^\pm R_1^\mu\right)\,
\mp\,{2i\over g}\left(H_2\,{{v}^\mu_2\over\rho_2}+g\kappa E_{23}^\pm R_2^\mu \right),
\label
{alternative-sol-}
\end{equation}                                       
where $H_l$ and $E^\pm_{mn}$ are generators of SU$(3)$ in the Cartan basis, which can be expressed in 
terms of the Gell-Mann matrices: 
\begin{eqnarray}
H_1= \frac12\left(\lambda_3+\frac{\lambda_8}{\sqrt{3}}\right),
\quad
H_2= -\frac12\left(\lambda_3-\frac{\lambda_8}{\sqrt{3}}\right),
\nonumber\\ 
E^+_{13}= \frac12\left(\lambda_4+i\lambda_5\right),
\quad
E^+_{23}= \frac12\left(\lambda_6+i\lambda_7\right).
\label
{H-E-in-terms-lambdas}
\end{eqnarray}                                       
$R_1^\mu=x^\mu-{z}^\mu_1(s_1)$ and $R_2^\mu=x^\mu-{z}^\mu_2(s_2)$ are, respectively, null four-vectors 
drawn from points ${z}^\mu_1(s_1)$ and ${z}^\mu_2(s_2)$ on the world lines of quarks 1 and 2, where 
the signals were emitted, to the point $x^\mu$, where the signals were received.

Expression (\ref{alternative-sol-}) is imaginary-valued in the color basis
\begin{eqnarray}
{\cal T}_1={\lambda_1}\,,\quad
{\cal T}_2=i{\lambda_2}\,,\quad
{\cal T}_3={\lambda_3}\,,\quad
{\cal T}_4={\lambda_4}\,,
\nonumber\\ 
{\cal T}_5=i{\lambda_5}\,,\quad
{\cal T}_6={\lambda_6}\,,\quad
{\cal T}_7=i{\lambda_7}\,,\quad
{\cal T}_8={\lambda_8}\,,
\label
{basis-Sec8.3}
\end{eqnarray}                                       
or in the Cartan basis spanned by $H_n$ and $E_{mn}^\pm$.
The ${\cal T}_n$'s are traceless real $3\times 3$ matrices satisfying the commutation relations of 
the Lie algebra sl$(3,{\mathbb R})$. 
Thus, the gauge group of the non-Abelian solution (\ref{alternative-sol-}) is actually SL$(3,{\mathbb R})$.

The structure of retarded non-Abelian solutions to the Yang--Mills equations with the source 
composed of $K$ quarks closely resembles that of (\ref{alternative-sol-}).
These solutions are the sum (not an arbitrary superposition) of single-quark terms, as 
exemplified by Eq.~(\ref{alternative-sol-}) in which two single-quark terms add up to give 
the retarded Yang--Mills field generated by two quarks in the cold phase.
The spontaneous symmetry deformation, responsible for the emergence of the gauge group which is a 
noncompact real form of the complex group SL$({\cal N},{\mathbb C})$, occurs universally in all 
systems of $K$ quarks governed by the action (\ref{Balachandr-action}). 

The corresponding retarded Abelian solutions
\begin{equation}
A_\mu=\sum_{I=1}^K\sum_{n=1}^{{\cal N}-1}\,q_I^n\,H_n\,{v_\mu^I\over\rho_I}\,
\label
{73}
\end{equation}                                   
are linear combinations of single-quark solutions (\ref{Abelian-1-Pauli-}) with arbitrary real 
coefficients $q_I^n$.
The gauge group of these configurations is the initial SU$({\cal N})$.

Varying $z_I^\mu$ in the action (\ref{Balachandr-action}) gives the equation of motion for $I$th 
bare quark 
\begin{equation}
\varepsilon_I^\lambda(z_I)=m_0^I\,{a}_I^\lambda-{v}^I_\mu\,Q^a_I\, G_a^{\lambda\mu}(z_I)=0\,.
\label
{Euler-Lagrange-z-YMW-}
\end{equation}                                          

Since we are to study the rearrangement in hot and cold phases, the retarded Abelian and 
non-Abelian solutions to the Yang--Mills equations with the source composed of $K$ quarks should be
substituted to Eq.~(\ref{Euler-Lagrange-z-YMW-}).
However, the singularities of these solutions on the world lines preclude the direct execution of 
this plan.
As before we invoke the Noether identity
\begin{equation}
\partial_\mu\!\left(\Theta^{\lambda\mu}+t^{\lambda\mu}\right)(x)={1\over 4\pi}\left({\cal E}^a_\mu G_a^{\lambda\mu}\right)(x)
+
\int^\infty_{-\infty}ds\,\varepsilon^\lambda(z)\,\delta^4\left[x-z(s)\right],
\label
{Noether-1-id-ym}
\end{equation}                         
where 
\begin{equation}
\Theta^{\mu\nu}={1\over 4\pi}\left(G_a^{\mu\sigma}G_\sigma^{a~\nu}+
{1\over 4}\,\eta^{\mu\nu}G^a_{\alpha\beta}G_a^{\alpha\beta}\right),
\label
{Theta-mu-nu-ym}
\end{equation}                           
$t^{\mu\nu}$ is given by (\ref{t-mu-nu}), ${\cal E}^a_\mu$ and $\varepsilon^\lambda$ are 
respectively the left-hand sides of (\ref{YM-eq-}) and (\ref{Euler-Lagrange-z-YMW-}).
 
In the hot phase, the Yang--Mills equations linearize, and become almost identical to Maxwell's 
equations.
Accordingly, all results of Sec.~\ref{rearrangement_ML} are reproduced with the only replacement 
$e^2\to q^2$.
The degrees of freedom appearing in the action (\ref{Balachandr-action}) are rearranged on the 
extremals subject to the retarded condition to give dressed quarks and Yang--Mills radiation 
closely resembling such entities in electrodynamics.
The behavior of a dressed quark is governed by the Abraham--Lorentz--Dirac equation (\ref{LD}),
\begin{equation} 
m\left[a^\mu-\tau_0\left({\dot a}^\mu +v^\mu a^2\right)\right]=f^{\mu}\,,
\label
{LD-quark}
\end{equation}                       
where $\tau_0$ is a characteristic time interval defined by (\ref{class-radius}) in which $e$ is 
substituted for $q$. 

As to the cold phase, three more points need to be made.
First, a special magnitude for the color charge of every quark of $K$-quark systems evolving in 
the non-Abelian regime is selected by the Yang--Mills equations:
\begin{equation} 
Q^2_I=-{4\over g^2}\left(1-\frac{1}{{\cal N}}\right),
\quad
{\cal N}\ge 3\,.
\label
{Q-2=4/g-2(1-1/N)}
\end{equation}                       
Our principal interest here is with the overall minus sign of expression (\ref{Q-2=4/g-2(1-1/N)}).

Second, the nonlinearity of the Yang--Mills equations is yet compatible with the fact that retarded 
non-Abelian solutions are given by the sum of single-quark terms.
Because all retarded non-Abelian solutions share this common property, the stress-energy tensor of the 
Yang--Mills field is written as
\begin{equation}
\Theta^{\mu\nu}=\sum_{I}\left(\Theta_{~I}^{\mu\nu}+\sum_{J}\Theta_{~IJ}^{\mu\nu}\right), 
\label
{Theta-self-mix-decomp-}
\end{equation}                        
where $\Theta_{I}^{\mu\nu}$ is comprised of the field generated by $I$th quark, and 
$\Theta_{~IJ}^{\mu\nu}$ contains mixed contributions of the fields due to $I$th and $J$th quarks.
Expression (\ref{Theta-self-mix-decomp-}) is similar to that of $\Theta^{\mu\nu}$ in the 
Maxwell--Lorentz theory.
We are thus entitled to reiterate the procedure used in Sec.~\ref{rearrangement_ML} to reveal the
rearrangement of Yang--Mills--Wong systems evolving in the non-Abelian regime.

Third, conformal invariance is apparently violated by the linearly rising terms of $A_\mu$ 
containing constants $\kappa$ which have dimension $({\rm length})^{-2}$.
Note, however, that 
\begin{equation}
{\rm tr}\left(H_l\,E^\pm_{mn}\right)=0\,,
\quad
{\rm tr}\left(E^\pm_{mn}\,E^\pm_{mn}\right)=0\,,
\label
{tr}
\end{equation}                                    
whence it follows that any color singlet, such as $\Theta^{\mu\nu}$, is free of the contributions 
violating conformal invariance. 
This is because the linearly rising terms depend upon either $E_{mn}^+$ or $E_{mn}^-$, but not both.
Although the linearly rising terms of $A_\mu$ contribute to the field strength $G_{\mu\nu}$, as 
viewed in Eq.~(\ref{W-df}), conformal invariance is recovered on the level of observables.
For instance, the force exerted on $I$th quark from all other quarks
\begin{equation}
f^\mu_I=v_\nu^I\,Q^a_I\, G_a^{\mu\nu}(z_I)\,
\label
{f-I-wong}
\end{equation}                                    
involves only the ``Li\'enard--Wiechert part'' of the field strength $G_{\mu\nu}^{{\rm LW}}$.
The explanation is simple.
The force (\ref{f-I-wong}) includes the scalar product of two color vectors $G_{\mu\nu}-G_{\mu\nu}^{{\rm LW}}$ and $Q_I$.
They are not arbitrary; the exact solutions constrain these vectors to be orthogonal to each other.

On this basis we can get the following conclusions \cite{Kosyakov1991}, \cite{Kosyakov1992}, 
\cite{Kosyakov1998}, \cite{Kosyakov2007}.
In the cold phase, an accelerated quark gains, rather than loses, energy by emitting the Yang--Mills 
radiation.
To see this, one has to derive the emitted four-momentum 
\begin{equation}
{\cal P}^{\mu}=m\ell\int^{{s}}_{-\infty}d\tau\,a^2\left(\tau\right) v^\mu\left(\tau\right),
\label
{P-rad-quark}
\end{equation}                        
making clear the fact that 
\begin{equation}
{\dot{\cal P}}\cdot v=m\ell a^2<0\,, 
\label
{P-rad-quark-cold}
\end{equation}                        
which is construed as absorbing convergent waves of positive energy, or, alternatively, as emitting 
divergent waves of negative energy.
Here, $m$ is the renormalized mass of $I$th quark (from here on, the label $I$ is omitted), 
and $\ell$ stands for a characteristic length
\begin{equation}
\ell=\frac{8}{3m g^2}\left(1-\frac{1}{{\cal N}}\right).
\label
{tau-0-quark}
\end{equation}

It is interesting to compare this parameter with the characteristic length $\tau_0$ in the 
Maxwell--Lorentz electrodynamics which is thought of as an effective theory resulting from 
quantum electrodynamics at long distances.
The validity of this effective theory is limited by a cutoff related to the Compton wave length of 
the electron, $\lambda_e=3.86\cdot 10^{-11}$ cm.
At shorter distances, the effect of pair creations becomes appreciable. 
Likewise, we may understood the Yang--Mills--Wong theory as an effective theory to low-energy quantum
chromodynamics, and associate the 
cutoff with the Compton wave length of the quark, $\lambda_q$. 
But unlike $\tau_0$, which is proportional to $e^2\approx 1/137$, the characteristic length in the
cold phase $\ell$ is inversely related to $g^2$.
In the strong coupling regime $g\sim 1$, the parameter $\ell$ is of order of the Compton wave length of 
the quark.
However, if $g\ll 1$, then  $\ell\gg \lambda_q$, so that all phenomena specified by $\ell$  fall in 
the range of validity of this classical theory being immune to the effect of pair creations.

The four-momentum of a dressed quark in the cold phase is
\begin{equation}
p^\mu=m\left(v^\mu+\ell a^\mu\right),
\label
{p-mu-dress-quark}
\end{equation}                                          
and therefore
\begin{equation}
p^2=m^2\left(1+\ell^2 a^2\right).
\label
{p-mu-dress-quark-sqr}
\end{equation}                                          
If the acceleration exceeds its critical value ${a}^2_c=-\ell^{-2}$ (which is another way of stating 
that the momentum transfer is greater than $\ell^{-1}$), the quark becomes a tachyon.
Note, however, that light constituent quarks have mass of about $300$ MeV which is close 
to the deconfinement transition temperature $T_c\approx 200\pm 50$ MeV.
With reference to what was said immediately after Eq.~(\ref{tau-0-quark}), this suggests that the 
attainment of ${a}^2_c$ may result 
in triggering between the cold and hot phases rather than producing a tachyonic state.  

The local energy-momentum balance 
\begin{equation}
{\dot p}^\mu+{\dot{\cal P}}^{\mu}+{\dot{\wp}}^{\mu}=0\,
\label
{local-balance-YMW}
\end{equation}                        
applies to the cold phase.
Here, $p^\mu$ and ${\cal P}^{\mu}$ are defined respectively by Eqs.~(\ref{p-mu-dress-quark}) and 
(\ref{P-rad-quark}), and ${\dot{\wp}}^{\mu}=-f^\mu$, with $f^\mu$ being given by (\ref{f-I-wong}).
According to this balance, the four-momentum $d{\wp}^{\mu}=-f^\mu ds$, extracted from an external 
field during an infinitesimal interval $ds$, is used for changing the four-momentum of a dressed 
cold quark $dp^{\mu}$ and emitting the Yang--Mills radiation four-momentum ${d{{\cal P}}}^\mu$
with negative energy content.

Equation (\ref{local-balance-YMW}) can be rewritten to give the equation of motion for a dressed quark
\begin{equation} 
m\left[ a^\mu+\ell\left({\dot a}^\mu+v^\mu a^2\right)\right]=f^{\mu}\,.
\label
{LD-quark-cold}
\end{equation}                       
The only qualitative difference between the equation of motion for a dressed quark in the cold phase,
Eq.~(\ref{LD-quark-cold}), and that in the hot phase, Eq.~(\ref{LD-quark}), is the overall sign of 
the parenthesized term. 
To appreciate this difference, let us assume that $f^\mu=0$.
While the general solution to equation (\ref{LD-quark}) takes the runaway form
(\ref{runaway-Abraham-Lorentz}), the general solution to equation (\ref{LD-quark-cold}) proves to be
\begin{equation} 
v^\mu(s)=V^\mu\cosh(\nu_0+w_0\ell\, e^{-s/\ell})+U^\mu\sinh(\nu_0+w_0\ell\,e^{-s/\ell})\,.
\label
{self-deceleration}
\end{equation}                         
Here, $V^\mu$ and $U^\mu$ are constant four-vectors such that $V\cdot U=0$,  $V^2=-U^2=1$, and 
$\nu_0$ and $w_0$ are arbitrary parameters.
This self-decelerated solution should be attended with the Haag asymptotic condition (\ref{a-to-0}); 
otherwise the radiated four-momentum (\ref{P-rad-quark-cold}) will be divergent. 
This requirement is fulfilled only for $w_0=0$. 
Therefore, a free dressed quark governed by Eq.~(\ref{LD-quark-cold}) moves along a
straight world line $v^\mu=$ const.

\section{CLASSICAL SELF-INTERACTING STRINGS}
\label
{string}
Do classical strings undergo self-interaction?
At first glance, strings are systems equipped with enough symmetry to be unstable.
This is indeed the case.
But the mechanism for revealing self-interaction is rather subtle.

For simplicity, we restrict our attention to bosonic strings.
We recall the reader some elements of their description \cite{Green1987}, \cite{Polchinski1999}.
The points of a string are specified by spacetime coordinates $X^\mu$.
During its motion, the string sweeps out a two-dimensional surface in Minkowski space, 
$X^\mu=X^\mu(\sigma, \tau)$, called the {world sheet}.
The coordinates $\tau$ and $\sigma$ parameterize the {world sheet}: $\sigma$ labels the position of 
a point on the string and $\tau$ measures its time evolution.
World sheets are assumed to be {timelike}, smooth surfaces, that is, a two-dimensional plane tangent 
to the world sheet is spanned by a timelike and a spacelike vectors, ${\dot X}_{\mu}={\partial X_\mu}/{\partial\tau}$ 
and ${X'}_{\!\mu}={\partial X_\mu}/{\partial\sigma}$.
By analogy with the action for a particle, Eq.~(\ref{S-Planck-general}), which is proportional to 
the length of the world line, the action for a string is taken to be proportional to the area of the 
world sheet:
\begin{equation}
S=\int_{\tau_1}^{\tau_2}d\tau\int_{0}^{l}d\sigma\,{\cal L}\left({\dot X}, X'\right)\,,
\qquad
{\cal L}=-\frac{T}{2\pi}\,\sqrt{({\dot X}\cdot X')^2-{\dot X}^2{X'}^2}\,.
\label
{Nambu-action}
\end{equation}                        

A classical string moves so as to minimize the area of the world sheet, with initial and final positions
of the string being fixed,
\begin{equation}
\delta S=\!-\!\int_{\tau_1}^{\tau_2}\!d\tau\!\int_{0}^{l}\!d\sigma\!
\left(\frac{\partial}{\partial\tau}\frac{\partial {\cal L}}{\partial{\dot X}_\mu}
+\frac{\partial}{\partial\sigma}\frac{\partial {\cal L}}{\partial{X'}_{\!\mu}}\right)\!\delta X_\mu
+\int_{\tau_1}^{\tau_2}\!d\tau\!
\left(\frac{\partial {\cal L}}{\partial{X'}_{\!\mu}}\delta X_\mu \right)\!\Bigl\vert
^{\sigma=l}_{\sigma\!=0}=0\,.
\label
{variation-action-Nambu}
\end{equation}                        
Taking the boundary conditions
\begin{equation}
\frac{\partial {\cal L}}{\partial{X'}_{\!\mu}} 
=0\,
\quad
{\rm at}
\quad
\sigma=0,l\,, 
\label
{bound-action-Nambu}
\end{equation}                        
we arrive at the Euler--Lagrange equations 
\begin{equation}
\frac{\partial}{\partial\tau}\left[\frac{{X'}^{\mu}({\dot X}\cdot X')-
{\dot X}^\mu{X'}^2}{\sqrt{({\dot X}\cdot X')^2-{\dot X}^2{X'}^2}}\right]
+
\frac{\partial}{\partial\sigma}\left[\frac{{\dot X}^\mu({\dot X}\cdot X')-{X'}^{\mu}{\dot X}^2}
{\sqrt{({\dot X}\cdot X')^2-{\dot X}^2{X'}^2}}\right]=0\,,
\label
{Euler-Lagrange-Nambu}
\end{equation}                        
nonlinear partial differential equations for $X^\mu$.
This is not the whole story, however.
The change of variables 
\begin{equation}
\tau= 
F({\bar\tau},{\bar\sigma})\,,
\quad
\sigma= 
G({\bar\tau},{\bar\sigma})\,,
\label
{reparam-string}
\end{equation}                        
where $F$ and $G$ are smooth functions, leaves the action (\ref{Nambu-action}) invariant.
Transformations (\ref{reparam-string}) form the {gauge} group of the string. 
To eliminate the gauge freedom of the string, one may impose two gauge fixing conditions.
A convenient choice is 
\begin{equation}
{\dot X}\cdot X'=0\,,
\quad
{\dot X}^2+{X'}^2=0\,,
\label
{orthonormal-gauge}
\end{equation}                        
whose geometrical significance is that the coordinate lines $\tau=$ const and $\sigma=$ const are 
orthogonal and uniformly parametrized, hence the name ``orthonormal gauge''.
With the gauge (\ref{orthonormal-gauge}), the Euler--Lagrange equations (\ref{Euler-Lagrange-Nambu}) simplify 
\begin{equation}
{X''}_{\!\mu}-{\ddot X}_\mu=0\,.
\label
{field-eq-Nambu-cone-gauge}
\end{equation}                        
String coordinates in the orthonormal gauge obey the wave equation.
This result makes it clear that the string dynamics at the world sheet is devoid of self-interaction.
The same is true for superstrings.

The boundary conditions (\ref{bound-action-Nambu}) become
\begin{equation}
{{X'}_{\!\mu}}(\tau,0)={{X'}_{\!\mu}}(\tau,l)=0\,. 
\label
{Neumann-Nambu}
\end{equation}                        
These are {Neumann boundary conditions}.
Using (\ref{orthonormal-gauge}) and (\ref{Neumann-Nambu}), we obtain
\begin{equation}
{\dot{X}}^2=0\quad {\rm at}\quad \sigma=0,l\,,
\label
{free-ends-string-move}
\end{equation}                        
that is, end points of strings obeying this boundary conditions move at the speed of light.

Alternatively, one may adopt {Dirichlet boundary conditions}
\begin{equation}
{{X}^{\mu}}(\tau,0)={{X}^{\mu}}(\tau,l)=C^{\mu}\,,
\label
{Dirichlet-Nambu}
\end{equation}                        
which imply that $\delta X^\mu=0$ in the last term of (\ref{variation-action-Nambu}).
This term can also be set to zero if we impose periodic boundary conditions
\begin{equation}
{{X}_{\mu}}(\tau, 0) 
=
{{X}_{\mu}}(\tau, l)\,. 
\label
{bound-closed-Nambu}
\end{equation}                        
These relations are suitable for {closed} strings in the orthonormal gauge. 

A free open string can be coupled to an external electromagnetic field by adding an interaction 
term to the free action.
This term is to be chosen in a form preserving most, or, better still, all symmetries of the free 
action. 
The only Poincar\'e and gauge invariant expression is
\begin{equation}
\frac{e}{\pi}\int_{\tau_1}^{\tau_2}d\tau\int^{l}_{0}d\sigma\,{\dot X}_\mu\,{X'}_{\!\nu}\,F^{\mu\nu}(X)\,,
\label
{action-string-ext-field-}
\end{equation}                        
where $e$ stands for the electric charge of the string, and $F_{\mu\nu}=\partial_\mu A_\nu-\partial_\nu A_\mu$ 
is the external electromagnetic field.
Because
\[
{\dot X}_\mu\,\frac{\partial}{\partial X_\mu}\,A^\nu=\frac{\partial}
{\partial\tau}\,A^\nu\,,
\quad
{X'}_{\!\nu}\,\frac{\partial}{\partial X_\nu}\,A^\mu=\frac{\partial}
{\partial\sigma}\,A^\mu\,,
\]
(\ref{action-string-ext-field-}) equals 
\begin{equation}
-\frac{e}{\pi}\int_{\tau_1}^{\tau_2}  d\tau\,{\dot X}_\mu(\tau)\,A^\mu(X)
\Bigl\vert^{\sigma=l}_{\sigma=0}\,,
\label
{action-string-ext-field}
\end{equation}                        
plus two terms at $\tau=\tau_1$ and $\tau=\tau_2$, which do not contribute to the Euler--Lagrange 
equations.
It is clear from (\ref{action-string-ext-field}) that the charge of an open string is located at its 
ends.
As for a closed string, the charge may be distributed over its entire length. 

Adding  (\ref{action-string-ext-field}) to the free action leaves the Euler--Lagrange equations 
unchanged, but the Neumann boundary conditions (\ref{Neumann-Nambu}) become
\begin{equation}
2{e}\,{\dot X}^\nu F_{\mu\nu}(X)={T}{{X'}_{\!\mu}}
\quad
{\rm at}
\quad
 {\sigma=0,l}\,.
\label
{Neumann-Nambu-em}
\end{equation}                        

This consideration illuminates a peculiar feature of string interactions: open strings interact with 
each other at their ends.
On the quantum level, strings interact locally, without mediation of long-range fields.
Open strings may join when their ends contact, a single open string may spontaneously split into two 
pieces, or become closed, or emit a closed string, {\it etc.}
Joining and splitting are the basic interactions of strings.
This form of interaction respects all symmetries of free strings.

It is significant that joining and splitting of strings change the topological structure of their world sheets.

The question now arises of whether joining and splitting are realizable in the classical picture. 
It is highly improbable that free open strings can move in such a way as to bring their ends into 
contact with each other at some point of a three-dimensional arena.
To be more specific, the probability measure of such events is zero.
That is why joining of open strings may be ignored in the classical context.                     
Furthermore, a classical string seems to be unable of splitting.
Strings can be indefinitely stretched without any evidence of being favourably disposed towards 
splitting.
There is no elastic limit for extended objects governed by the action  (\ref{Nambu-action}).
The only dimensional parameter in the action is an overall factor $T/2\pi$ which defines the scale 
of length.
Classical strings are thus immune from compulsory splittings.
It remains to see whether classical strings can split at random.
A close look at the realization of Laplace's determinism in the classical picture shows that this 
is the case.

The classical is associated with Laplace's determinism.
Of course, classical statistical mechanics invokes probability theory, but the reason for this is 
that uncertainties of this description may be attributed to lack of knowledge of actual 
deterministic histories of macroscopic systems which have too many degrees of freedom to be 
completely controlled. 
 
Worthy of mention are also chaotic systems.
Although  chaotic dynamics is formulated by means of probability theory, ``chaotic'' is not to
be confused with ``random''.
Classical chaotic systems are governed by deterministic laws, but their histories are given by 
highly {tangled} trajectories.
Motions displaying extreme sensitivity to initial conditions are taken to be chaotic.
Complexity effects in the behavior of unstable classical systems are a major manifestation of the 
state of chaos.
The apparent indeterminism in the behavior of chaotic systems is then fictitious; it is due to 
imperfect knowledge of initial conditions.

The methods of statistical physics and chaotic dynamics have considerable utility in string problems 
of experimental interest, notably in cosmic strings  \cite{VilenkinShellard1994}, \cite{Anderson2003}.
However, in the discussion that follows, we omit these topics, and focus on the amenability of 
classical strings to splitting at random.

Laplace's determinism holds for a given classical system if the Cauchy problem for the dynamical equations of 
this system has a {unique} solution.
This requirement is generally believed to be imperative in classical physics. 
Nevertheless, there are classical systems that run counter to Laplace's determinism.
Examples of systems whose behavior can be regarded as truly random are given below \cite{Kosyakov2008c}.

Let two particles be moving towards each other along a straight line.
Having spent kinetic energy for overcoming the interparticle repulsion by the instant of their 
meeting, these particles merge into a single point aggregate.

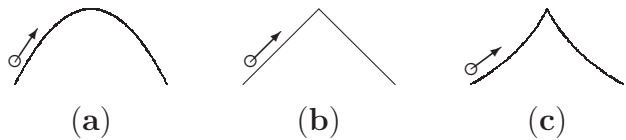
\begin{figure}[htb]
\begin{center}
\unitlength=1.00mm
\special{em:linewidth 0.4pt}
\linethickness{0.4pt}
\begin{picture}(90.00,30.00)
\bezier{180}(10.00,10.00)(20.00,30.00)(30.00,10.00)
\emline{40.00}{10.00}{1}{50.00}{20.00}{2}
\emline{60.00}{10.00}{3}{50.00}{20.00}{4}
\bezier{64}(70.00,10.00)(77.00,14.00)(80.00,20.00)
\bezier{64}(90.00,10.00)(83.00,14.00)(80.00,20.00)
\put(10.00,13.00){\vector(2,3){3.00}}
\put(10.00,13.00){\makebox(0,0)[cc]{$\circ$}}
\put(41.00,13.00){\vector(1,1){4.00}}
\put(41.00,13.00){\makebox(0,0)[cc]{$\circ$}}
\put(70.00,12.00){\vector(4,3){4.00}}
\put(70.00,12.00){\makebox(0,0)[cc]{$\circ$}}
\put(20.00,5.00){\makebox(0,0)[cc]{$({\bf a})$}}
\put(50.00,5.00){\makebox(0,0)[cc]{$({\bf b})$}}
\put(80.00,5.00){\makebox(0,0)[cc]{$({\bf c})$}}
\end{picture}
\caption{A particle moving to the top of a potential hill}
\label
{hill-g}
\end{center}
\end{figure}

Since our interest is with final stage of this head-on collision when the particles move slowly, 
the use of nonrelativistic approximation seems to be accurate.
The two-particle problem can be brought to a one-particle problem by introducing the relative 
coordinate $q=x_2-x_1$, the reduced mass $m=m_1 m_2/(m_1+m_2)$, and the potential energy $U(q)$.
The problem is then to describe the motion of this particle in $U(q)$, so that its velocity vanishes 
on its arrival at the top of the potential hill, see Fig.~\ref{hill-g}.
Let the top be located at $q=0$, $U_{\rm max}=U(0)$, and the instant that the particle comes to this
point be  $t=0$.
The fact that the velocity is vanishing at $q=0$ implies that the total energy is zero, 
\begin{equation}
E={\frac12}\,{m}{\dot q}^2+U(q)=0\,.
\label
{hill}
\end{equation}                                            
The time it takes for the particle to arrive at the top is therefore
\begin{equation}
t(q)=-\sqrt{\frac{m}{2}}\int_q^0 \frac{dx}{\sqrt{-U(x)}}\,.
\label
{t}
\end{equation}                                            
What is the behavior of the particle after its arrival at the top?
If the integral in (\ref{t}) diverges, then the question of the subsequent evolution does not arise 
because the ascent takes an infinite period. 
Such is the case for an analytical at $q=0$ function $U(q)$, say for $U(q)=-U_0\,q^2$.
However, the integral is finite for 
\begin{equation}
U(q)= -U_0\, q^{2(1-\alpha)}\,,\quad 0<\alpha<1\,,
\label
{deg}
\end{equation}                                            
\begin{equation}
U(q)= -U_0\,q^2\left(\ln q^2\right)^{2(1+\beta)}\,,\quad \beta>0\,,
\label
{ln}
\end{equation}                                            
\begin{equation}
U(q)= -U_0\,q^2\left(\ln q^2\right)^2\left[\ln\left(\ln q^2\right)^2\right]^{2(1+\gamma)}\,,
\quad\gamma>0\,,
\label
{lnln}
\end{equation}                                            
and the like.

Equation (\ref{hill}) is invariant under time reversal.
Furthermore, $q(t)=0$ is another solution to (\ref{hill}).
Therefore, if the climb takes a finite period of time, then an infinity of options is available: 
after staying at the top for an arbitrary period of time ${\cal T}$, the particle can start to 
descend, realizing the reversed order of events.
Analytically, 
\begin{equation}
q(t)=\cases{{\tt Q}\,(t) & $t<0$\,,\cr
0 & $0\le t\le {\cal T}$,\cr
{\tt Q}\,({\cal T}-t) & $t>{\cal T}$\,,\cr}
\label
{e-}
\end{equation}                                           
where ${\tt Q}\,(t)$ is the inverse of $t(q)$ defined in (\ref{t}). 
Going back to the initial two-particle problem, we see that the aggregate of two merged particles 
spontaneously disintegrates into its constituents after a lapse of an arbitrary period ${\cal T}$, 
and the particles move apart.

By the Picard theorem, the solution to the Cauchy problem for the ordinary differential equation 
(\ref{hill}) with the initial condition $q(0)=0$ is unique if the Lipschitz condition holds: 
$\sqrt{-U(q)}<C\,|q|$.
Clearly, this inequality fails for $U(q)$ given by any of Eqs.~(\ref{deg})--(\ref{lnln}). 

The potentials $U(q)$ which can be visualized as hills are thus divided into two classes: the $U$'s 
whose top is a point of unstable equilibrium in the conventional sense, and the $U$'s whose top is a 
point of an ``over-unstable'' equilibrium.
The state of equilibrium on the former top is kept until a small external perturbation occurs.
By contrast, the state of equilibrium on the latter top can be violated {spontaneously}, with no 
external cause.
The Lipschitz condition is sufficient but not necessary for stability against spontaneous violation.
Convergence of the integral in Eq.~(\ref{t}) may serve as a necessary condition for an 
unstable equilibrium to be classified as over-unstable.

One may disregard this argument for at least two reasons.
First, the potentials $U(q)$ shown in (\ref{deg})--(\ref{lnln}) are unlikely to bear on physical 
reality.
Second, time reversal is crucial for the spontaneous breakdown of equilibrium to occur.
But the dynamics of an accelerated charged particle is dissipative because it radiates electromagnetic 
energy.
A similar situation holds for particles carrying non-Abelian charges whose dynamics is also 
irreversible.
Therefore, the exact solution (\ref{e-}) is no longer valid for such particles.

Both objections are disproved if we turn to ${\mathbb R}_{1,1}$.
First, Eq.~(\ref{potential-2D}) shows that the time component of the retarded vector potential 
is given by $A_0=-e\,\vert\,q\,\vert$ which falls into the type of Eq.~(\ref{deg}) on putting 
$\alpha=\frac12$, see Fig.~\ref{hill-g}~(b).
Second, it was established in Sec.~\ref{d=2} that charged particles in ${\mathbb R}_{1,1}$ do not 
radiate, and all processes are locally reversible.
Therefore, the above indeterministic phenomenon is feasible here.

Indeed, let us choose the barycentric frame, assume that the particles have equal masses $m$ 
and charges $e$, and use the notation $w=e^2/m$.
Then the exact solution to the Cauchy problem for the set of equations governing a closed system  
``two charged particles plus electromagnetic field in ${\mathbb R}_{1,1}$'', having zero total 
energy, is given by 
\begin{equation}
z^\mu_1(s)
=\cases
{w^{-1}\left({\sinh}\,w(s-s^\ast),1-{\cosh}\,w(s-s^\ast)\right) & $s<s^{\ast}$\,,\cr 
\left(s-s^{\ast},0\right) & $s^{\ast}\le s<s^{\ast\ast}$\,,\cr 
w^{-1}\left(w{\cal T}+{\sinh}\,w(s-s^{\ast\ast}),1-{\cosh}\,w(s-s^{\ast\ast})\right)
&$s\ge s^{\ast\ast}$\,,\cr}  
\label
{z}
\end{equation}
\begin{equation}
z^\mu_2(s)
=\cases
{w^{-1}\left({\sinh}\,w(s-s^\ast),{\cosh}\,w(s-s^\ast)-1\right) & $s<s^{\ast}$\,,\cr 
z^\mu_1(s) & $s^{\ast}\le s<s^{\ast\ast}$\,,\cr 
w^{-1}\left(w{\cal T}+{\sinh}\,w(s-s^{\ast\ast}),{\cosh}\,w(s-s^{\ast\ast})-1\right)
&$s\ge s^{\ast\ast}$\,,\cr}  
\label
{-z}
\end{equation}
which describes two world lines $z^\mu_1(s)$ and ${\ z}^\mu_2(s)$ that coalesce at $s=s^\ast$, and 
separate at $s=s^{\ast\ast}=s^{\ast}+{\cal T}$. 
Here, $s^{\ast}$ and ${\cal T}$ are arbitrary positive constants. 
If $s^{\ast}$ and $s^{\ast\ast}$ are different and finite, then (\ref{z}) and (\ref{-z}) 
describe the history of the aggregate with finite life time. 
If $s^{\ast\ast}\to\infty$, we obtain the history of a stable aggregate formed at a finite instant.
In the limit $s^{\ast}\to -\infty$, the solution tells us that the aggregate arose at the infinitely 
remote past, and its decay occurs at a finite instant. 
For $s^{\ast}\to -\infty$ and $s^{\ast\ast}\to\infty$, the solution becomes a straight line
corresponding to an absolutely stable aggregate, and for $s^{\ast}=s^{\ast\ast}$, the solution 
describes an aggregate existing during a single moment. 
We thus have a continuum of solutions because ${\cal T}$ is arbitrary. 
In physical terms, the aggregate can disintegrate quite accidentally at any instant after its 
formation.

It is significant that the solution (\ref{z}) and (\ref{-z}) can be accomplished on condition that 
the total energy of the system is zero.
Clearly the initial data of the Cauchy problem describing the collision of two charged particles 
which merge into a single aggregate after their meeting constitute a measure zero set.
In contrast, the spontaneous disintegration of the aggregate is a physically tangible event.
The global dynamics of the electromagnetic realm in ${\mathbb R}_{1,1}$ is thus effectively 
irreversible: the formation of the discussed aggregates is highly improbable, whereas their 
disintegration may well take place.
The usual outcome of electromagnetic self-interaction in the classical picture, the violation of 
invariance under time reversal, has again been emerged.

This analysis gives us an inkling of the self-interaction mechanism peculiar to classical strings.
A similarity of the above two-particle system living in ${\mathbb R}_{1,1}$ to strings lies in the 
fact that the potential energy of a string varies linearly with distance between two somehow 
labelled points of this string.
A closed charged string may then be thought of as a chain whose elements are point aggregates of 
merged charged particles.
We assume that this assemblage is feasible, but drop out of sight a particular way for its making.  
This enables us to avoid the foregoing conundrum: to ensure that two colliding particles amalgamate 
in a single aggregate, their total energy must be exactly zero, and so the initial data of the 
corresponding Cauchy problem constitute a measure zero set.
To settle this question, we switch from a particular element of the chain to a continual set of 
identical aggregates constituting a closed string, and assume that the availability of this set is 
afforded by its cardinality.
We thus reason in opposite way by saying that if it is granted that a free closed charged string is 
capable of spontaneous splitting into two closed charged strings, then extending analytically the 
history of this disintegration back in time, according to Eqs.~(\ref{z}) and (\ref{-z}), we would 
restore the prehistory, and this would reinforce the statement that the rest energy of the aggregate 
whose splitting is responsible for the occurrence of a new string is indeed zero. 

Therefore, classical self-interaction of a closed charged string manifests itself in its capability 
of spontaneous creating other closed charged strings.

On the other hand, the mechanism underlying the history shown in (\ref{z}) and (\ref{-z}) is
unsound for joining of classical strings. 
To see this, imagine two closed charged strings coming in contact at some point $x^\mu$.
Consider the colliding points of the strings just before their meeting at $x^\mu$ in the 
barycentric frame.
The condition that zero total energy is necessary for these points to merge appears extremely 
exotic. 
To aggravate matters, the collision happens in ${\mathbb R}_{1,3}$ to which the exact solution 
(\ref{z}) and (\ref{-z}) is unrelated.

The vast disparity between joining and splitting of strings in the classical picture can be 
regarded as an effective violation of the property of time reversal invariance: the emission of closed charged 
strings is a unidirectional process.

\section{SELF-INTERACTION IN GENERAL RELATIVITY
}
\label
{GR}
Self-interaction of gravitating systems is much different from that of other classical gauge field 
systems.
Restricting our discussion to General Relativity (GR), we note that a closed gravitating system has 
infinitely large number of topological phases which unceasingly change, as exemplified by the 
occurrence of more and more black holes during the history of the Universe.
An important point is that these transmutations are irreversible.
Once converted into a black hole, a massive star can never regain its previous state.
It seems likely that the irreversibility is the only common property which is shared by gravitating 
and other gauge field systems.   

The other manifestation of self-interaction, the rearrangement, is missing from GR. 
Indeed, the effect of the rearrangement can be summarized in the local energy-momentum balance, 
Eq.~(\ref{local-balance-dr}).
Of particular interest is the case $f^\mu=0$, in which the rate of change of the energy-momentum of 
a dressed particle equals the negative of the emission rate, Eq.~(\ref{local-balance-f=0}).
This is tantamount to the statement that the action--reaction principle holds for the given system.
But the action--reaction principle is just the one incompatible with the equivalence principle 
underlying GR.
To see this, one should observe that a particle of mass $m$ is governed by the geodesic equation 
\begin{equation}
{\ddot z}^\lambda+\Gamma^\lambda_{~\mu\nu}{\dot z}^\mu{\dot z}^\nu=0\,,
\label
{geodesics}
\end{equation}
which is independent of $m$, while the field equation with the delta-function source 
\begin{equation}     
\left(R^{\mu\nu}-\frac12\,g^{\mu\nu}R\right)(x)=8\pi G_{\rm N} m\int_{-\infty}^\infty ds\,
{\dot z}^\mu(s)\,{\dot z}^\nu(s)\,\delta^{4}[x-z(s)]\,
\label
{Hilbert-Einstein} 
\end{equation}
shows that the greater is $m$, the stronger is the generated gravitational field.
Although the gravitational field exerts on every particle in a uniform way, no matter what is $m$,
the influence of particles on the state of the gravitational field is different for different $m$.
This is contrary to the action--reaction principle  \cite{Chubykalo2017}.

The failure of the action--reaction principle can be clarified using the old reasoning by Planck  
\cite{Planck1908} who regarded this principle as the rationale of momentum conservation.
By Noether's first theorem  \cite{Noether1918}, the energy-momentum is conserved due to invariance of the action 
under spacetime translations. 
But the requirement of translational invariance does not apply to curved manifolds, and so
gravitating systems are generally devoid of conserved momentum and energy.
Furthermore, the very construction of momentum and energy inspired by Noether's first theorem is 
no longer defined.

To avoid this conclusion, one normally turns to field-theoretic treatments of gravity, which go back 
to Rosen \cite{Rosen1940}. 
Such treatments are feasible if the gravitational field can be granted to be ``sufficiently weak'',
\begin{equation}
g_{\mu\nu}=\eta_{\mu\nu}+\phi_{\mu\nu}\,. 
\label
{weak gravity}
\end{equation}
Here, a second-rank tensor field $\phi_{\mu\nu}$ is assumed to be defined in a flat background ${\mathbb R}_{1,3}$ 
with the metric tensor $\eta_{\mu\nu}$, and has small components in every Lorentz frame, $|\phi_{\mu\nu}|\ll 1$.
The symmetry properties of ${\mathbb R}_{1,3}$ afford energy-momentum conservation through the 
standard Noether's argument.  
The field-theoretic treatment is valid until the mapping of $g_{\mu\nu}$ into $\phi_{\mu\nu}$ shown 
in (\ref{weak gravity}) is bijective and smooth, that is, every curved spacetime configuration, 
associated with a gravitational effect, can be smoothly covered with a single coordinate patch.
Bimetric theories of gravitation have been the objective of much recent research (for a review see
 \cite{Schmidt-MayStrauss2016}). 

GR leaves room for both weak and strong gravity. 
Strong gravitational effects are associated with great warpings of spacetime, that is, with great 
values of components of $R^\lambda_{~\mu\nu\rho}$.
However, a characteristic curvature whereby such warpings might be rated 
as ``drastic'' is absent from GR.
It seems reasonable to take a criterion for discriminating between strong and weak gravity as that
related to whether it is possible or impossible to alter the {topology} of spacetime  
\cite{Chubykalo2017}.  
According to this criterion, the gravitational field is weak if the topology of spacetime is 
identical to that of ${\mathbb R}_{1,3}$, otherwise it is strong.
The weakness of the gravitational field is thus a qualitative rather than quantitative concept.

The field equations (\ref{Hilbert-Einstein}) tell nothing about the topological properties of their 
solutions because differential equations are local in character.
A global solution can be formed by gathering its infinitesimal pseudoeuclidean fragments.
The topology of the resulting solution may differ from that of ${\mathbb R}_{1,3}$ if the assembly 
is subject to a restrictive  boundary condition.
To illustrate, we refer to the Schwarzschild metric  \cite{Schwarzschild1916},
\begin{equation}
ds^2=\left(1-\frac{r_{\rm S}}{r}\right)dt^2-\left(1-\frac{r_{\rm S}}{r}\right)^{-1}dr^2-r^2d{\Omega}\,,
\quad
\label
{Schwarzschild-metric}
\end{equation}                                          
where $r_{\rm S}={2G_{\rm N}M}$ is the Schwarzschild radius, and $d{\Omega}$ the round metric in a 
sphere $S^2$. 
A three-dimensional spacelike surface $\Sigma_3$ endowed with this metric has a twofold geometric 
interpretation.
First, it looks like a ``bridge'' between two otherwise Euclidean spaces, and, second, it may be 
regarded as the ``throat of a wormhole'' connecting two distant regions in one Euclidean space in 
the limit when this separation of the wormhole mouths is very large compared to the circumference 
of the throat  \cite{FullerWheeler1962}.

The price for the multiplicity of topologically distinct phases in GR is the absence of the Killing 
vector fields responsible for energy and momentum conservation.
Nevertheless, some models of the Universe can be equipped with the total energy-momentum $P^\mu$.
An ``island of matter surrounded by emptiness'' is a good case in point \cite{Misner1973}.
Whatever the topological setup of the island, spacetime is asymptotically flat, which provides a 
means for the Hamiltonian formulation of this system \cite{Arnowitt1960a}, \cite{Arnowitt1960b}, 
\cite{Arnowitt1961}. 
More specifically, one supposes that $g_{\mu\nu}$ approaches $\eta_{\mu\nu}$  at spatial 
infinity sufficiently rapidly, namely
\begin{equation}
g_{\mu\nu}=\eta_{\mu\nu}+O\left(\frac{1}{r}\right),
\quad
\partial_{\lambda}g_{\mu\nu}=O\left(\frac{1}{r^2}\right),
\quad
r\to\infty\,.
\label
{g-eta=1/r}
\end{equation}                                          
The second condition shown in Eq.~(\ref{g-eta=1/r}) is needed for the Lagrangian function 
\begin{equation}
L=\int d^3{\bf x}\,{\cal L}(t,{\bf x})\,
\label
{Lagr-}
\end{equation}                                          
with the Lagrangian density 
\begin{equation}
{\cal L}=\sqrt{-g}\,g^{\mu\nu}\left(\Gamma_{~\mu\sigma}^{\lambda}\,\Gamma_{~\nu\lambda}^{\sigma}
-\Gamma_{~\mu\nu}^{\sigma}\,\Gamma_{~\sigma\lambda}^{\lambda}\right)\,
\label
{Lagr-function}
\end{equation}                                          
should be convergent.
Note that the volume integral in (\ref{Lagr-}) diverges for the Schwarzschild solution expressed in 
terms of the original Schwarzschild 
coordinates, appearing in (\ref{Schwarzschild-metric}), because ${\cal L}=O(1)$ as $r\to\infty$.
But the use of isotropic coordinates (see, {\it e.~g.}, \cite{LandauLifshitz1971}) which convert the 
Schwarzschild metric (\ref{Schwarzschild-metric}) to   
\begin{equation}
ds^2= 
\left({1-\frac{r_{\rm S}}{4\varrho}}\right)^2\left({1+\frac{r_{\rm S}}{4\varrho}}\right)^{-2}dt^2-
\left(1+\frac{r_{\rm S}}{4\varrho}\right)^4\left(d\varrho^2+\varrho^2 d\Omega\right)
\quad
\label
{Schwarzschild-isotropic}
\end{equation} 
gives ${\cal L}=O(1/\varrho^4)$.
This ensures the convergence of the Lagrangian (\ref{Lagr-}). 
This example shows that both the asymptotic flatness  
\begin{equation}
R^\alpha_{~\beta\gamma\delta}\to 0\,, 
\quad 
r\to \infty\,,
\label
{flat}
\end{equation}                      
and a good choice of coordinates share in the responsibility for the convergence of additive 
quantities such as the Lagrangian and Hamiltonian.                                          

Being armed with the Hamiltonian formulation of Arnowitt--Deser--Misner \cite{Arnowitt1960a}, 
\cite{Arnowitt1960b}, \cite{Arnowitt1961}, one 
would expect that the total energy-momentum $P^\mu$ is unambiguously defined.
Let us verify if this expectation is correct restricting ourselves to $E=P^0$ for simplicity.

The total energy   
\begin{equation}
E=\int d^3{\bf x}\,{\cal H}\left(t,{\bf x}\right),
\label
{Mass-}
\end{equation}                                          
with ${\cal H}$ being a cumbersome construction immaterial for the present discussion (see, 
{\it e.~g.,} \cite{Faddeev1982}) can be 
transformed \cite{ReggeTeitelboim1974} to a simple expression,  
\begin{equation}
E=\frac{1}{16\pi}\oint dS_j\left(\frac{\partial}{\partial x_i}\,g_{ij}-
\frac{\partial}{\partial x_j}\,g_{ii}\right).
\label
{Total-energy}
\end{equation}                                          
Here, the integral is taken over a 2-dimensional surface at spatial infinity.

Schoen and Yau \cite{SchoenYau1979} and Witten \cite{Witten1981} were able to demonstrate that an 
isolated gravitating 
system having non-negative local mass density has non-negative total energy $E$.
Applying the surface integral (\ref{Total-energy}) to the Schwarzschild 
configuration generated by a point particle of mass $m$ they get the conclusion that 
\begin{equation}
E=m\,.
\label
{Mass-Schwarzschild}
\end{equation}                                          

Meanwhile we should take proper account of the freedom to foliate spacetime into spacelike
sections consistent with asymptotic symmetries of the manifold.
Is it possible to relax the asymptotic condition (\ref{g-eta=1/r}) in such a way as to preserve 
the asymptotic flatness (\ref{flat}) and the convergence of pertinent additive quantities in this 
Hamiltonian formulation?
To be more precise, let us proceed from the metric $g_{\mu\nu}$ with the asymptotic behavior 
(\ref{g-eta=1/r}), and map the initial grid of spatial coordinates, $\{x_i\}$, into a new one, 
$\{{\tilde x}_i\}$,
\begin{equation}
x_i={\tilde x}_i\left[1+ f({\tilde r})\right],
\label
{diffeomorphism-gen}
\end{equation}                                          
where $f$ is an arbitrary regular function subject to the following conditions:
\begin{equation}
f({\tilde r})\ge 0\,,
\quad
\lim_{{\tilde r}\to\infty} f({\tilde r})=0\,,
\quad
\lim_{{\tilde r}\to\infty} {\tilde r}\,f'({\tilde r})=0\,,
\label
{conditions_on_f}
\end{equation}                                          
and
\begin{equation}
\frac{\partial r}{\partial{\tilde r}}=1+f({\tilde r})+{\tilde r}f'({\tilde r})>0\,.
\quad
\label
{reversibl}
\end{equation}                                          
Condition (\ref{reversibl}) is necessary and sufficient for the mapping (\ref{diffeomorphism-gen}) to 
be invertible because
\begin{equation}
J=\det \left(\frac{\partial x}{\partial{\tilde x}}\right)=
\left[1+f({\tilde r})\right]^2\frac{\partial r}{\partial{\tilde r}}\ne 0\,.
\quad
\label
{Jacobian_ne_0}
\end{equation}                                          

To illustrate, we refer to a mapping proposed in \cite{DenisovSolov'ev1983}, 
\begin{equation}
f({\tilde r})=2\,\alpha^2\,\sqrt{\frac{l}{{\tilde r}}} 
\left[1-\exp\left(-\frac{\epsilon^2\,{\tilde r}}{l}\right)\right],
\label
{diffeomorph}
\end{equation} 
where $\alpha$ and $\epsilon$ are arbitrary nonzero numbers, and $l$ is an arbitrary parameter of 
dimension of length.
This is a bijective monotonic regular mapping $r\to{\tilde r}$ 
which becomes $1$ as $\epsilon\to 0$.
The leading asymptotic terms of spatial components of the metric and those of 
the Christoffel symbols are 
\begin{equation} 
g_{ij}=\delta_{ij}+O\left({\tilde r}^{-1/2}\right),
\quad 
\Gamma_{~jk}^{i}=O\left({\tilde r}^{-3/2}\right),
\quad {\tilde r}\to\infty\,.
\label
{g-eta=1/sqrt_r}
\end{equation} 
It follows that the Lagrangian density behaves as
\begin{equation}
{\cal L}=O\left({\tilde r}^{-7/2}\right),
\quad 
{\tilde r}\to \infty\,,
\label
{Hamiltonian-transf-behavior}
\end{equation}                                          
which provides the convergence of the volume integral (\ref{Lagr-}) and other additive quantities of
this kind.
Note also that the asymptotic flatness, Eq.~(\ref{flat}), is still the case.

The mapping (\ref{diffeomorphism-gen}) with $f$ defined in (\ref{diffeomorph}) is instructive to 
apply to the Schwarzschild metric written in terms of isotropic coordinates (\ref{Schwarzschild-isotropic}).
Denisov and Solov'ev \cite{DenisovSolov'ev1983} showed that the total energy of the Schwarzschild 
configuration generated by a point particle of mass $m$ takes any positive values, greater 
than, or equal to $m$, when $\alpha^2$ runs through ${\mathbb R}_{+}$,
\begin{equation}
E=m\left(1+\alpha^4\right).
\label
{Mass}
\end{equation}                                          

We thus see that the total energy of gravitational systems with nontrivial topological contents 
depends on a particular foliation of spacetime.
The same is true for the total momentum of such gravitational systems. 

The occurrence of ill-defined additive quantities in GR closely parallels that in the 
Banach and Tarski theorem \cite{BanachTarski1924}.
The theorem states that a ball in ${\mathbb R}_3$ can be split into a finite disjoint 
subsets which can then be put back together through continuous movements of the pieces, without 
changing their shape and without running into one another, to yield a ball twice as large as the 
original, or, in more abstract terms, arbitrary bounded sets with nonempty interiors in  
${\mathbb R}_3$ are equidecomposable. 
Both the Banach--Tarski decomposition-reassembly and the formation of a black hole derive from a 
topological reorganization by which the three-dimensional measures of the geometrical layouts 
become poorly defined. 
The measure appearing in the Banach--Tarski theorem is the ordinary volume of the balls (more 
precisely, Lebesgue measure), while the measure in the gravitational energy-momentum problem is 
the measure of integral quantities like that given by Eq.~(\ref{Mass-}).
When turning to the surface integral for calculation of the total energy, Eq.~(\ref{Total-energy}), 
there arises the situation which may be likened to that concerning paradoxical duplicating or 
enlarging spheres discovered by Hausdorff \cite{Hausdorff1914}.

The usual objection against addressing our concern with the Banach--Tarski paradox in the physical 
context is that material bodies are made of atoms; the partitioning procedure of a mathematically 
continuous ball is unrelated to their actual disintegration.
Therefore, it is impossible to cut up a 
pea into finitely many pieces and then reassemble them to form a Sun-sized ball.
However, this objection overlooks one important instance, black holes.
Each isolated stationary black hole is completely specified by three parameters: its mass $m$, 
angular momentum $J$, and electric charge $e$.
Whatever the content of a system which collapses under its own gravitational field, the exterior of 
the resulting black hole is described by a Kerr--Newman solution.
All individual geometric features of the collapsing system disappear in the black hole state 
\cite{Heusler1996}. 
Furthermore, the event horizon which is meant for personifying the black hole is stripped of the 
grain structure that was inherent in its precursor, the collapsing system, and the black hole
appears as a perfect object. 

It is therefore interesting to enquire into why the values of the functionals (\ref{Mass-}) and (\ref{Total-energy}) are 
foliation-dependent for black holes in the light of the analyses which are lumped together as the 
``Banach--Tarski theorem'' (the subject has been detailed in \cite{Wagon1994}). 

A central idea in obtaining a paradoxical decomposition of a set is to get such a decomposition in 
an isometry group acting on the set, and then transfer it to the set. 
If a bounded set can be decomposed in a paradoxical way with respect to a group $G$, then $G$ 
contains {free} subgroups.
In particular, a ball in ${\mathbb R}_3$ is SO$(3)$-paradoxical because SO$(3)$ acts as a free 
non-Abelian isometry group.
On the other hand, Banach's theorem \cite{Banach1923} states that no paradoxical decompositions 
exist in ${\mathbb R}$ and ${\mathbb R}_2$.   
The class of groups whose actions preserve isometry-invariant, finitely additive measures of the 
bounded sets, the so-called amenable groups, are found to be fairly extensive, containing all 
solvable groups.  
A subclass of this class of particular interest is comprised of Abelian groups. 

The interplay between measure theory and group theory discovered in the framework of the 
Banach--Tarski theorem mirrors that in GR in the following way \cite{Chubykalo2017}.
The failure of the action-reaction principle suggests that the dynamics encoded by Eqs.~(\ref{geodesics}) 
and (\ref{Hilbert-Einstein}) is unstable.
This instability is a prerequisite to the formation of topologically nontrivial manifolds, in
particular those associated with black holes.
Much of the current interest in physics of black holes refers to the model of island Universe 
satisfying the requirement of asymptotic flatness, Eq.~(\ref{flat}). 
However, the asymptotic flatness leaves room for a wide range of foliations of a curved 
spacetime manifold compatible with the condition that all additive observables be convergent.
The grids of coordinates realizing these foliations can be interconverted by diffeomorphisms which 
respect asymptotic symmetry of the manifold.
The effect of rearranging gravitational degrees of freedom is entrusted to the properties of the 
group of asymptotic symmetry.
It is generally believed that the Bondi--Metzner--Sachs group \cite{Bondi1962}, \cite{Sachs1962a},
\cite{Sachs1962b} 
 is just this group.
What counts is that the Bondi--Metzner--Sachs group is a non-Abelian isometry group. 
This is the reason why the energy--momentum of gravitational systems with nontrivial topological 
contents is not uniquely defined, in particular the total energy of the Schwarzschild configuration 
is controlled by the foliation parameter $\alpha$, Eq.~(\ref{Mass}).

\section*{ACKNOWLEDGMENTS}
\addcontentsline{toc}{section}{ACKNOWLEDGEMENTS}
\label
{ACKNOWLEDGMENTS}
The first impetus to summarize the state of the art in the self-interaction problem of classical 
gauge theories came to me from Asim Barut in 1994.
The idea of writing a review of this kind came up again in my discussions with Rudolf Haag in 1998.
He advised me to extend the project to cover all attendant issues so that the text would meet the 
needs of senior students.
Furthermore, he wrote a letter of support to the International Science and Technology Center with 
recommendations to allot funds for writing my book \cite{Kosyakov2007}.
The Springer Verlag published the book in 2007, which was a partial implementation of the initial 
plan. 
Both Barut and Haag found the notion of rearrangement particularly promising.
I would like to dedicate the present paper to the memory of these remarkable personalities and 
outstanding theorists.

\end{document}